\documentclass[aps,prx,twocolumn,english,showpacs,letterpaper,superscriptaddress]{revtex4-1}

\usepackage[dvips]{epsfig}
\usepackage[colorlinks=true,citecolor=blue,linkcolor=blue]{hyperref}

\usepackage{amssymb}
\usepackage{graphicx}
\usepackage{subfigure}
\usepackage{array}

\usepackage{epsfig}

\usepackage{amsmath}
\usepackage{xcolor}
\usepackage{float}
\usepackage{mathrsfs}
\usepackage{indentfirst}
\usepackage{textcomp}
\usepackage{comment}
\usepackage{mathtools}

\usepackage{multirow}   

\newcommand{\red}[1]{\textcolor{red}{#1}}

\newcommand{\gray}[1]{\textcolor{gray}{#1}}
\newcommand{\ignore}[1]{}
\newcommand{\mater}{{EuZnSb$_2$}}
\newcommand{\zhat}{{$\hat{z}$}}
\newcommand{\xhat}{{$\hat{x}$}}
\newcommand{\Neel}{N\'{e}el~}
\newcommand{\Pcal}{\mathcal{P}}
\newcommand{\Tcal}{\mathcal{T}}

\newcommand{\COMMENTED}[1]{}
\newcommand{\REMARKS}[1]
{
{ \color{red}{\textbf{ {[#1]} }} }
}

\newcommand{\be}{\begin{equation}}
\newcommand{\ee}{\end{equation}}
\newcommand{\bea}{\begin{eqnarray}}
\newcommand{\eea}{\end{eqnarray}}

\newcommand{\s}{\sigma}

\newcommand{\ri}{\mbox{i}}
\newcommand{\re}{\mbox{e}}

\newcommand{\titledef}{Topological Antiferromagnetic Semimetal for Spintronics: A Case Study of a Layered Square Net System \mater}

\begin{document}

\title{\titledef}

\author{Niraj Aryal}
\email{naryal@bnl.gov}
\affiliation{Condensed Matter Physics and Materials Science Division, Brookhaven National Laboratory, Upton, New York 11973, USA}
\author{Qiang Li}
\affiliation{Condensed Matter Physics and Materials Science Division, Brookhaven National Laboratory, Upton, New York 11973, USA}
\affiliation{Department of Physics, Stony Brook University, Stony Brook, New York 11794, USA}
\author{A. M. Tsvelik}
\affiliation{Condensed Matter Physics and Materials Science Division, Brookhaven National Laboratory, Upton, New York 11973, USA}
\author{Weiguo Yin}
\email{wyin@bnl.gov}
\affiliation{Condensed Matter Physics and Materials Science Division, Brookhaven National Laboratory, Upton, New York 11973, USA}

\date{\today}

\begin{abstract}
 
    We use the first principles and effective Hamiltonian methods to study the electronic structure and magnetic properties of a recently synthesized layered antiferromagnetic square net topological semimetal \mater ~\cite{WangEuZnSb2}. The main message of the paper is that effects of small changes in the band structure produced by the  magnetic ordering and changes in the  orientation of the \Neel vector are amplified in such transport properties as the spin Hall conductivity. We predict that the effects of the broken symmetry introduced by the ordering of the \Neel vector, being very weak in the bulk,  are pronounced in the surface electronic dispersion,  suggesting that surface probes may be more suited to measure them.
    The coexistence of the magnetism with  many other competing phases make this material interesting and possibly useful for quantum spintronics applications.

\end{abstract}


\maketitle

\section{Introduction}

Interplay of the topological bands and competing magnetic orders could result in novel physical properties such as large anomalous Hall effect~\cite{MagneticWeyl_Wang_NatureComm2018} and  axion electrodynamics~\cite{AxionMTI_SCZhang_NaturePhys2010}. It also presents a possibility to use magnetic reordering to manipulate electronic transport~\cite{TAFMSpin_Smejkal_NaturePhysics2018}.  The recent discoveries of magnetic topological insulators~\cite{AFMTI_Otrokov_Nature2019}
and magnetic Dirac~\cite{DiracAFM_Maca_JMMM2012,DiracAFM_Zhang_NaturePhys_2016,AFM2D_Niu_PRL2020} and Weyl~\cite{PyrochloreIridates_Wan_PRB2011,MagneticWeylHgCrSe_Xu_PRL2011,Mn3Sn_AFMWeyl_NJP2017,HighThroughput_Xu_Nature2020} semimetals have triggered a flurry of research activities on this topic~\cite{ReviewMTSM_Zho_npjCM2019}. In particular, antiferromagnetic (AFM) systems with broken parity ($\Pcal$) and time reversal ($\Tcal$) symmetries but unbroken $\Pcal\Tcal$ symmetry have attracted a lot of interests recently for novel effects such as the electrical control
of AFM magnetization \cite{SwitchingAFM_Wadley_Science2016,RashbaEdelsteinEffect_NatureComm_Salemi2019} and the Dirac band topology \cite{SpinOrbitTorqueCuMnAs_Jungwirth_PRL2017,DNLSpintronics_Shao_PRL2019}.

After Young and Kane proposed the existence of topological nodal Fermions in square net motifs~\cite{2DDirac_YoungKane_PRL2015}, different variants of the square net topological materials have been studied extensively ~\cite{ZrSiS_Schoop_NatureComm2016,TopologySquareNets_AnnualReview_Schoop2019}. 
Recently, antiferromagnetic semimetals consisting of  strongly correlated 4\textit{f}-electrons in the 111 family of type \textit{Ln}SbTe [\textit{Ln}=lanthanide elements]~\cite{CeSbTe_AFM_Schoop_Science2018,GdSbTe_HosenNeupane_SciReports2018} 
and 112 family of type \textit{Ln}Mn(Bi,Sb)$_2$ have been reported to host topological Fermi surface similar to the well known ZrSiS family of materials
~\cite{YbMnBiDiracAFM_Wang_PRB2016,YbMnSbAFMDirac_PRB2018,EuMnBiDiracAFM_May_PRB2014,EuMnSbDiracAFM_Yi_PRB2017}.
 It is known that when the magnetic atoms directly contribute to the formation of the conduction bands as in such transition metal antiferromagnetic Dirac semimetal (AFM-DSM) systems as CuMnAs, Mn$_3$Ge {\it etc}., the magnetic ordering and in particular, the orientation of the magnetic moments can bring subtle changes in the electronic structure and the related transport properties~\cite{DNLSpintronics_Shao_PRL2019,NLAHENeel_Shao_PRL2020}.
Such subtle changes in the band topology and associated transport signatures  in AFM-DSM caused by   changes in  the orientation of the magnetic moments may be useful for spintronics applications such as  low power electronics and magnetic memory devices~\cite{AFMSpintronics_Nunez_PRB2006, AFMSpintronics_Shick_PRB2010}.

An influence of magnetism on the electronic dispersion and transport is less explored in the  afore-mentioned $f$-electron square net systems, where it may also lead to interesting effects. 
However, the conduction bands in these systems are formed by the $p_x$-$p_y$ orbitals of the (Bi, Sb) square nets weakly hybridized with the \textit{f}-electron bands lying far away from the Fermi level.
This makes a connection between the $f$-electron magnetism  and the conduction bands as well as the associated transport anomalies less  obvious. Given the potential advantages of \textit{f}-electron systems over conventional semiconductors for spintronics applications~\cite{SHE_Kontani_JPSJ2008,Edelstein_Robert_PRB2018} and the availability of rich material pool and magnetic properties obtained by varying \textit{Ln} elements~\cite{CDWMag_Lei_AQT2019}, it is important to study the systems where the itinerant electrons coexist with the localized ones. 

Recently, Wang \textit{et. al.} have reported a discovery of a layered \textit{4f} square net material {\mater}~\cite{WangEuZnSb2}. 
This material is the zintl cousin of the more famous Mn-based 112 phases.
Because of the unpaired \textit{4f}-electrons, it orders antiferromagnetically  with a \Neel temperature (T$_N$) of 20K.
Density functional theory calculations showed the presence of the extended Fermi surface formed by the $p$-electrons. 
In this paper, we study in detail the electronic structure of various 
AFM phases of this material. We aim to establish if and how the orientation of the \Neel vector influences the band topology and related transport properties.
We find that 
magnetic orderings introduce small but non-negligible corrections in the gap size across the Fermi surface which  have consequences in the Berry curvature related 
transport properties.
More importantly, we find that depending on the orientation of the \Neel vector, different crystalline symmetries are broken globally. 
Such broken symmetries are manifested in the electronic structure and transport properties. 
Our findings suggest that systems with coexisting itinerant and localized $f$-electrons  can be useful platforms for topological spintronics applications and more studies along this direction are necessary.

The organization of this paper is as follows.
In Sec.~\ref{Sec:Computation}, we present the details of our computational methods.
In Sec.~\ref{Sec:Results}, we present our results and discuss them in detail.
Finally, in Sec.~\ref{Sec:Conclusion}, we present our conclusion
and future outlook. The derivation of the tight-binding model and the effective Kondo exchange Hamiltonian is relegated to the Appendix.

\section{Computational Details}
\label{Sec:Computation}
The density-functional-theory (DFT) calculations were done using Wien2k DFT package~\cite{Wien2k2021}. The basis size was determined
by $R_\textrm{mt}K_\textrm{max}$ = 7 and the Brillouin zone was sampled with a regular $18 \times 18 \times 3$ mesh containing 162 irreducible $k$ points to achieve energy convergence of 1 meV. A 10,000 $k$-point
mesh was used for the Fermi surface calculations.
Some of the calculations, especially in the paramagnetic phase were verified using Quantum Espresso (QE)~\cite{QE} package.
Perdew-Burke-Ernzerhof (PBE) exchange-correlation functional~\cite{PBE} within the generalized gradient approximation (GGA) were used in all the calculations.  The GGA + $U_{\textrm{eff}}$ method was used to handle the Eu $4f$ orbitals.
$U_{\textrm{eff}}$ of 6 eV was chosen in our calculations~\cite{EuN_Pickett_PRB2005,EuExchange_Pickett_JPSJ2005,Eu_Larson_IOP2006,Ce_PRM_Zhang2017}; however we have also verified that the results presented here remain robust for a large range of $U_{\textrm{eff}}$ values. The spin-orbit coupling (SOC) was treated in the second variation method.
The spin Hall conductivity calculations were done using  Wannier90 software~\cite{Wannier902014,SHCWannier_Qiao_Zhao_PRB2018} by taking 80$\times$80 Wannierised Hamiltonian. All Eu $\textit{4d, 4f}$ orbitals and Sb $\textit{5s, 5p}$ orbitals were used in the Wannierisation procedure in order to accurately reproduce the DFT bands in the energy window from $-1$ to 1 eV.


\section{Results and Discussion}
\label{Sec:Results}
\subsection{Crystal Structure}

\begin{figure}[t]
    \begin{center}
\hskip -0.05 in
        \subfigure[]{
            \includegraphics[width=0.20\textwidth]{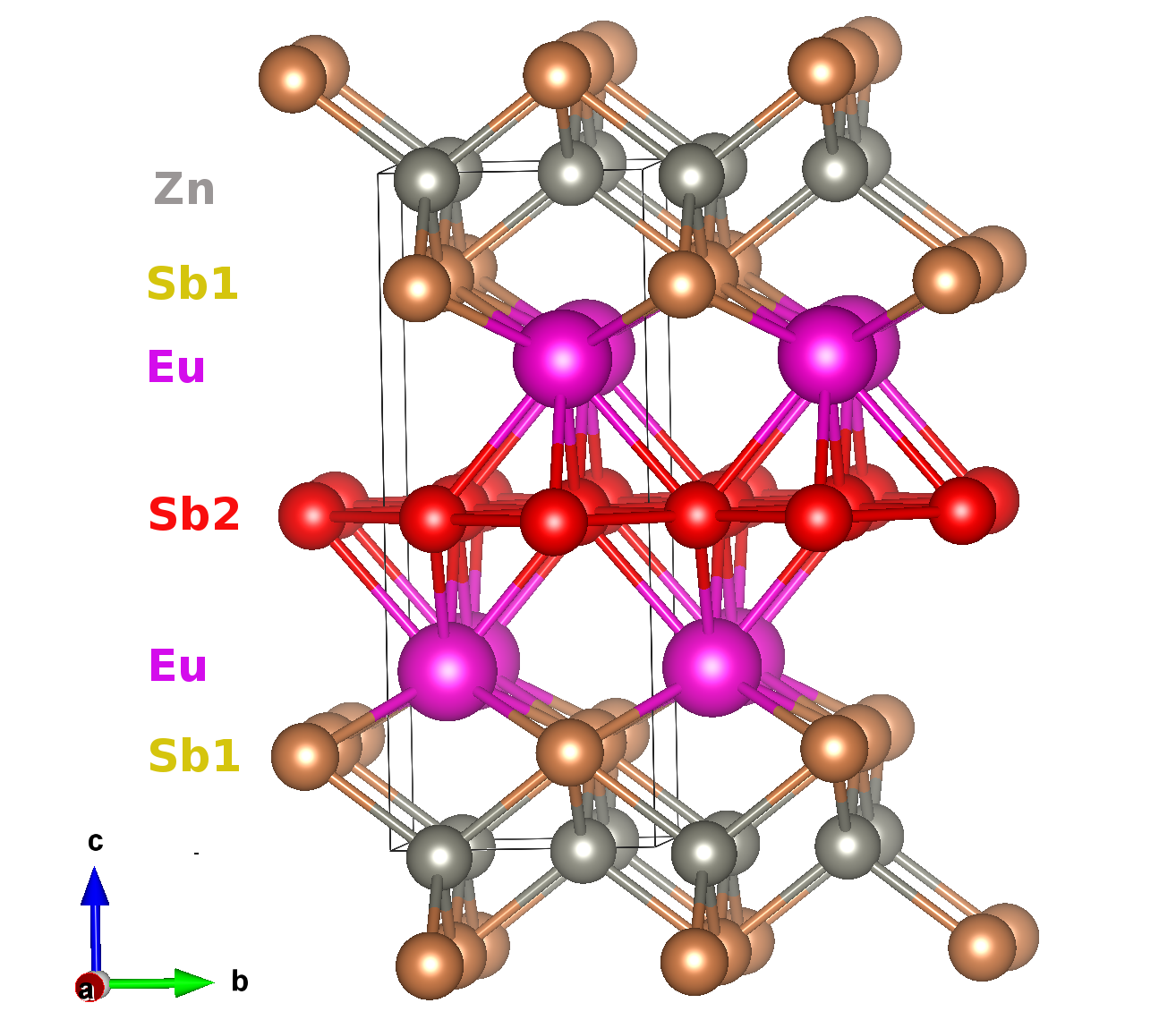}
        } 
\hskip -0.05 in
        \subfigure[]{
            \includegraphics[width=0.20\textwidth]{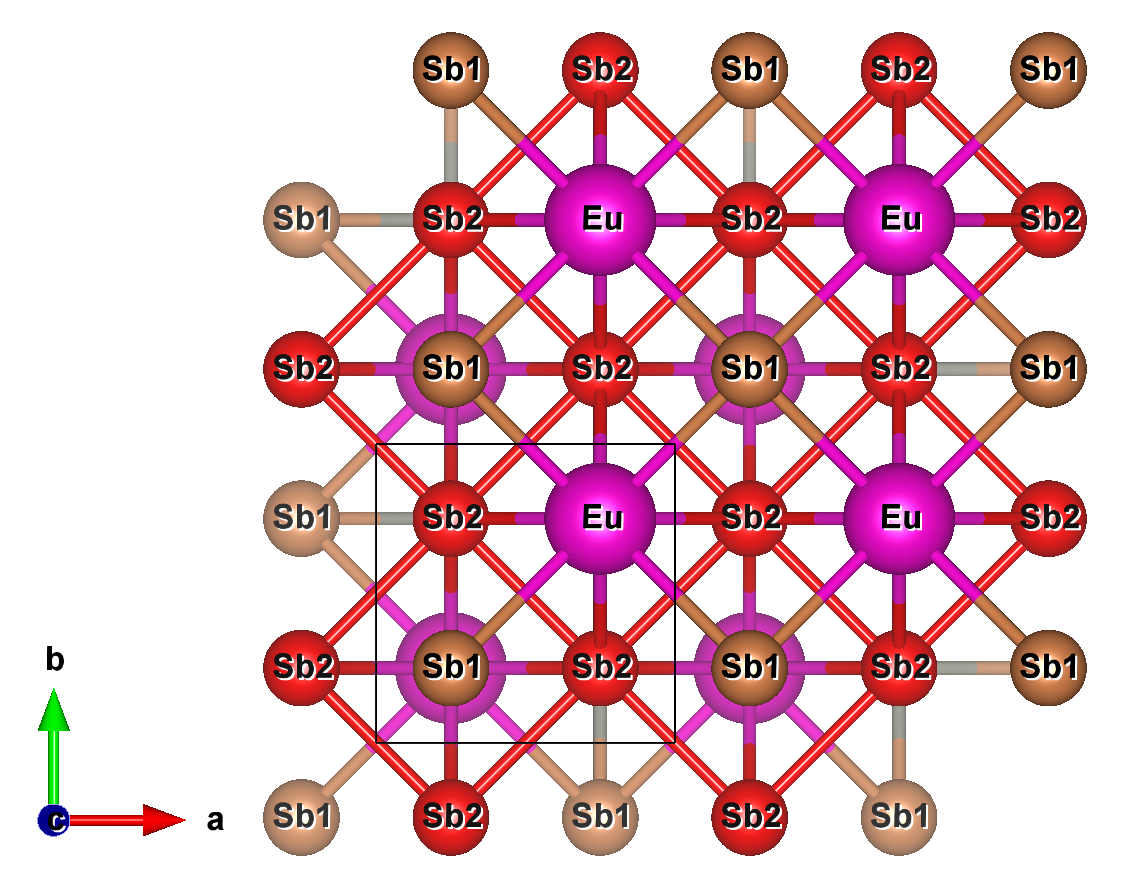}
        }
            \end{center}
            \vspace{-0.2in}
            \caption{The unit cell of {\mater}. Fig. (a) shows the stackings of the layered square lattices along the \textbf{c}-axis and Fig. (b) shows the projection onto the \textbf{a}-\textbf{b} plane and the denser ($\sqrt 2 \times \sqrt 2$) square lattice of Sb2 atoms. 
            Eu atoms above and below Sb2 atoms occupy the interstitial site of the Sb2 square lattice and are related by the glide (or inversion) symmetry; Zn atoms occupy the same site as Sb2 atoms.
        }
        \label{fig:UnitCell} 
\end{figure}

\mater~ is a layered square net material in the space group $P4/nmm$ (no. 129) similar to the well-known nodal-line family of materials of ZrSiS~\cite{WangEuZnSb2}.
The crystal structure of \mater~ is shown in Fig.~\ref{fig:UnitCell}.
The unit cell consists of stacking of square lattices of Eu, Zn and two types of Sb atoms (called Sb1 and Sb2 here) along the \textbf{c}-direction in the arrangement of ---Zn-Sb1-Eu-Sb2-Eu-Sb1-Zn---.
Sb2 and Zn atoms form a denser  ($\sqrt 2 \times \sqrt 2$) square lattice (also known as $4^4$ square lattice in the crystallographic community~\cite{squarenets_TremelJACS1987}), with 2 atoms in each 2D square plane whereas Sb1 and Eu atoms form a less denser square lattice with just 1 atom in each 2D square plane.
Sb2 and Zn atoms occupy the same site when projected on the \textbf{a}-\textbf{b} plane 
whereas Eu (and Sb1) atoms above and below Sb2 atoms occupy the interstitial site of the 4$^4$ lattice and are related by inversion (or a glide ) symmetry.

\subsection{Non-magnetic phase and Wannier tight-binding analysis}

\begin{figure}[htb]
    \begin{center}
        \subfigure[]{
            \includegraphics[width=0.22\textwidth]{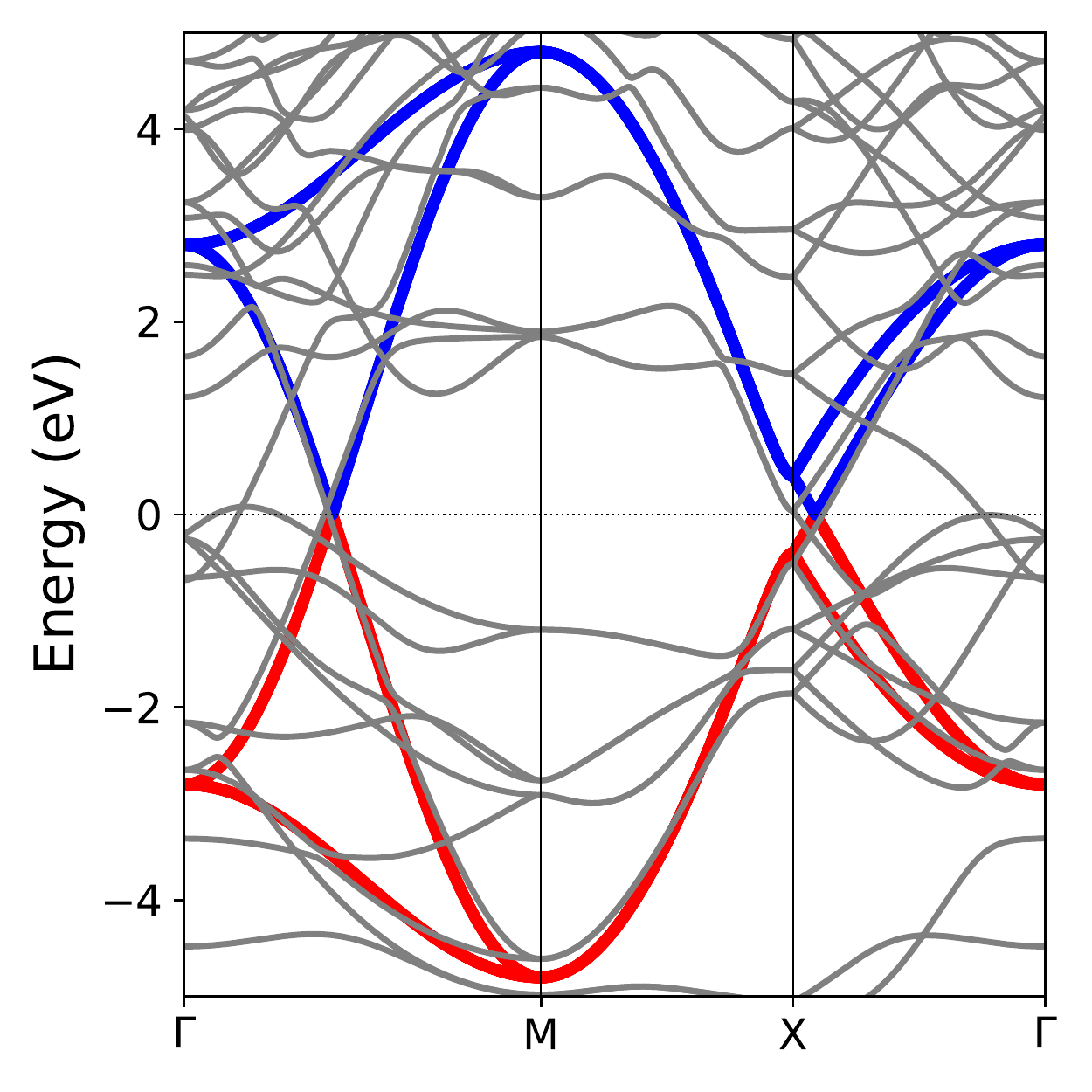}
        } 
\hskip -0.05 in
        \subfigure[]{
            \includegraphics[width=0.22\textwidth]{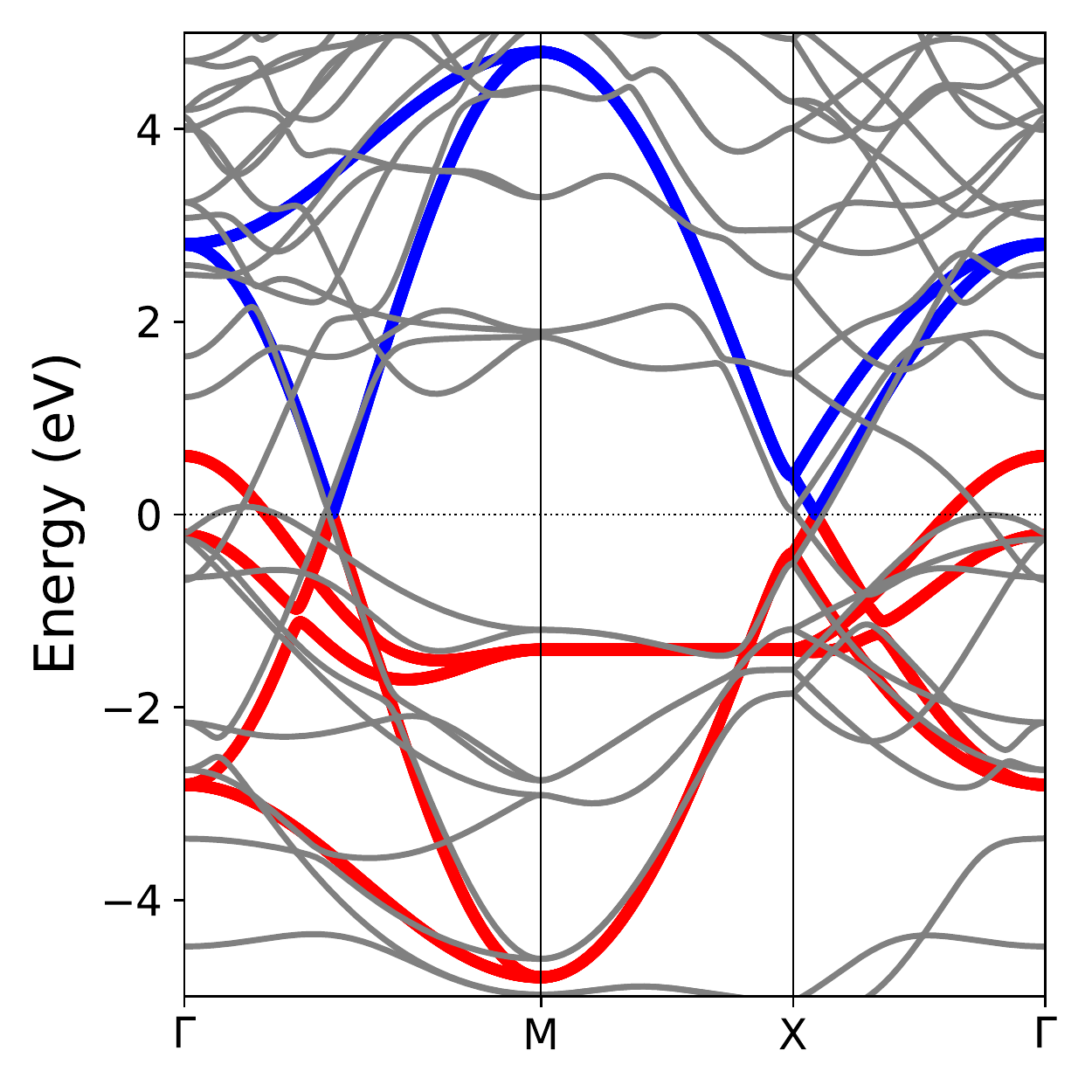}
        } 
\vskip -0.05 in
        \subfigure[]{
            \includegraphics[width=0.45\textwidth]{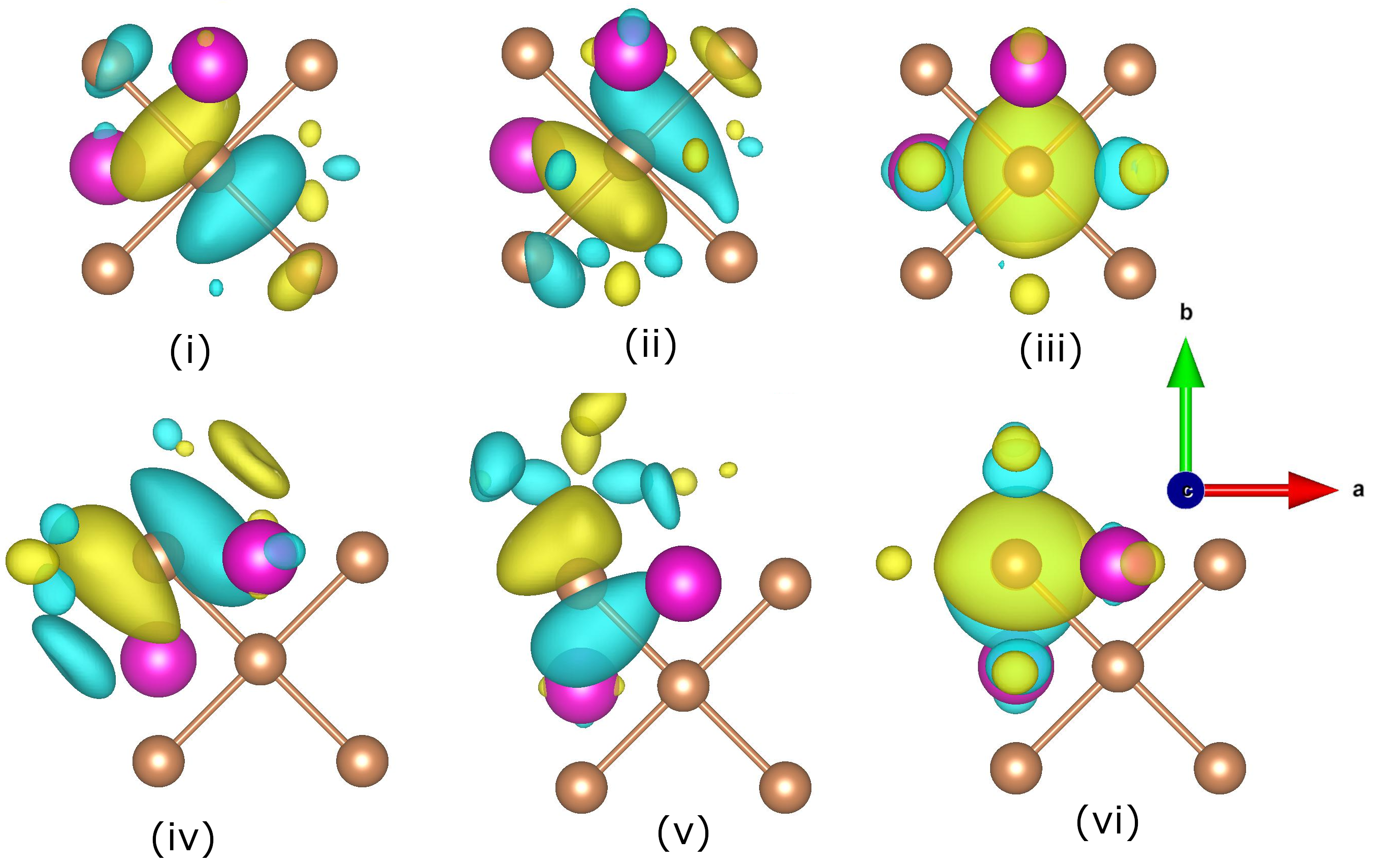}
        } 
            \end{center}
            \caption{ Comparison between the DFT calculated bands (gray lines) with (a) 4 and (b) 6 band TB Hamiltonian eigenvalues (red-blue dots) for the paramagnetic phase without SOC obtained using the parameters in Table~\ref{table:TightBindingParams}.
            In (c), we show the 6 $p_x$, $p_y$ and $p_z$-like Wannier orbitals used in the description of the 6-band TB Hamiltonian.
        }
        \label{fig:WannierComparison} 
\end{figure}

It was shown in Ref.~\onlinecite{WangEuZnSb2} that the lowest energy antiferromagnetic phase of \mater~ hosts broad band dispersion close to the Fermi level.
Before investigating how the band topology changes with the change in the magnetic texture, we would like to understand the origin of the conduction bands.
 For this end we first study the non-magnetic phase in the absence of the Eu-\textit{4f} electrons within the opencore approximation to simplify the problem.
Since we are interested in the electronic dispersion in the vicinity of the Fermi level, such opencore electron description
is approximately equivalent to the application of the Hubbard $U$ within the mean-field approximation that pushes the localized states away from the Fermi level.

\begin{figure}[htb]
    \begin{center}
        \subfigure[]{
            \includegraphics[width=0.23\textwidth]{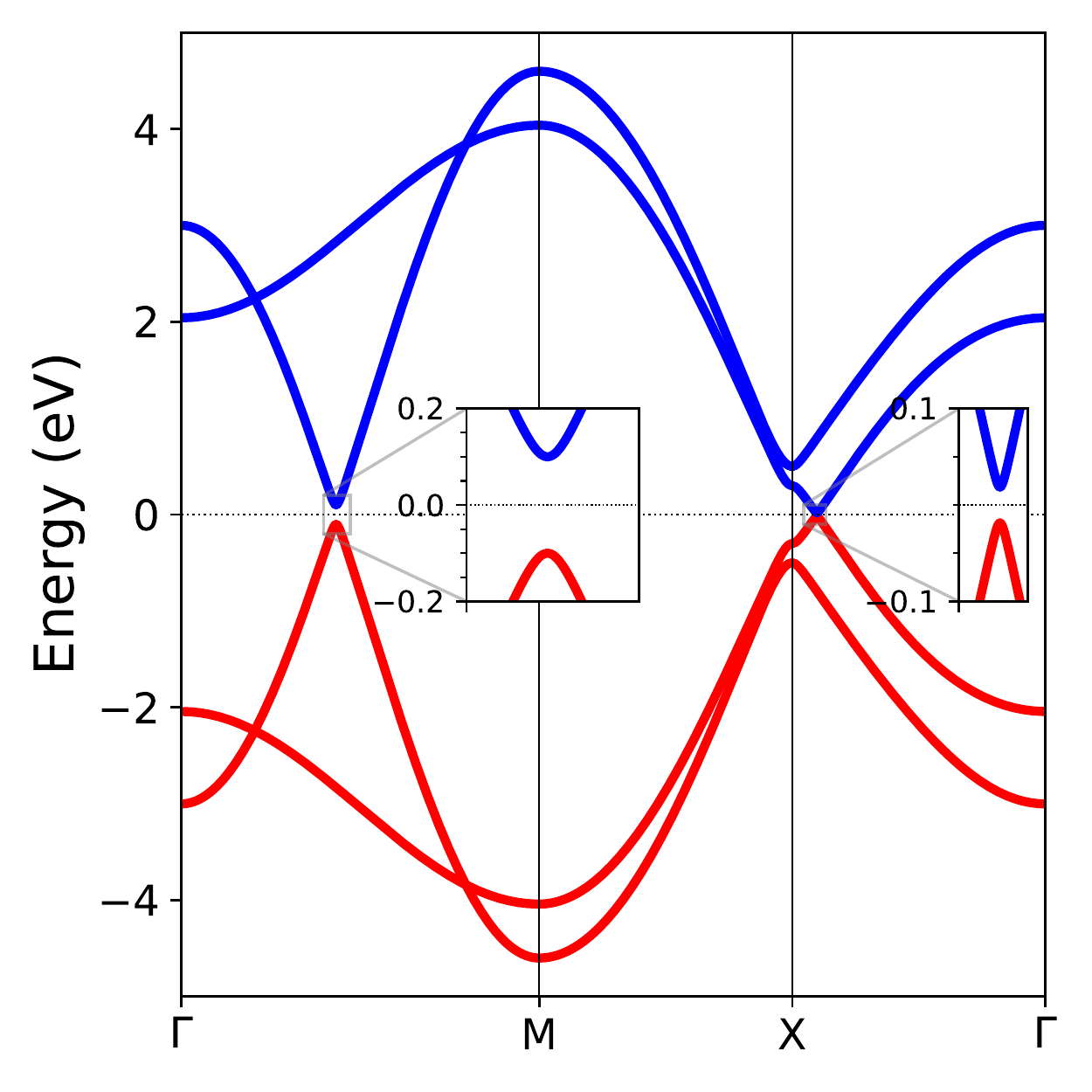}
        }
\hskip -0.1 in
        \subfigure[]{
            \includegraphics[width=0.23\textwidth]{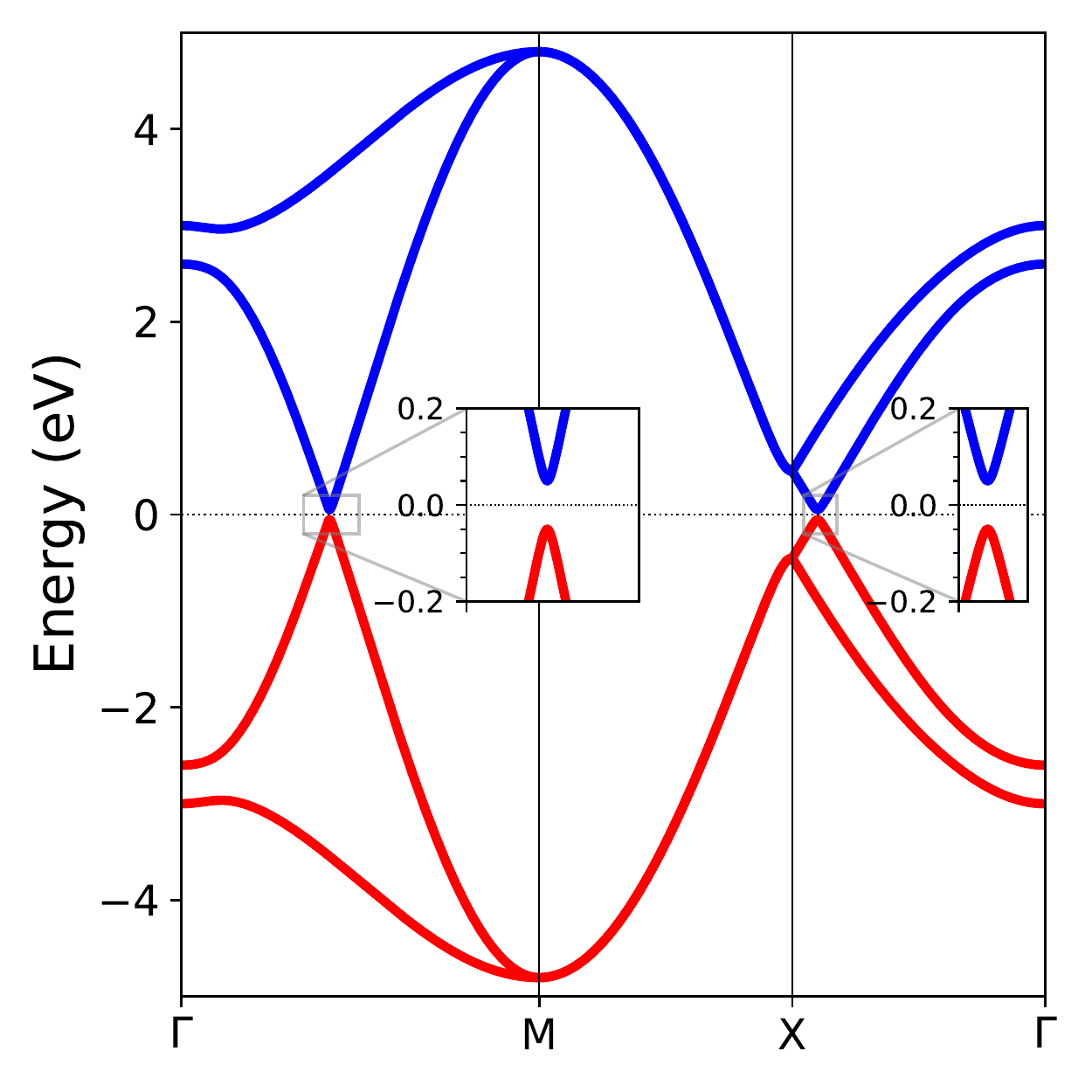}
        } 
            \end{center}
            \caption{ Band structure from nnn 4$\times$4 Hamiltonian using the tight-binding parameters in Table~\ref{table:TightBindingParams}.
            In (a), sublattice and $\mathcal{R}_4$ is broken explicitly in the 4$\times$4 model by introducing onsite anisotropy and making  the $\sigma$ and $\pi$ hopping different along $\hat{x}$ and $\hat{y}$ direction. 
            In (b), the effect of SOC is included.
            The insets show the gap openings in the band crossings.
        }
        \label{fig:WannierComparison2} 
\end{figure}

The DFT calculated band dispersion for non-magnetic calculation in the absence of SOC are shown in Fig.~\ref{fig:WannierComparison}(a) (gray lines). 
The band diagram shows crossings along $\Gamma$-M and $\Gamma$-X directions close to the Fermi level. Such crossings are the consequence of the band folding due to the doubling of the unitcell and 
are formed between the dispersive $p_x$-$p_y$ orbitals from the denser square nets of Sb2 atoms. The $p_z$ states from Sb2 atoms have smaller band width and are almost completely filled. Sb1 $p$ orbitals also contribute to the states close to the Fermi level (see band character plots in Appendix~\ref{appendixA} Fig. ~\ref{fig:appen-bandcharacter}).

The red and blue dots overlapped onto the gray lines in Figs.~\ref{fig:WannierComparison}(a) and (b) are the results from 4 and 6 band Wannier function description, respectively, obtained by using the method of disentanglement~\cite{Wannier902014}.
Hopping parameters only up to the next-nearest neighbour (nnn) given in Table.~\ref{table:TightBindingParams} are used for this comparison.
The 4-band model reproduces the crossings between the dispersive $p_x$-$p_y$ bands along the $\Gamma$-M and X-$\Gamma$ directions but it can not describe other features in the vicinity of the Fermi level.
It appears that 6 band tight-binding (TB) model is the minimal model to describe the band structure in the vicinity of the Fermi surface.
Even with the nnn TB model, one can describe approximately the generic features as seen in Fig.~\ref{fig:WannierComparison}(b).
In fact, by including the long range hopping terms, we can reproduce all the features in the periphery of the Fermi level exactly (see Fig.~\ref{fig:appen-WannierComparison-6band} in the Appendix ~\ref{appendixA} where all long range hopping matrix elements  are included in the 6-band model).
The Wannier orbitals are formed by the bond directed $p_x$, $p_y$ and $p_z$-like orbitals centered on the Sb2 atoms forming the square lattice as shown in Fig.~\ref{fig:WannierComparison}(c).

Qualitatively, 4-band model Hamiltonian in the basis of the $p_x$ and $p_y$ orbitals is sufficient  to understand the origin of the band  dispersion along the $\Gamma$-M and X-$\Gamma$~\cite{StackedSqNets_Klemenz_PRB2020}.
Hence, in the following, we only discuss the simple 4-band model to understand the effect of $\mathcal{R}_4$ symmetry breaking introduced by the magnetic Eu-atoms and the action of the spin orbit coupling.
Notice that in the absence of the nnn term, four fold degenerate bands are present right at the X high symmetry point whereas nnn term makes the bands two fold degenerate slightly away from the X point [see Fig.~\ref{fig:appen-WannierComparison-4band}].
We refer the reader to Appendix~\ref{appendixA} for exact analytical expression of the eigenvalues from the 4$\times$4 Hamiltonian and other effective Hamiltonian analysis.

\begin{table}[!tht]
\caption{\label{table:TightBindingParams}
Tight binding parameters (in eV) for the 6 band Wannier functions in the basis of the $p_x$, $p_y$ and $p_z$-like Wannier orbitals centered on the Sb2 atoms forming square lattice. The nearest neighbour (nn) $p_i$-$p_i$ hopping integrals are  $t_{\sigma}=1.9$ and $t_{\pi}=-0.5$ for the $\sigma$- and $\pi$-bonding-like orbital overlaps, respectively.
}
\begin{center}
\begin{tabular}{p{75pt}p{40pt}p{40pt}p{40pt}} 
\hline\hline
 $\langle WFs|H|WFs\rangle$ & Sb $p_x$ & $p_y$ & $p_z$ \\ \hline 
 Sb $\epsilon -\mu$ & 0.12 & 0.12 & -1.0    \\ 
 Sb$_{nn}$ $p_x$  & t$_{\sigma}$;t$_{\pi}$ & 0.00 & -0.05   \\ 
 Sb$_{nn}$ $p_y$  & 0.00 & t$_{\sigma}$;t$_{\pi}$ & -0.05   \\ 
 Sb$_{nn}$ $p_z$  & -0.05 & -0.05 & -0.10   \\ 
 Sb$_{nnn}$ $p_x$  & 0.00 & $\pm$0.10 & 0.00   \\ 
 Sb$_{nnn}$ $p_y$  & $\pm$0.1 & 0.00 & 0.00   \\ 
 Sb$_{nnn}$ $p_z$  & 0.00 & 0.00 & 0.20   \\ 
\hline\hline 
\end{tabular}
\end{center}
\end{table}

\textit{Gap on the Fermi surface--}
The Fermi surface (FS) can be gapped partially along $\Gamma$-M by breaking the sublattice symmetry  (or equivalently inversion or glide symmetry of the 2D square plane). This can be done easily in our TB Hamiltonian by introducing the onsite anisotropy term.
However, in order to open a gap at the X-$\Gamma$ direction, one needs to break both sublattice  and $\mathcal{R}_4$ symmetry; the latter can be introduced in our TB model by making the $\sigma$ and $\pi$ hoppings asymmetric along the $\hat{x}$ and $\hat{y}$ directions. Fig.~\ref{fig:WannierComparison2}(a) presents the case of opening gaps in the band crossings by introducing such terms in our TB model.

In the presence of the SOC, the entire Fermi surface gaps out. 
We find the 
 form of SOC in our TB Hamiltonian to be $(\lambda \sigma_z\tau_y + \delta \gamma_z\sigma_z\tau_z)$ where $\lambda$ and $\delta$ are constants and $\gamma$, $\sigma$, $\tau$ are the Pauli matrices acting on the site, spin, and orbital indices, respectively.
 The first term introduces coupling between the $p_x$ and $p_y$ orbitals of same site and same spin whereas the second term introduces the on-site asymmetry.
Fig.~\ref{fig:WannierComparison2}(b) shows the action of SOC on the nnn tight-binding model using $\delta=0.1$ and $\lambda=0.2$ which are extracted from the \textit{ab-initio} Wannier function analysis.
In addition to gapping out the entire Fermi surface, the introduction of SOC breaks 2 fold degeneracy at the $\Gamma$ point but preserves crossings at M and X points away from the Fermi level. 
These observations are consistent with the DFT calculation.

\subsection{Magnetic Phases and Electronic Structure}
\begin{figure}[t]
    \begin{center}
        \includegraphics[width=0.45\textwidth]{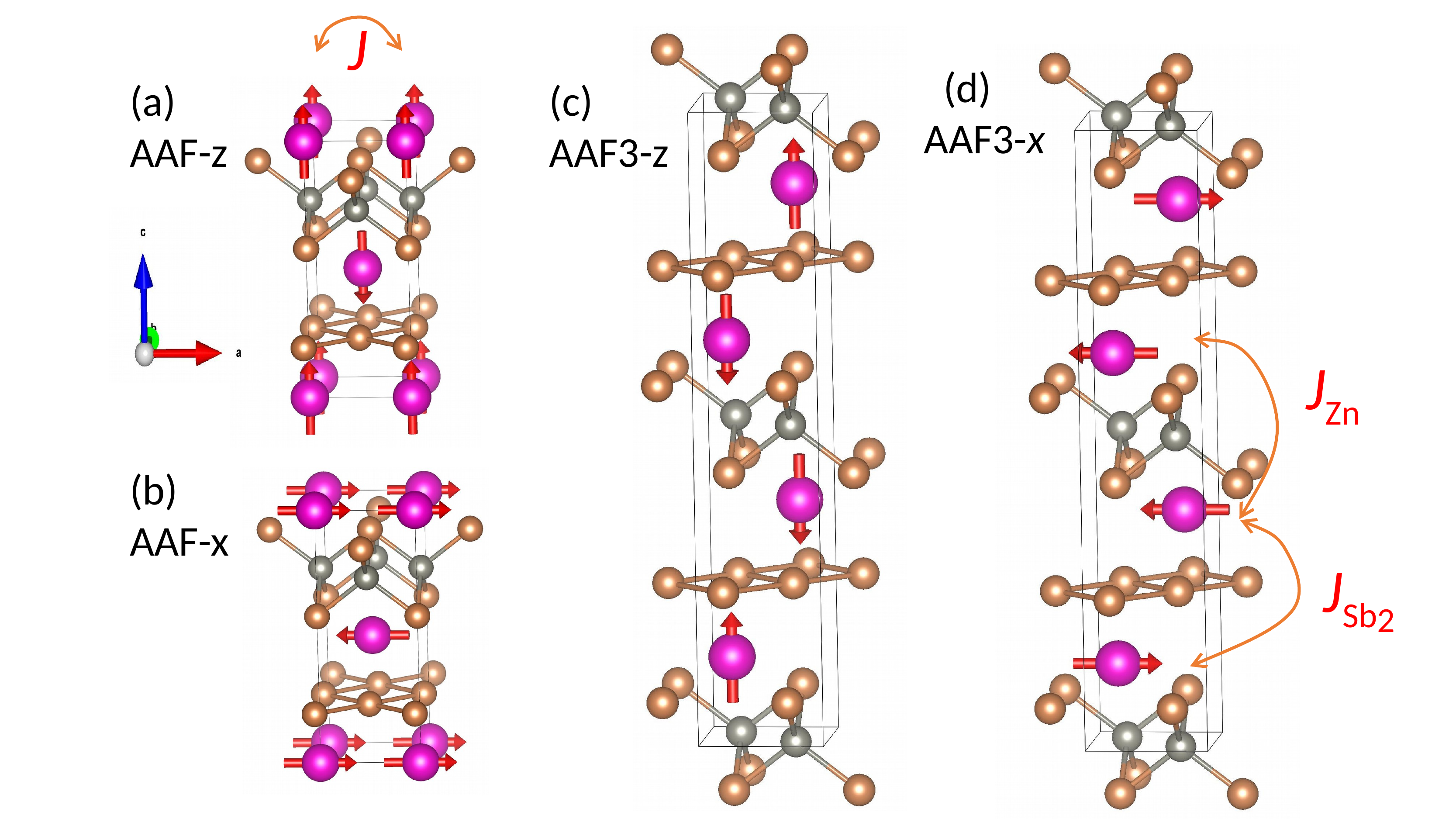}
            \end{center}
            \vspace{-0.2in}
            \caption{\label{fig:MagneticPatterns} (a, b) AAF and (c, d) AAF3 magnetic patterns of {\mater} with magnetization along the {\xhat} and {\zhat} axes, respectively.}
        
\end{figure}

Having explored the electronic dispersion of the non-magnetic phase in detail, we are now in the position to understand the influence of magnetic texture in the electronic properties. Ref.~\onlinecite{WangEuZnSb2}  found that two AFM patterns (referred to as AAF and AAF3) of {\mater} are competing for the ground state, i.e., having very small energy differences well within the error bar of the calculations. The different magnetic patterns studied in this work are shown in Fig.~\ref{fig:MagneticPatterns}. 
Throughout this manuscript, the \zhat~ and \xhat~ magnetic phase (alternatively called soz and sox phases in the manuscript) imply the direction of the \Neel vector ($\vec{n})$ parallel to the [001] and [100] direction respectively.
Our DFT calculations find that the energy differences between the magnetic phases depend on the value of the Hubbard interaction $U$. As shown in Table~\ref{table:MagneticPatternEnergy}, for $U=0$, AAF3 has lower energy, whereas as $U$ is increased, AAF becomes lower in energy. Also the magnetic anisotropy energy is very small i.e the energy differences between the \zhat~ and \xhat~ AFM pattern are almost the same. Note that the ferromagnetic (FM) state is also closer in energy, but the nonmagnetic (NM) phase is a very high-energy state, due to the Eu$^{2+}$ $4f^7$ high-spin electronic configuration yet with weak coupling between well-localized $f$ electrons. The relevant information about magnetic space groups and symmetry relationships is listed in Table~\ref{table:MagneticPatternSymmetry}.

\begin{table}[t]
\vspace{-0.15in}
\begin{center}
\caption{\label{table:MagneticPatternEnergy}Calculated energy difference per formula unit in meV for different magnetic patterns (see text).}
\begin{tabular}{l|p{30pt}p{40pt}p{55pt}p{55pt}}
\hline\hline
 Pattern & GGA & GGA+SO &  GGA+SO+U (3eV) & GGA+SO+U (6eV) \\ \hline
 AAF-x & 0 & 0 & 0 & 0 \\ 
 AAF-z & - & 0.804 & 0.24 & 0.017 \\ 
 AAF3-x & - & $-2.85$ &$-0.41$ & 1.29 \\ 
 AAF3-z & $-6.46$ & $-6.18$ & $-0.53$ & 1.35 \\ 
 FM & 1.13 & $-0.53$ & - & 3.71 \\ 
 NM & 7297 & 5795 & - & - \\ 
 \hline\hline
\end{tabular}
\end{center}
\end{table}

\begin{table} [hb]
\begin{center}
\caption{\label{table:MagneticPatternSymmetry}The magnetic symmetry group and symmetry operations of different magnetic patterns.
In the first and last 2 rows, the operations which are different are colored in gray. Other symmetry operations can be generated by the action of $\Pcal\Tcal$ for the AAF phase and by the action of $\Pcal$, $\Tcal'$ and $\Pcal\Tcal$ for the AAF3 phase, hence not mentioned here for clarity.}
\begin{tabular}{|l|p{48pt}|p{150pt}|} \hline
    Pattern & MSG & Symmetry Operations \\ \hline
    AAF-z & 129.419 (P$4/n'm'm$) & $\gray{\{4^+_{001}|\frac{1}{2}00\}}$, $\gray{\{4^-_{001}| 0\frac{1}{2}0\}}$, $\gray{\{2_{100} |\frac{1}{2}00\}}$, $\{2_{010}|0\frac{1}{2}0\}$, $\gray{\{2_{001}|\frac{1}{2}\frac{1}{2} 0\}}$, $\gray{\{2_{110}| \frac{1}{2} \frac{1}{2} 0\}}$, $\gray{\{2_{1-10}|0\}}$, $\{-1'|0\}$  \\ \hline 
    AAF-x & 59.407 (P$_{m'mn}$) & $\gray{\{m_{100}|\frac{1}{2}00\}}$, \{2$_{010}|0\frac{1}{2}0$\}, $\gray{\{m_{001}|\frac{1}{2} \frac{1}{2} 0\}}$, $\{-1'|0\}$  \\ \hline \hline 
    AAF3-z & 130.432 (P$_c$4/ncc) & $\gray{\{ 4^+_{001} | \frac{1}{2} 0 0 \}}$, $\gray{\{ 4^-_{001} | 0 \frac{1}{2} 0 \}}$, $\gray{\{ 2_{110} |\frac{1}{2} \frac{1}{2} \frac{1}{2} \}}$, $\gray{\{ 2_{1-10} | 0 0 \frac{1}{2} \}}$,  $\{ 2_{001} | \frac{1}{2} \frac{1}{2} 0 \}$, $\{ 2_{100} | \frac{1}{2} 0 \frac{1}{2} \}$, $\{ 2_{010} | 0 \frac{1}{2} \frac{1}{2} \}$, $\{ -1 | 0 \}$, $\{ 1' | 0 0 \frac{1}{2} \}$    \\ \hline 
    AAF3-x & 62.450 (P$_a$nma) & $\{ 2_{001} | \frac{1}{2} \frac{1}{2} \gray{\frac{1}{2}} \}$, $\{ 2_{100} | \frac{1}{2} 0 \gray{0} \}$, $\{ 2_{010} | 0 \frac{1}{2} \frac{1}{2} \}$, $\{ -1 | 0 \}$, $\{ 1' | 0 0 \frac{1}{2} \}$ \\ \hline 
\end{tabular}
\end{center}
\end{table}

\subsubsection{AAF phase}

\begin{figure*}[htb]
\hskip -0.05 in
    \includegraphics[width=0.95\textwidth]{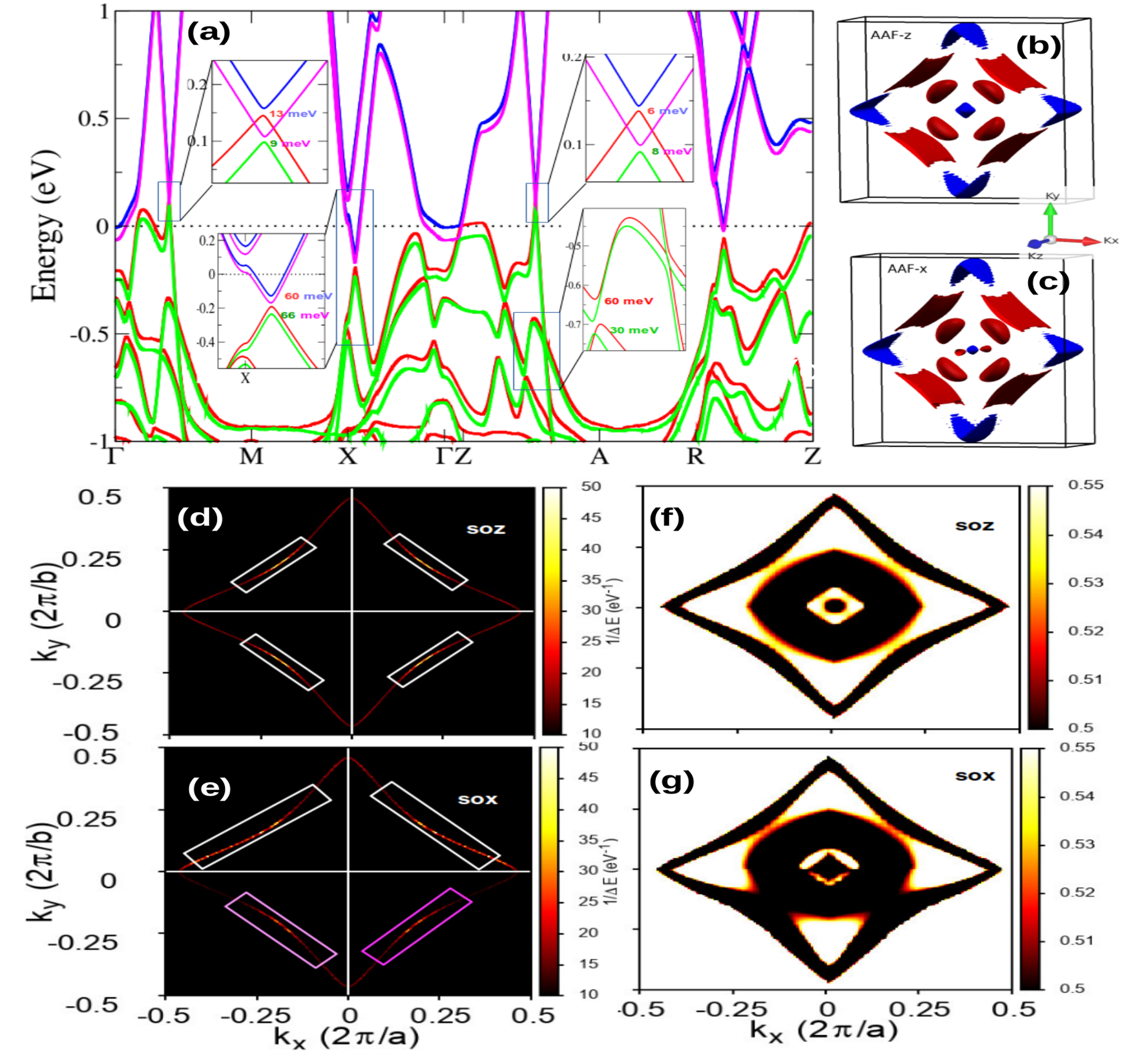} 
\caption{Comparison of the electronic structure between the AAF-\zhat~ and AAF-\xhat~ phases  with the inclusion of SOC and U of 6 eV.
    (a) Band structure along $\Gamma$-M-X-$\Gamma$-Z-A-R-Z high symmetry direction for the two phases. The red-blue (green-magenta) lines are for magnetization along  $\hat{x}$ ($\hat{z}$) direction.
    (b, c) Fermi surface plots on the 3D Brillouin zone.
    (d, e) Intensity plot of the inverse band gap value between the valence and conduction band on the $k_x$-$k_y$ plane at $k_z$=0 which shows the formation of the gapped nodal line feature around the $\Gamma$ point. The rectangular boxes are drawn to highlight the fact that the {\xhat}-phase has a broken $\mathcal{R}_4$ symmetry whereas {\zhat}-phase obeys $\mathcal{R}_4$ perfectly.
    (f, g) Contour plot of the Eu-\textit{4f} orbital distribution on the valence band.
    The intensity range (color-bar) is shown in a narrow to amplify the small differences between the phases.
    These figures again highlight that $\mathcal{R}_4$ is weakly broken for the {\xhat} phase.
    See text for details.
        }
        \label{fig:BandsAAF} 
\end{figure*}

The AFM arrangement of the 2 Eu atoms
in the primitive unitcell located above and below the Sb2 square lattice gives AAF phase.
The two antiferromagnetically aligned Eu atoms are no longer inversion symmetric irrepective of any direction of the \Neel vector. 
This is because inversion symmetry does not operate on the spin degree of freedom.
 However, because of the broken inversion ($\Pcal$) and broken time reversal symmetry ($\Tcal$), their product $\Pcal\Tcal$ is a conserved quantity which makes the bands doubly degenerate throughout the BZ. Such magnetic space group (MSG) falls into type-II MSG.

Unlike inversion symmetry, rotation (or screw) and mirror (or glide) symmetries  act on the spin degree of freedom.
Hence, depending on the orientation of the \Neel vector, some of the symmetries could be broken.
For example, $\hat{x}$ ($\hat{z}$) direction of the \Neel vector preserves (breaks) glide symmetry $\mathcal{G}_z=\{m_{001}\vert \frac{1}{2} \frac{1}{2} 0\}$ but breaks (preserves) 2-fold rotation symmetry $\mathcal{R}_{2z}=\{2_{001}\vert \frac{1}{2} \frac{1}{2} 0\}$ which can be seen from the following action of the symmetry operation on the space and spin degree of freedom:
\begin{equation}
\begin{gathered}
(x,y,z) \xrightarrow{\mathcal{G}_z} (x+\frac{1}{2},y+\frac{1}{2},-z),\\ 
(x,y,z) \xrightarrow{\mathcal{R}_z} (-x+\frac{1}{2},-y+\frac{1}{2},z),\\ 
(m_x,m_y,m_z) \xrightarrow{\mathcal{G}_z/\mathcal{R}_z} (-m_x,-m_y,m_z) 
\end{gathered}
\label{eqn:Gz_symmetry}
\end{equation}
Similarly, other symmetries absent in the AAF-\xhat~ magnetic phase are four fold rotation symmetries along the \textbf{c}-axis $\{4^{\pm}_{001}\vert\frac{1}{2}00\}$, 2-fold rotation along the [110] axes $\{2_{110}\vert\frac{1}{2}\frac{1}{2}0\}, \{2_{1-10}\vert0\}$ and 
the product of these symmetries with $\Pcal\Tcal$ symmetry.
The details of the symmetry operations present for different magnetic patterns are shown in Table.~\ref{table:MagneticPatternSymmetry}.

In the following paragraphs, we will first present results from the DFT calculations which show clear differences in the bulk and surface electronic dispersion between the \zhat~ and \xhat~ phase. 
We will then derive effective Kondo exchange Hamiltonian using parameters extracted from DFT calculations to understand some of the major findings. 

\paragraph{Bulk states:}
The bulk band structure comparison between the AAF-\xhat~ and AAF-\zhat~ phases in a narrow energy window of $\pm$1 eV is shown in Fig.~\ref{fig:BandsAAF}(a). 
Fig.~\ref{fig:appen-bandsdos_soz} in  Appendix~\ref{appendixC} shows the electronic dispersion and density of states (DOS) in a wider energy range. 
The Eu-\textit{4f} states are 1 eV below the Fermi level and the small DOS at the Fermi level mainly comes from the Sb2 $p_x$-$p_y$ orbitals.

We find that despite the absence of some symmetries in the AAF-\xhat~ phase as mentioned in the previous section, the bulk band structure for both phases in the vicinity of the Fermi level looks almost identical except small momentum dependent shifts.
This is expected as the localized Eu-$4f$ orbitals are pushed away from the Fermi level due to the application of the Hubbard $U$ correction.
The Fermi surface plots in Fig.~\ref{fig:BandsAAF}(b \&c) highlight this fact which show similar features except slight difference at the $\Gamma$-point.
In fact, the electronic dispersion  from the non-magnetic phase [Fig.~\ref{fig:WannierComparison}] is not very different from the AAF phase which shows that the effect of magnetism in the electronic dispersion is small in this system.
In the absence of SOC, the valence and conduction bands form a gapped nodal line on the $k_x$-$k_y$ plane with 
band crossings along the $\Gamma$-M direction.
All the crossings are gapped by the action of SOC for both phases. This is similar to the case of nodal line semimetal ZrSiS where SOC has been found to open gap in the Dirac crossings~\cite{ZrSiS_Schoop_NatureComm2016}. 

Although there are no protected crossings for both phases, the magnitude of the band gap is different for the two magnetic phases.
In Fig.~\ref{fig:BandsAAF}(d \& e), we show the inverse of the eigenvalue difference between the valence and the conduction bands on the $k_x$-$k_y$ plane at $k_z=0$ to highlight the presence of the gapped nodal line feature.
In the $\hat{z}$ phase, the gap distribution is identical in all 4-quadrants, however that is not the case in the $\hat{x}$ phase. This is a clear signature of the broken global $\mathcal{R}_4$ symmetry on the $\hat{x}$ phase only [Fig.~\ref{fig:BandsAAF}(e)].
To understand further the origin of the $\mathcal{R}_4$ symmetry breaking, we also looked at the $\textbf{k}$-dependent valence band occupancy of the $4f$-electrons on the $k_x$-$k_y$ plane [Figs.~\ref{fig:BandsAAF}(f \& g)].
In the vicinity of the $\Gamma$-point, we find that the $4f$-electron contribution to the valence band is not symmetric across the 4-quadrants unlike the $\hat{z}$ phase indicating that $\mathcal{R}_4$ symmetry is weakly broken due to the orientation of the Eu-$4f$ magnetic moments. 

\paragraph{Surface states:}

\begin{figure}[htb]
    \begin{center}
            \includegraphics[width=0.43\textwidth]{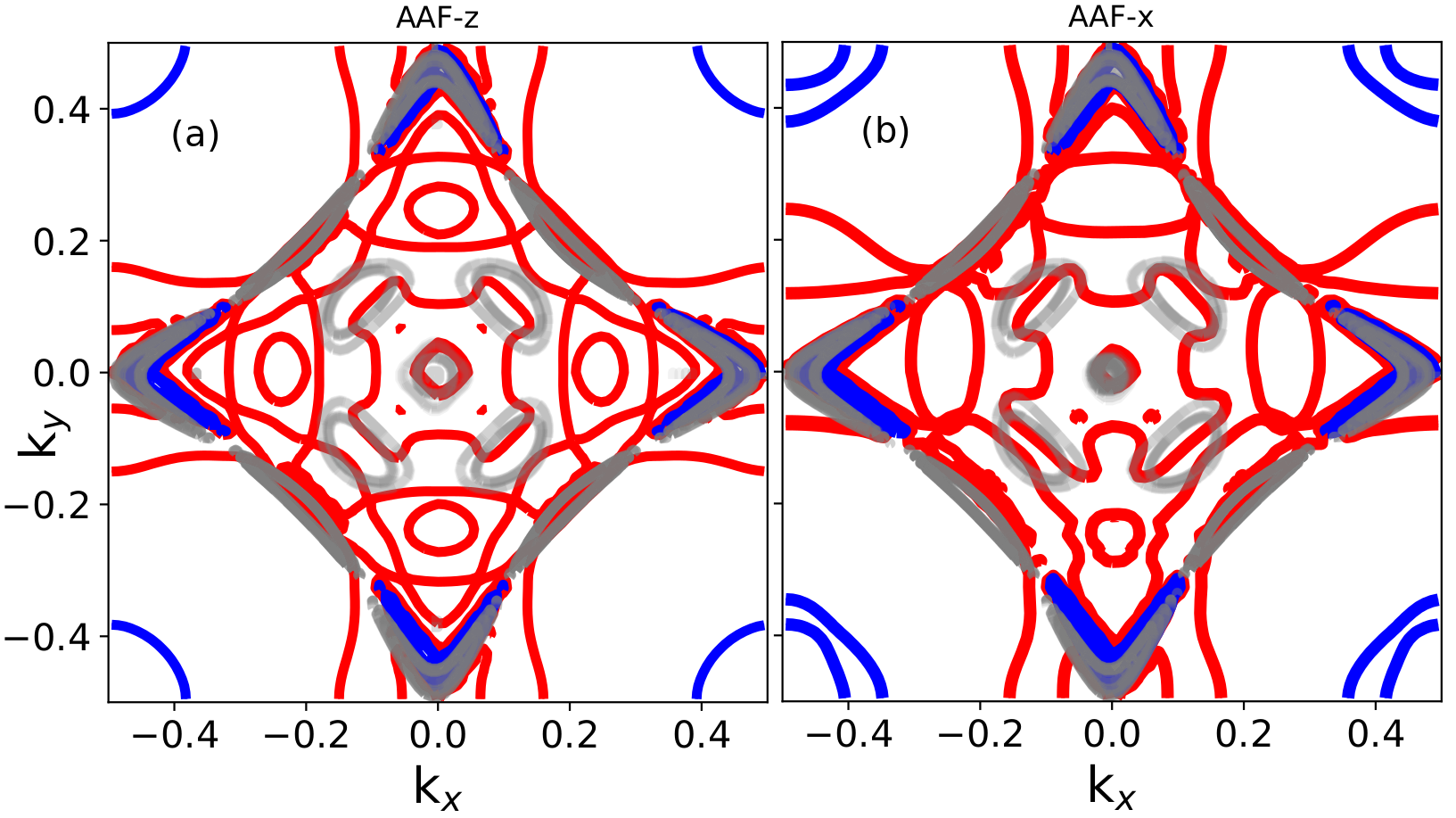}
            \end{center}
            \caption{Comparison of the slab Fermi surface between the (a) $\hat{z}$ and (b) $\hat{x}$ phase.
                The gray lines are the bulk-derived states and the red-blue lines denote the hole and electron Fermi pockets obtained from the 3 layer slab calculation.
                Notice that the broken $\mathcal{R}_4$ symmetry is amplified in the surface dispersion in panel (b).
        }
        \label{fig:SlabBandsAAF} 
\end{figure}
We also performed slab calculation to see how the orientation of the \Neel vector affects the surface states. Fig.~\ref{fig:SlabBandsAAF} shows the slab Fermi surface of the AAF-\zhat~ and \xhat~ phases obtained from a 3 layer slab calculation.
The surface is terminated on the Zn and Sb1 layers.
The slab calculations show a number of additional features compared to the bulk states which are shown as a gray background.
The most notable feature which is absent in the bulk dispersion is the closed loop (red lines) state around the $\Gamma$ point connecting the gray ellipsoids. 
Remarkably, the surface states are different for the two phases with the  breaking of the $\mathcal{R}_4$ symmetry clearly visible now for the $\hat{x}$ phase unlike the subtle differences we found in in the bulk-band features.
For example, along the $\Gamma$-X as well as $\Gamma$-M directions, the surface states, especially the hole ones (red colored lines), are related by the mirror symmetry $\mathcal{M}_{x}$  but are asymmetric with respect to the $\mathcal{R}_4$ symmetry i.e. the pockets are not identical along the 4-quadrants only for the \xhat~ phase.  
This is amazing given the fact that the slab is not terminated on the Eu-atoms and the Eu-\textit{4f} states have negligible contribution to the Fermi surface. 
This demonstrates that the magnitude of the broken symmetry in the electronic dispersion are more amplified on the surface compared to the bulk states.
Hence, we anticipate that surface probes could be more suitable for resolving the broken symmetry phases.
%

In order to understand further how Eu-$f$ states can affect the Fermi surface properties, we have derived an effective Kondo exchange Hamiltonian for the $p_x$-$p_y$ electrons in the presence of the Eu-$f$ spins in Appendix~\ref{appendixB}.
From such effective Hamiltonian analysis, we find that the Eu magnetic moments can affect the dispersion of the itinerant $p_x$-$p_y$ electrons through exchange coupling which introduces hopping between the $p$-electrons through Eu-sites.  
In addition, we show that depending on the orientation of the Eu-$f$ spins, the band spectrum as well as their spin-texture will be different. This will have consequences in the spin transport properties which will be discussed later.

\subsubsection{AAF3 phase}

\begin{figure}[htb]
    \begin{center}
\hskip -0.15 in
            \includegraphics[width=0.45\textwidth]{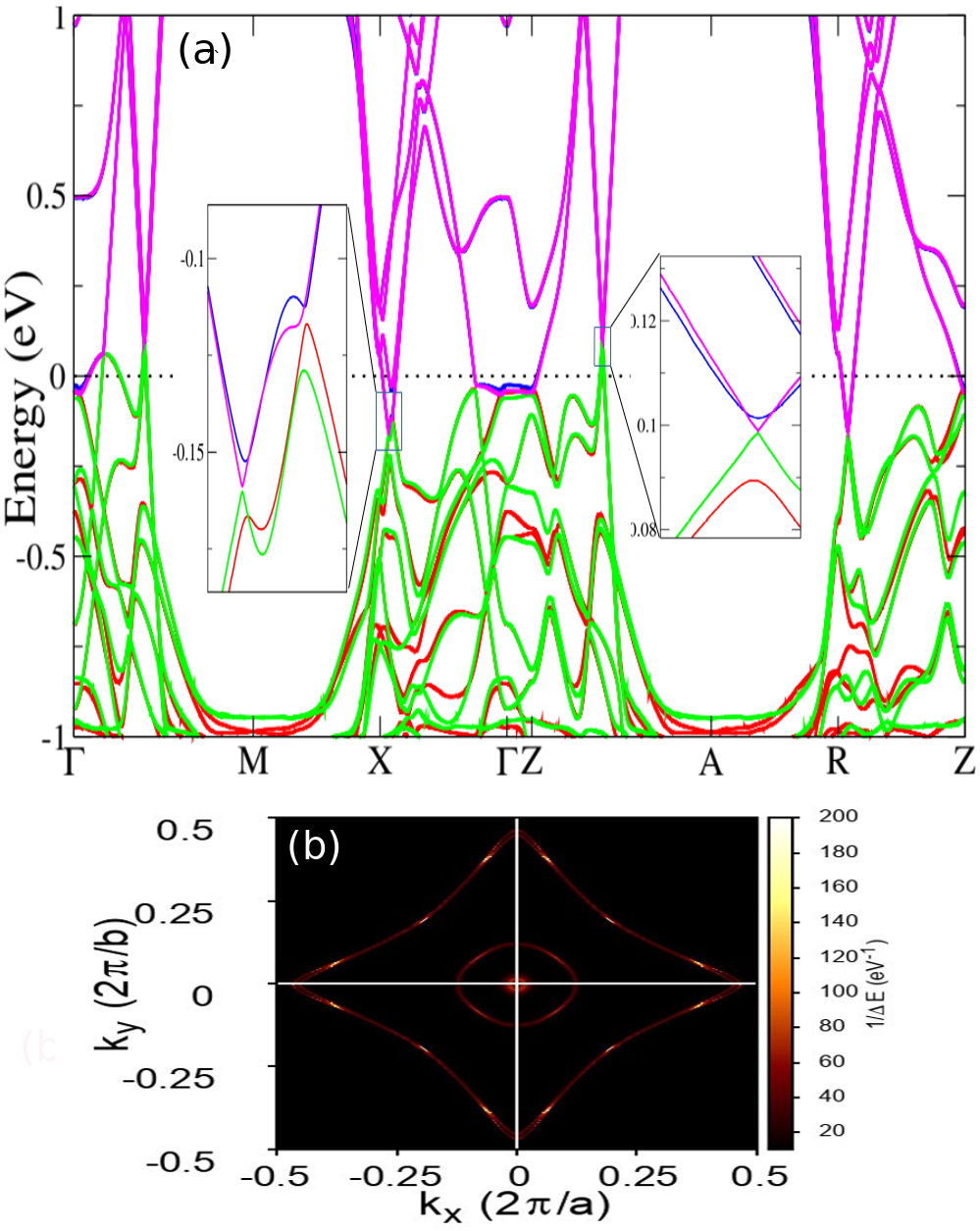}
            \end{center}
            \caption{(a) Comparison of the electronic dispersion between the AAF3-$\hat{x}$ (red-blue) and $\hat{z}$ (green-magenta) magnetic phases.
            The inset shows that \zhat-phase has band crossings near the Fermi level whereas they are avoided for the \xhat-phase.
               (b) Intensity plot of the inverse band gap between the valence and conduction bands for the \zhat~phase showing crossings along the $\Gamma$-M directions and in the vicinity of the X point. The small gap of $\sim$ 10 meV seen in the  plot is due to the finite $\textbf{k}$-mesh.
        }
        \label{fig:BandsAAF3} 
\end{figure}

We also studied another competing AFM phase AAF3 in detail.
AAF3 phase requires the doubling of the unitcell along the \textbf{c}-axis as seen in Fig.~\ref{fig:MagneticPatterns} (c \& d).
Unlike the AAF phase, $\Pcal$ is preserved here.
Moreover, despite breaking of the $\Tcal$ symmetry, 
$\Tcal' = \Tcal\tau$ i.e. $\Tcal$ followed by translation $\tau=\{0,0,\frac{1}{2}\}$ is the symmetry of the system.
Such MSG with non-symmorphic  $\Tcal$ falls into  type IV category and has been found in another AFM Dirac material EuCdAs$_2$~\cite{TypeIVMSG_Hau_PRB2018}.
 $\Pcal\Tcal'$ symmetry makes the bands doubly degenerate throughout the Brillouin zone.
In addition, because of the half translation along with $\Tcal$, there are interesting symmetry properties.

Similar to the AAF magnetic pattern, the \zhat-phase here has more symmetry compared to the \xhat-phase due to the direction of the spins.
For example, $\mathcal{R}^+_4$ symmetry is preserved for magnetization along $\hat{z}$ direction whereas it is not preserved for magnetization along $\hat{x}$ or  
$\hat{y}$ directions. The complete list of symmetry operations are tabulated in Table.~\ref{table:MagneticPatternSymmetry}.

Following closely the arguments given in Refs.~\cite{DiracAFM_Zhang_NaturePhys_2016,DNLSpintronics_Shao_PRL2019}, we prove that due to the extra rotational symmetries in the AAF3-$\hat{z}$ phase, 
band crossings along different high symmetry directions are preserved whereas 
they are avoided (or gapped) in the AAF3-$\hat{x}$ phase. 
First, we find the eigenvalues of the 4-fold rotoinversion symmetry operator $\bar{\mathcal{R}}^+_4=\{\pm -4^+_{001}\vert\frac{1}{2},0,0\}$:
\begin{equation}
\begin{gathered}
	(x,y,z) \xrightarrow{\bar{\mathcal{R}}^+_4} (y+\frac{1}{2},-x,-z) \xrightarrow{\bar{\mathcal{R}}^+_4} (-x+\frac{1}{2},-y-\frac{1}{2},z)\\ 
	\xrightarrow{\bar{\mathcal{R}}^+_4} (y,x-\frac{1}{2},-z) \xrightarrow{\bar{\mathcal{R}}^+_4} (x,y,z) ,\\ 
\end{gathered}
\label{eqn:R4_symmetry2}
\end{equation}
Hence, $(\bar{\mathcal{R}}^+_4)^4 = -1$ and the eigenvalues are $J_m = e^{i\frac{(2m+1)\pi}{4}}$,
where the minus sign is from the spin rotation and $m$=0, 1, 2, 3 such that:
\begin{equation}
	J_0=e^{i\pi/4}=J^*_3, 
	J_1=e^{i3\pi/4}=J^*_2
\end{equation}
Because of the $\Pcal\Tcal'$ symmetry, the bands are 2 fold degenerate throughout the BZ.
If $|\psi\rangle$ is the simultaneous eigenstate of the Hamiltonian operator and the $\bar{\mathcal{R}}^+_4$, 
we would like to find the $PT'$ partner of $|\psi\rangle$.
For this, we need to find the commutation of $\bar{\mathcal{R}}^+_4$ with $\Pcal\Tcal'$.
\begin{equation}
\begin{gathered}
	(x,y,z) \xrightarrow{\bar{\mathcal{R}}^+_4} (y+\frac{1}{2},-x,-z) \xrightarrow{\Pcal\Tcal'} (-y-\frac{1}{2},x,z-\frac{1}{2}),\\ 
	(x,y,z) \xrightarrow{\Pcal\Tcal'} (-x,-y,-z-\frac{1}{2}) \xrightarrow{\bar{\mathcal{R}}^+_4} (-y+\frac{1}{2},x,z+\frac{1}{2}).
\end{gathered}
\end{equation}
i.e. 
\begin{eqnarray}
\bar{\mathcal{R}}^+_4 PT' &=& \tau(1,0,1)  \Pcal\Tcal'\bar{\mathcal{R}}^+_4,\nonumber \\
\implies \bar{\mathcal{R}}^+_4 \Pcal\Tcal' |\psi\rangle &=& e^{-i(k_x+k_z)} PT' \bar{\mathcal{R}}^+_4 |\psi\rangle,\nonumber \\
\mathrm{RHS},
	&=&e^{-i(k_x+k_z)} \Pcal\Tcal' \bar{R}^+_4 |\psi\rangle,\nonumber \\
	&=&e^{-i(k_x+k_z)} \Pcal\Tcal' J_m |\psi\rangle,\nonumber \\
	&=&e^{-i(k_x+k_z)} J^*_m e^{(ik_z/2)} \Pcal\Tcal' |\psi\rangle,\nonumber \\
	&=&e^{-i(k_x)}  e^{-(ik_z/2)} J^*_m \Pcal\Tcal' |\psi\rangle.
\end{eqnarray}
This implies that if $|\psi\rangle$ is an eigenstate of $\bar{\mathcal{R}}^+_4$ operator with eigenvalue $J_m$, then $\Pcal\Tcal'|\psi\rangle$ is also an eigenstate of $\bar{R}^+_4$ with eigenvalue $e^{-i(k_x-k_z/2)} J^*_m$.
Now, let's examine the $\bar{\mathcal{R}}^+_4$ eigenvalues of $|\psi\rangle$ and its $\Pcal\Tcal'$ partner at different high symmetry points which are invariant under the $\bar{\mathcal{R}}^+_4$ operation.

At the $\Gamma$-point where $\mathbf{k}=\mathbf{0}$, states with $\bar{\mathcal{R}}^+_4$ eigenvalues of $(J_0$, $J_3)$ and $(J_1$, $J_2)$ form degenerate pairs.
Similarly, at the M high symmetry point $(\pi,\pi,0)$,
because of the extra $e^{-ik_x}$ factor, $(J_0, J_2)$ and $(J_1, J_3)$ form  degenerate pairs
i.e. the $\bar{\mathcal{R}}^+_4$ eigenvalues of the Kramer's pairs switches partner compared to the  $\Gamma$-point.
This makes the crossing unavoidable along this line.
Following similar arguments, we find that there is an unavoidable crossing along the Z-A line. 

In Fig.~\ref{fig:BandsAAF3}(a), we compare the DFT calculated bands between the AAF3-\xhat~ and AAF3-\zhat~ phases.
The differences between the two are very small but most importantly, there are some crossings in the AAF3-\zhat~ patterns which are absent in the \xhat~ pattern.
For example, along the $\Gamma$-M and Z-A directions, \zhat-pattern shows crossings whereas there is a small gap of $\sim$10 meV in the \xhat-pattern. 
Similar feature is seen along the X-$\Gamma$ and R-Z line.
On the contrary, at the Z point, \xhat-phase shows 4-fold degeneracy whereas \zhat-phase has a gap of few meVs.

In Fig.~\ref{fig:BandsAAF3}(b), we plot the 2D bands of the AAF3-\zhat~ pattern on the $k_x$-$k_y$ plane at $k_z$=0 plane.
Similar to the AAF pattern, we see a gapped nodal line feature with a small difference that point nodes survive along $\Gamma$-M and $\Gamma$-X direction for the AAF3-\zhat~ phase only.

\subsection{Calculation of the Spin Hall Conductivity and spin current manipulation}

\begin{figure}[htb]
    \begin{center}
        \subfigure[]{
            \includegraphics[width=0.48\textwidth]{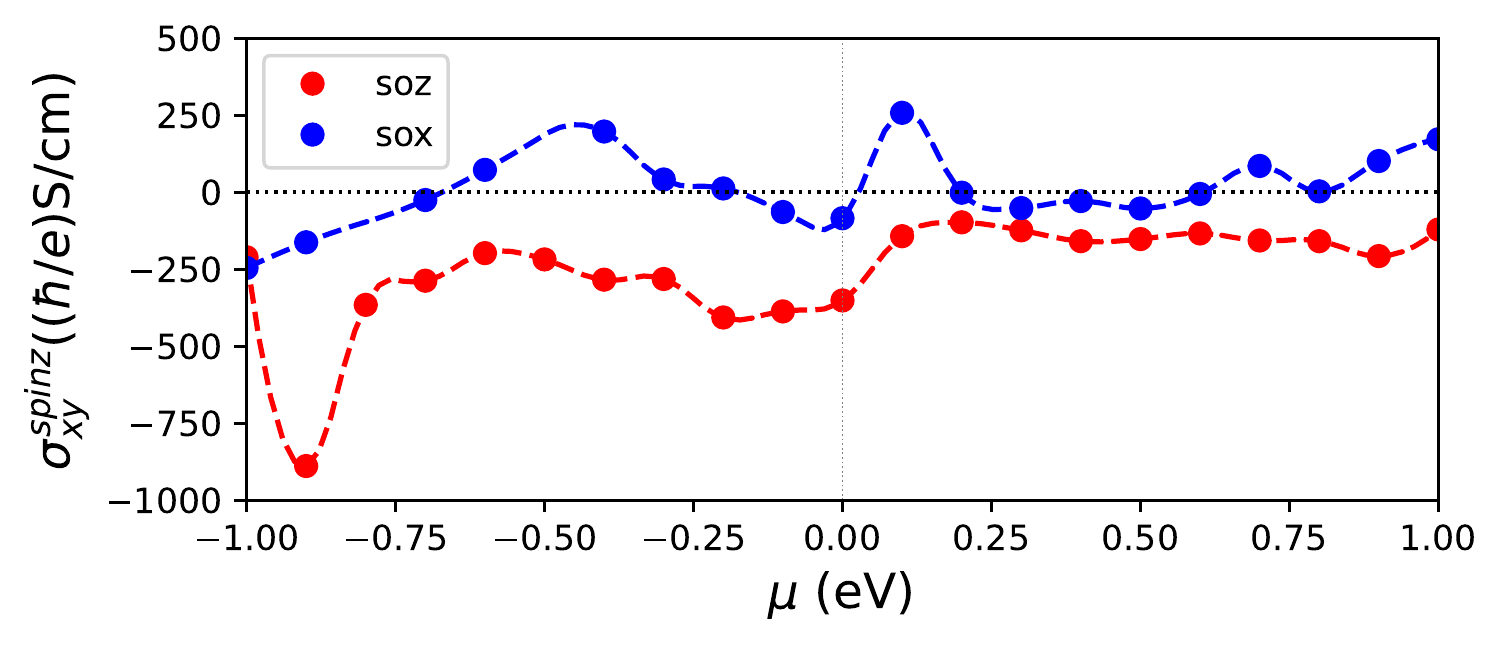}
        } 
\vskip -0.10 in
        \subfigure[]{
            \includegraphics[width=0.48\textwidth]{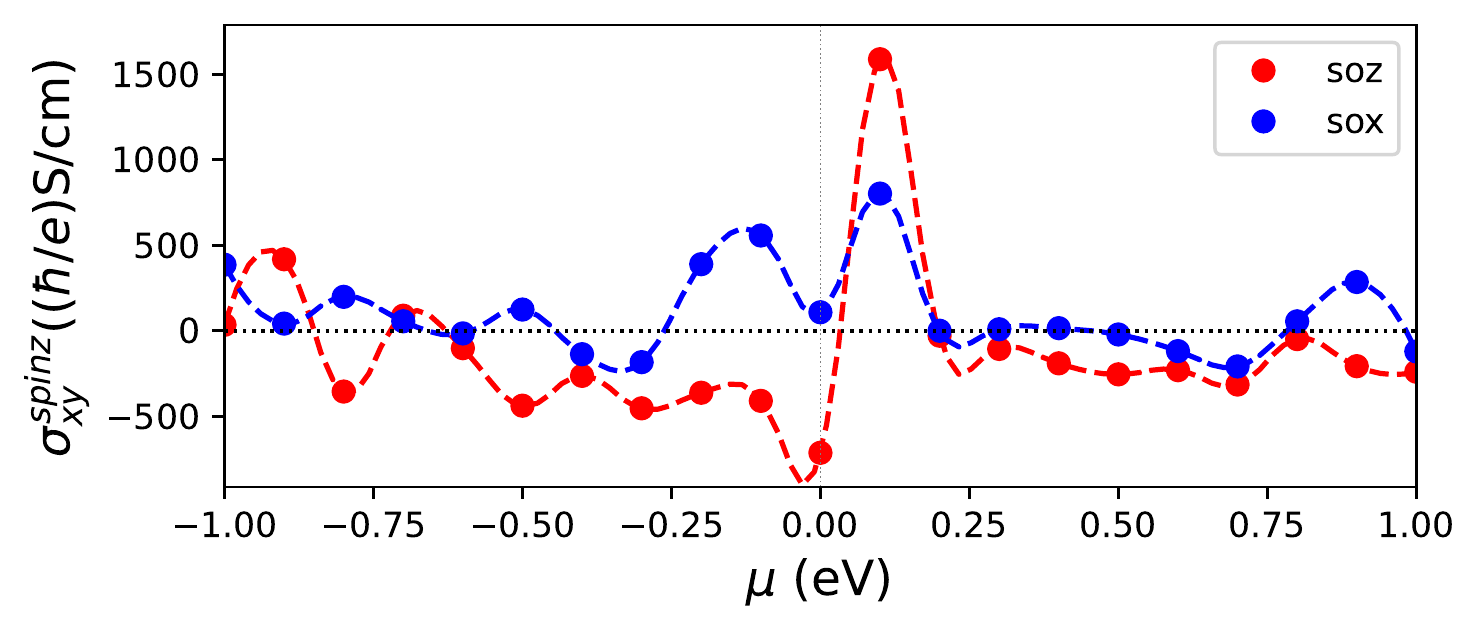}
        } 
     \end{center}
            \caption{Comparison of the $\sigma_{xy}^z$ component of the SHC tensor as a function of the chemical potential between the \xhat~ and \zhat~phases of the (a) AAF and (b) AAF3 magnetic patterns.
        }
        \label{fig:AAF-SHC} 
\end{figure}

\begin{figure*}[htb]
    \begin{center}
        \subfigure[]{
            \includegraphics[width=0.47\textwidth]{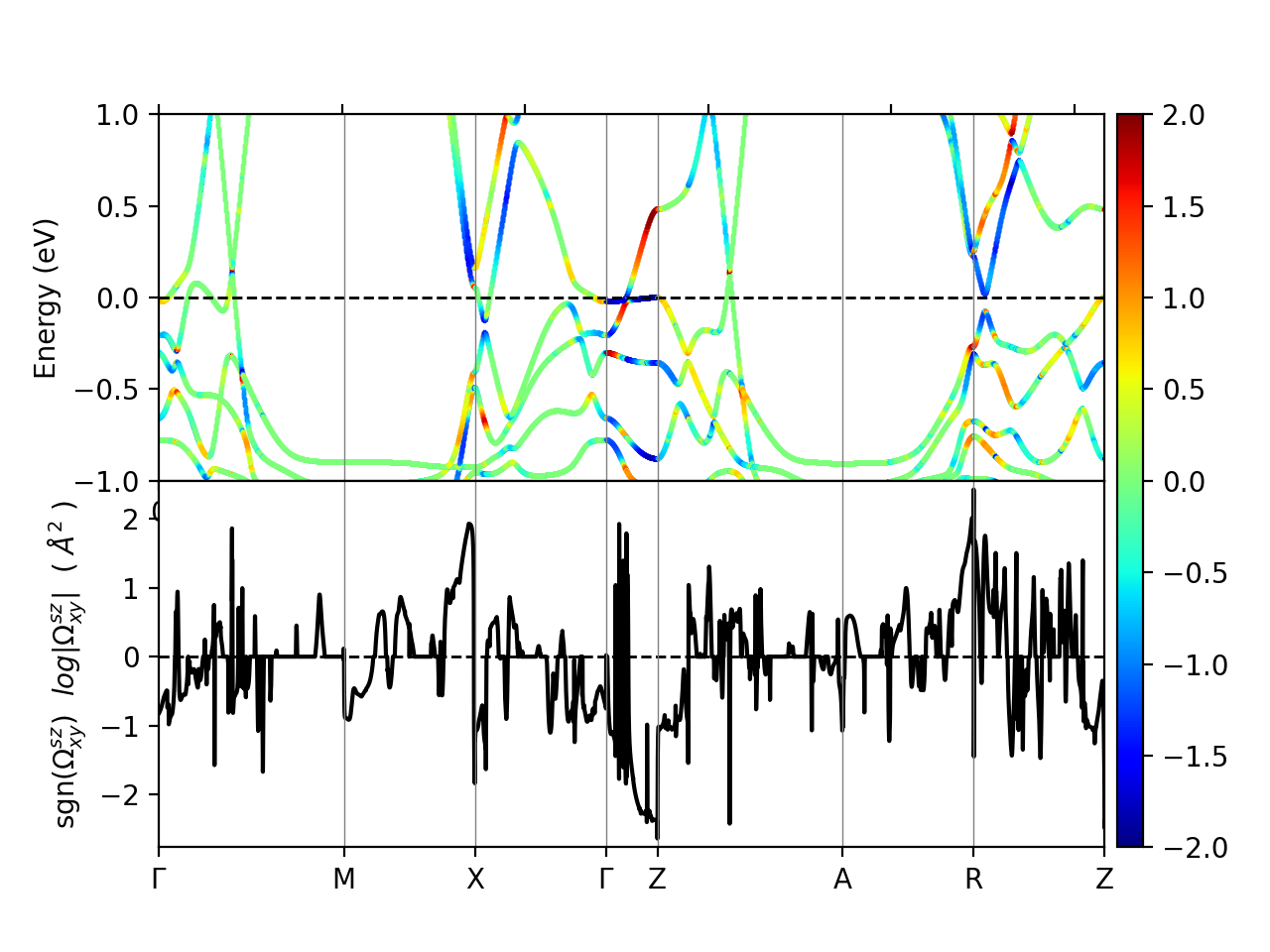}
        } 
\hskip -0.05 in
        \subfigure[]{
            \includegraphics[width=0.47\textwidth]{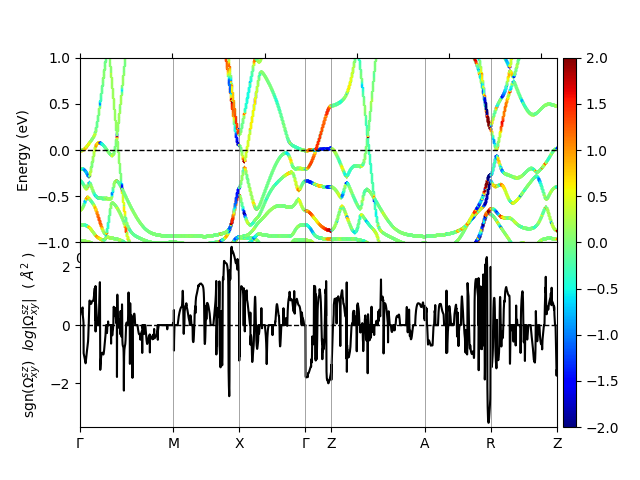}
         }
            \end{center}
            \caption{Comparison of the \textbf{k}-resolved spin berry curvature $\Omega_{xy}^{sz}$ between (a) AAF-\zhat~ and (b) AAF-\xhat~ patterns.
                   Top panels: Band structure along high symmetry directions colored by the  spin Berry curvature of individual bands $\Omega_{n,xy}^{sz}(\mathbf{k})$. The color bar is in the log scale and sign indicates the sign of the spin Berry curvature.
                    Bottom panels: Sum of the $\Omega_{n,xy}^{z}(\mathbf{k})$ upto the Fermi level along the high symmetry directions.
                    In these plots, the values of  $\Omega_{xy}^{z}(\mathbf{k})$ below 1 \AA$^2$ are considered 0.
        }
        \label{fig:AAF-spinberry} 
\end{figure*}

In the previous section, we saw that the orientation of the \Neel vector gives very small changes in the electronic structure.
In this section, we show that even though the electronic dispersion appears similar, the difference in the spin dependent transport property is appreciable.
To highlight this difference we will study spin Hall conductivity (SHC).

The phenomenon of Spin Hall effect corresponds to  generation of a purely transverse spin current by the  applied electric field~\cite{SHE_Sinova_JungwirthRMP2015}.
Out of the three mechanisms (intrinsic, skew scattering and side-jump) contributing to the SHC, 
the intrinsic component of the SHC is a direct consequence of the band topology similar to the phenomenon of anomalous Hall conductivity when the applied electric field generates a  transverse charge current.
SHC is a third rank tensor and is defined as~\cite{SHCWannier_Qiao_Zhao_PRB2018}:
\begin{equation}
    \sigma_{ab}^{c} (\mu) = -\frac{e^2}{\hbar} \int_{BZ} \frac{d^3 {\bf k}}{(2\pi)^3}\sum_n  f_{n\bf{k}}(\mu) \Omega_{n,ab}^{c}({\bf k}),
    \label{eqn:conductivity}
\end{equation}
where,
$f_{nk}(\mu)$ is the $\bf k$-dependent equilibrium occupation factor of $n^{th}$ band at the chemical potential of $\mu$. 
 $\Omega_{n,ab}^{c}$ is the band resolved spin Berry curvature which is, in general, a function of $\bf k$ and frequency $\omega$ and is given by:
\begin{eqnarray}
  && \Omega_{n,ab}^{c}(\omega,\mathbf{k}) = \label{eqn:SpinBerryCurvature}\\
 && -2\hbar^2\sum_{m \neq n} \mathrm{Im} \frac{\langle n\mathbf{k} |\{\hat{\sigma}_c,\hat{v}_a \} |m\mathbf{k}\rangle \langle m\mathbf{k} | \hat{v}_b|n\mathbf{k}\rangle}{\Delta^2_{nm}(k) - (\hbar\omega +i\eta)^2},
    \nonumber
\end{eqnarray}
where, $\hat{v}_i = \frac{\partial }{\partial k_b}$
is the velocity operator, $\hat{\sigma}$ is the Pauli spin matrix,  
and 
$\Delta_{nm}(k)=E_{nk} - E_{mk}$.
Here, we ignore the $\omega$ dependence of the spin Berry curvature.


\begin{table}
    \caption{The form of the spin Hall conductivity tensor for the \xhat~ and \zhat~ magnetic phases for both AAF and AAF3 magnetic patterns.}
    \label{table:shc_tensor_form}
\begin{center}
\begin{tabular}{ |p{25pt}|p{75pt}p{65pt}p{65pt}|}
 \hline \hline
& ~~~$\sigma^x$ & ~~~$\sigma^y$ & ~~~$\sigma^z$\\
 \hline
 \zhat-phase  & $\left(\begin{matrix}0 & 0 & 0\\0 & 0 & - \sigma^y_{xz}\\0 & - \sigma^y_{zx} & 0\end{matrix}\right)$ & $\left(\begin{matrix}0 & 0 & \sigma^y_{xz}\\ 0 & 0 & 0\\ \sigma^y_{zx} & 0 & 0\end{matrix}\right)$ & $\left(\begin{matrix} 0 & - \sigma^z_{yx} & 0\\ \sigma^z_{yx} & 0 & 0\\0 & 0 & 0\end{matrix}\right)$ \\ \hline
 \xhat-phase  & $\left(\begin{matrix}0 & 0 & 0\\0 & 0 & \sigma^x_{yz}\\0 & \sigma^x_{zy} & 0\end{matrix}\right)$ & $\left(\begin{matrix}0 & 0 & \sigma^y_{xz}\\ 0 & 0 & 0\\ \sigma^y_{zx} & 0 & 0\end{matrix}\right)$ & $\left(\begin{matrix} 0 & \sigma^z_{xy} & 0\\ \sigma^z_{yx} & 0 & 0\\0 & 0 & 0\end{matrix}\right)$ \\
 \hline
\end{tabular}
\end{center}
\end{table}
Eq.~(\ref{eqn:SpinBerryCurvature}) is deceivingly similar to the formula for the normal (charge) Berry curvature; the only difference is that here one has to evaluate the matrix element of the anti-commutator between the velocity operator and Pauli matrix instead of just the velocity operator in the normal Berry curvature.

SHC, being a 3$^{rd}$ rank tensor, has 27 components; the magnetic symmetry determines which of them are non-zero.
From the symmetry analysis, we find that there are only 3 (6) independent components for the \zhat~(\xhat) phase. The exact form of the SHC tensor for each of these phases is presented explicitly in Table ~\ref{table:shc_tensor_form}.
The reason \xhat -phase has twice the number of independent components compared to the \zhat -phase is a consequence of the broken  tetragonal symmetry due to the broken $\mathcal{R}^4$ symmetry as mentioned before.
Note that due to the $\Pcal\Tcal$ symmetry, the anomalous Hall conductivity is identically zero in the AFM phase. 

In Fig.~\ref{fig:AAF-SHC}, we compare the $\sigma_{xy}^z$ component 
for the \xhat~ and \zhat~ \Neel vector patterns for both AAF and AAF3 magnetic phases as a function of the chemical potential ($\mu$).
Other non-zero components of SHC $\sigma_{yz}^x$ and $\sigma_{zx}^y$ are shown in Appendix \ref{appendixB} Fig.~\ref{fig:appen-AAF-SHC}.
$\sigma_{xy}^z$ measures the \zhat~ component of the spin current along the \xhat~ direction in the presence of the external field in the $\hat{y}$  direction.
We find that the differences in the SHC value between the \xhat~ and \zhat~ patterns are appreciable.
Interestingly, the magnitude of SHC increases by more than 2-fold for the AAF3 pattern compared to the AAF pattern in the vicinity of the Fermi level. This is likely due to the presence of band crossings or small band gaps in the AAF3 pattern. 


Fig.~\ref{fig:AAF-spinberry} shows momentum resolved spin Berry curvature along the high symmetry directions for the \zhat- and \xhat-phases of the AAF magnetic pattern.
The intensity on the top panels shows the magnitude of the band resolved spin Berry curvature for each  $\textbf{k}$-value whereas
the bottom panels show the band summation of the \textbf{k}-dependent spin Berry curvature up to the occupied states (E$_F$).
These figures highlight the fact that despite the bulk band structure features being similar, the distribution and magnitude of the spin Berry curvature can be different due to the small differences in the band eigenvalues and spin texture of the bands. Our analysis from effective Kondo exchange Hamiltonian presented in Appendix~\ref{appendixB} indeed finds that the band dispersion and their spin texture will be different depending on the orientation of the \Neel vector.

\section{Conclusions and Outlook}
\label{Sec:Conclusion}
In summary, we have studied the electronic structure and magnetic properties of an antiferromagnetic square net topological semimetal \mater~ by employing the first-principles and effective Hamiltonian methods.
We have found that the effect of magnetism on the bulk low energy spectrum, especially that introduced by the orientation of the \Neel vector, is weak. 
Despite such weak effects in the bulk dispersion, we find that there are consequences for the transport properties and surface electronic dispersion.
For example, our calculations predict that the broken symmetry introduced by the direction of the \Neel vector is amplified in the surface electronic dispersion.
Similarly, the differences in the spin Hall conductivity response between different magnetic phases is appreciable.
We derived an effective Kondo exchange Hamiltonian to understand our main findings.
It will be interesting to confirm some of the predictions made in this study by experiments like ARPES  in conjunction with the spin orbit torque experiments that can control the orientation of the \Neel vector~\cite{SpinOrbitTorqueCuMnAs_Jungwirth_PRL2017}.
Because of the presence of the competing magnetic states which depend on the strength of the interaction term in our calculations, it will also be intriguing to study the possibilities of controlling the magnetic ground state and the band topology by means of small external perturbations like pressure, doping, intercalation, chemical substitution etc.
Similarly, study of the surface magnetism,  surface transport properties etc. could be other directions in these investigations.

To conclude, our work  provides a compelling  evidence
that study of the \textit{f}-electron AFM  can be a promising field for  band engineering and spintronics applications. 
Similar investigations are necessary for other predicted \textit{f}-electron square net systems in order to make systematic comparisons and predictions. 
We believe that our study will motivate future works in this direction, especially towards prediction and search of Dirac materials showing large electronic response to the magnetic texture, using more sophisticated numerical techniques.

\section{Acknowledgements}
This work was supported by U.S. Department of Energy (DOE) the Office of Basic Energy Sciences, Materials Sciences and Engineering Division under Contract No. DE-SC0012704. 


\begin{thebibliography}{46}%
\makeatletter
\providecommand \@ifxundefined [1]{%
 \@ifx{#1\undefined}
}%
\providecommand \@ifnum [1]{%
 \ifnum #1\expandafter \@firstoftwo
 \else \expandafter \@secondoftwo
 \fi
}%
\providecommand \@ifx [1]{%
 \ifx #1\expandafter \@firstoftwo
 \else \expandafter \@secondoftwo
 \fi
}%
\providecommand \natexlab [1]{#1}%
\providecommand \enquote  [1]{``#1''}%
\providecommand \bibnamefont  [1]{#1}%
\providecommand \bibfnamefont [1]{#1}%
\providecommand \citenamefont [1]{#1}%
\providecommand \href@noop [0]{\@secondoftwo}%
\providecommand \href [0]{\begingroup \@sanitize@url \@href}%
\providecommand \@href[1]{\@@startlink{#1}\@@href}%
\providecommand \@@href[1]{\endgroup#1\@@endlink}%
\providecommand \@sanitize@url [0]{\catcode `\\12\catcode `\$12\catcode
  `\&12\catcode `\#12\catcode `\^12\catcode `\_12\catcode `\%12\relax}%
\providecommand \@@startlink[1]{}%
\providecommand \@@endlink[0]{}%
\providecommand \url  [0]{\begingroup\@sanitize@url \@url }%
\providecommand \@url [1]{\endgroup\@href {#1}{\urlprefix }}%
\providecommand \urlprefix  [0]{URL }%
\providecommand \Eprint [0]{\href }%
\providecommand \doibase [0]{http://dx.doi.org/}%
\providecommand \selectlanguage [0]{\@gobble}%
\providecommand \bibinfo  [0]{\@secondoftwo}%
\providecommand \bibfield  [0]{\@secondoftwo}%
\providecommand \translation [1]{[#1]}%
\providecommand \BibitemOpen [0]{}%
\providecommand \bibitemStop [0]{}%
\providecommand \bibitemNoStop [0]{.\EOS\space}%
\providecommand \EOS [0]{\spacefactor3000\relax}%
\providecommand \BibitemShut  [1]{\csname bibitem#1\endcsname}%
\let\auto@bib@innerbib\@empty
\bibitem [{\citenamefont {Wang}\ \emph {et~al.}(2020)\citenamefont {Wang},
  \citenamefont {Baranets}, \citenamefont {Liu}, \citenamefont {Tong},
  \citenamefont {Stavitski}, \citenamefont {Zhang}, \citenamefont {Chai},
  \citenamefont {Yin}, \citenamefont {Bobev},\ and\ \citenamefont
  {Petrovic}}]{WangEuZnSb2}%
  \BibitemOpen
  \bibfield  {author} {\bibinfo {author} {\bibfnamefont {A.}~\bibnamefont
  {Wang}}, \bibinfo {author} {\bibfnamefont {S.}~\bibnamefont {Baranets}},
  \bibinfo {author} {\bibfnamefont {Y.}~\bibnamefont {Liu}}, \bibinfo {author}
  {\bibfnamefont {X.}~\bibnamefont {Tong}}, \bibinfo {author} {\bibfnamefont
  {E.}~\bibnamefont {Stavitski}}, \bibinfo {author} {\bibfnamefont
  {J.}~\bibnamefont {Zhang}}, \bibinfo {author} {\bibfnamefont
  {Y.}~\bibnamefont {Chai}}, \bibinfo {author} {\bibfnamefont {W.-G.}\
  \bibnamefont {Yin}}, \bibinfo {author} {\bibfnamefont {S.}~\bibnamefont
  {Bobev}}, \ and\ \bibinfo {author} {\bibfnamefont {C.}~\bibnamefont
  {Petrovic}},\ }\href {\doibase 10.1103/PhysRevResearch.2.033462} {\bibfield
  {journal} {\bibinfo  {journal} {Phys. Rev. Research}\ }\textbf {\bibinfo
  {volume} {2}},\ \bibinfo {pages} {033462} (\bibinfo {year}
  {2020})}\BibitemShut {NoStop}%
\bibitem [{\citenamefont {Wang}\ \emph {et~al.}(2018)\citenamefont {Wang},
  \citenamefont {Xu}, \citenamefont {Lou}, \citenamefont {Liu}, \citenamefont
  {Li}, \citenamefont {Huang}, \citenamefont {Shen}, \citenamefont {Weng},
  \citenamefont {Wang},\ and\ \citenamefont
  {Lei}}]{MagneticWeyl_Wang_NatureComm2018}%
  \BibitemOpen
  \bibfield  {author} {\bibinfo {author} {\bibfnamefont {Q.}~\bibnamefont
  {Wang}}, \bibinfo {author} {\bibfnamefont {Y.}~\bibnamefont {Xu}}, \bibinfo
  {author} {\bibfnamefont {R.}~\bibnamefont {Lou}}, \bibinfo {author}
  {\bibfnamefont {Z.}~\bibnamefont {Liu}}, \bibinfo {author} {\bibfnamefont
  {M.}~\bibnamefont {Li}}, \bibinfo {author} {\bibfnamefont {Y.}~\bibnamefont
  {Huang}}, \bibinfo {author} {\bibfnamefont {D.}~\bibnamefont {Shen}},
  \bibinfo {author} {\bibfnamefont {H.}~\bibnamefont {Weng}}, \bibinfo {author}
  {\bibfnamefont {S.}~\bibnamefont {Wang}}, \ and\ \bibinfo {author}
  {\bibfnamefont {H.}~\bibnamefont {Lei}},\ }\href
  {https://doi.org/10.1038/s41467-018-06088-2} {\bibfield  {journal} {\bibinfo
  {journal} {Nature Communications}\ }\textbf {\bibinfo {volume} {9}},\
  \bibinfo {pages} {3681} (\bibinfo {year} {2018})}\BibitemShut {NoStop}%
\bibitem [{\citenamefont {Li}\ \emph {et~al.}(2010)\citenamefont {Li},
  \citenamefont {Wang}, \citenamefont {Qi},\ and\ \citenamefont
  {Zhang}}]{AxionMTI_SCZhang_NaturePhys2010}%
  \BibitemOpen
  \bibfield  {author} {\bibinfo {author} {\bibfnamefont {R.}~\bibnamefont
  {Li}}, \bibinfo {author} {\bibfnamefont {J.}~\bibnamefont {Wang}}, \bibinfo
  {author} {\bibfnamefont {X.-L.}\ \bibnamefont {Qi}}, \ and\ \bibinfo {author}
  {\bibfnamefont {S.-C.}\ \bibnamefont {Zhang}},\ }\href
  {https://doi.org/10.1038/nphys1534} {\bibfield  {journal} {\bibinfo
  {journal} {Nature Physics}\ }\textbf {\bibinfo {volume} {6}},\ \bibinfo
  {pages} {284} (\bibinfo {year} {2010})}\BibitemShut {NoStop}%
\bibitem [{\citenamefont {Šmejkal}\ \emph {et~al.}(2018)\citenamefont
  {Šmejkal}, \citenamefont {Mokrousov}, \citenamefont {Yan},\ and\
  \citenamefont {MacDonald}}]{TAFMSpin_Smejkal_NaturePhysics2018}%
  \BibitemOpen
  \bibfield  {author} {\bibinfo {author} {\bibfnamefont {L.}~\bibnamefont
  {Šmejkal}}, \bibinfo {author} {\bibfnamefont {Y.}~\bibnamefont {Mokrousov}},
  \bibinfo {author} {\bibfnamefont {B.}~\bibnamefont {Yan}}, \ and\ \bibinfo
  {author} {\bibfnamefont {A.~H.}\ \bibnamefont {MacDonald}},\ }\href
  {https://doi.org/10.1038/s41567-018-0064-5} {\bibfield  {journal} {\bibinfo
  {journal} {Nature Physics}\ }\textbf {\bibinfo {volume} {14}},\ \bibinfo
  {pages} {242} (\bibinfo {year} {2018})}\BibitemShut {NoStop}%
\bibitem [{\citenamefont {Otrokov}\ \emph {et~al.}(2019)\citenamefont
  {Otrokov}, \citenamefont {Klimovskikh}, \citenamefont {Bentmann},
  \citenamefont {Estyunin}, \citenamefont {Zeugner}, \citenamefont {Aliev},
  \citenamefont {Gaß}, \citenamefont {Wolter}, \citenamefont {Koroleva},
  \citenamefont {Shikin}, \citenamefont {Blanco-Rey}, \citenamefont {Hoffmann},
  \citenamefont {Rusinov}, \citenamefont {Vyazovskaya}, \citenamefont
  {Eremeev}, \citenamefont {Koroteev}, \citenamefont {Kuznetsov}, \citenamefont
  {Freyse}, \citenamefont {Sánchez-Barriga}, \citenamefont {Amiraslanov},
  \citenamefont {Babanly}, \citenamefont {Mamedov}, \citenamefont {Abdullayev},
  \citenamefont {Zverev}, \citenamefont {Alfonsov}, \citenamefont {Kataev},
  \citenamefont {Büchner}, \citenamefont {Schwier}, \citenamefont {Kumar},
  \citenamefont {Kimura}, \citenamefont {Petaccia}, \citenamefont {Di~Santo},
  \citenamefont {Vidal}, \citenamefont {Schatz}, \citenamefont {Kißner},
  \citenamefont {Ünzelmann}, \citenamefont {Min}, \citenamefont {Moser},
  \citenamefont {Peixoto}, \citenamefont {Reinert}, \citenamefont {Ernst},
  \citenamefont {Echenique}, \citenamefont {Isaeva},\ and\ \citenamefont
  {Chulkov}}]{AFMTI_Otrokov_Nature2019}%
  \BibitemOpen
  \bibfield  {author} {\bibinfo {author} {\bibfnamefont {M.~M.}\ \bibnamefont
  {Otrokov}}, \bibinfo {author} {\bibfnamefont {I.~I.}\ \bibnamefont
  {Klimovskikh}}, \bibinfo {author} {\bibfnamefont {H.}~\bibnamefont
  {Bentmann}}, \bibinfo {author} {\bibfnamefont {D.}~\bibnamefont {Estyunin}},
  \bibinfo {author} {\bibfnamefont {A.}~\bibnamefont {Zeugner}}, \bibinfo
  {author} {\bibfnamefont {Z.~S.}\ \bibnamefont {Aliev}}, \bibinfo {author}
  {\bibfnamefont {S.}~\bibnamefont {Gaß}}, \bibinfo {author} {\bibfnamefont
  {A.~U.~B.}\ \bibnamefont {Wolter}}, \bibinfo {author} {\bibfnamefont {A.~V.}\
  \bibnamefont {Koroleva}}, \bibinfo {author} {\bibfnamefont {A.~M.}\
  \bibnamefont {Shikin}}, \bibinfo {author} {\bibfnamefont {M.}~\bibnamefont
  {Blanco-Rey}}, \bibinfo {author} {\bibfnamefont {M.}~\bibnamefont
  {Hoffmann}}, \bibinfo {author} {\bibfnamefont {I.~P.}\ \bibnamefont
  {Rusinov}}, \bibinfo {author} {\bibfnamefont {A.~Y.}\ \bibnamefont
  {Vyazovskaya}}, \bibinfo {author} {\bibfnamefont {S.~V.}\ \bibnamefont
  {Eremeev}}, \bibinfo {author} {\bibfnamefont {Y.~M.}\ \bibnamefont
  {Koroteev}}, \bibinfo {author} {\bibfnamefont {V.~M.}\ \bibnamefont
  {Kuznetsov}}, \bibinfo {author} {\bibfnamefont {F.}~\bibnamefont {Freyse}},
  \bibinfo {author} {\bibfnamefont {J.}~\bibnamefont {Sánchez-Barriga}},
  \bibinfo {author} {\bibfnamefont {I.~R.}\ \bibnamefont {Amiraslanov}},
  \bibinfo {author} {\bibfnamefont {M.~B.}\ \bibnamefont {Babanly}}, \bibinfo
  {author} {\bibfnamefont {N.~T.}\ \bibnamefont {Mamedov}}, \bibinfo {author}
  {\bibfnamefont {N.~A.}\ \bibnamefont {Abdullayev}}, \bibinfo {author}
  {\bibfnamefont {V.~N.}\ \bibnamefont {Zverev}}, \bibinfo {author}
  {\bibfnamefont {A.}~\bibnamefont {Alfonsov}}, \bibinfo {author}
  {\bibfnamefont {V.}~\bibnamefont {Kataev}}, \bibinfo {author} {\bibfnamefont
  {B.}~\bibnamefont {Büchner}}, \bibinfo {author} {\bibfnamefont {E.~F.}\
  \bibnamefont {Schwier}}, \bibinfo {author} {\bibfnamefont {S.}~\bibnamefont
  {Kumar}}, \bibinfo {author} {\bibfnamefont {A.}~\bibnamefont {Kimura}},
  \bibinfo {author} {\bibfnamefont {L.}~\bibnamefont {Petaccia}}, \bibinfo
  {author} {\bibfnamefont {G.}~\bibnamefont {Di~Santo}}, \bibinfo {author}
  {\bibfnamefont {R.~C.}\ \bibnamefont {Vidal}}, \bibinfo {author}
  {\bibfnamefont {S.}~\bibnamefont {Schatz}}, \bibinfo {author} {\bibfnamefont
  {K.}~\bibnamefont {Kißner}}, \bibinfo {author} {\bibfnamefont
  {M.}~\bibnamefont {Ünzelmann}}, \bibinfo {author} {\bibfnamefont {C.~H.}\
  \bibnamefont {Min}}, \bibinfo {author} {\bibfnamefont {S.}~\bibnamefont
  {Moser}}, \bibinfo {author} {\bibfnamefont {T.~R.~F.}\ \bibnamefont
  {Peixoto}}, \bibinfo {author} {\bibfnamefont {F.}~\bibnamefont {Reinert}},
  \bibinfo {author} {\bibfnamefont {A.}~\bibnamefont {Ernst}}, \bibinfo
  {author} {\bibfnamefont {P.~M.}\ \bibnamefont {Echenique}}, \bibinfo {author}
  {\bibfnamefont {A.}~\bibnamefont {Isaeva}}, \ and\ \bibinfo {author}
  {\bibfnamefont {E.~V.}\ \bibnamefont {Chulkov}},\ }\href
  {https://doi.org/10.1038/s41586-019-1840-9} {\bibfield  {journal} {\bibinfo
  {journal} {Nature}\ }\textbf {\bibinfo {volume} {576}},\ \bibinfo {pages}
  {416} (\bibinfo {year} {2019})}\BibitemShut {NoStop}%
\bibitem [{\citenamefont {Máca}\ \emph {et~al.}(2012)\citenamefont {Máca},
  \citenamefont {Mašek}, \citenamefont {Stelmakhovych}, \citenamefont
  {Martí}, \citenamefont {Reichlová}, \citenamefont {Uhlířová},
  \citenamefont {Beran}, \citenamefont {Wadley}, \citenamefont {Novák},\ and\
  \citenamefont {Jungwirth}}]{DiracAFM_Maca_JMMM2012}%
  \BibitemOpen
  \bibfield  {author} {\bibinfo {author} {\bibfnamefont {F.}~\bibnamefont
  {Máca}}, \bibinfo {author} {\bibfnamefont {J.}~\bibnamefont {Mašek}},
  \bibinfo {author} {\bibfnamefont {O.}~\bibnamefont {Stelmakhovych}}, \bibinfo
  {author} {\bibfnamefont {X.}~\bibnamefont {Martí}}, \bibinfo {author}
  {\bibfnamefont {H.}~\bibnamefont {Reichlová}}, \bibinfo {author}
  {\bibfnamefont {K.}~\bibnamefont {Uhlířová}}, \bibinfo {author}
  {\bibfnamefont {P.}~\bibnamefont {Beran}}, \bibinfo {author} {\bibfnamefont
  {P.}~\bibnamefont {Wadley}}, \bibinfo {author} {\bibfnamefont
  {V.}~\bibnamefont {Novák}}, \ and\ \bibinfo {author} {\bibfnamefont
  {T.}~\bibnamefont {Jungwirth}},\ }\href {\doibase
  https://doi.org/10.1016/j.jmmm.2011.12.017} {\bibfield  {journal} {\bibinfo
  {journal} {Journal of Magnetism and Magnetic Materials}\ }\textbf {\bibinfo
  {volume} {324}},\ \bibinfo {pages} {1606} (\bibinfo {year}
  {2012})}\BibitemShut {NoStop}%
\bibitem [{\citenamefont {Tang}\ \emph {et~al.}(2016)\citenamefont {Tang},
  \citenamefont {Zhou}, \citenamefont {Xu},\ and\ \citenamefont
  {Zhang}}]{DiracAFM_Zhang_NaturePhys_2016}%
  \BibitemOpen
  \bibfield  {author} {\bibinfo {author} {\bibfnamefont {P.}~\bibnamefont
  {Tang}}, \bibinfo {author} {\bibfnamefont {Q.}~\bibnamefont {Zhou}}, \bibinfo
  {author} {\bibfnamefont {G.}~\bibnamefont {Xu}}, \ and\ \bibinfo {author}
  {\bibfnamefont {S.-C.}\ \bibnamefont {Zhang}},\ }\href
  {https://doi.org/10.1038/nphys3839} {\bibfield  {journal} {\bibinfo
  {journal} {Nature Physics}\ }\textbf {\bibinfo {volume} {12}},\ \bibinfo
  {pages} {1100} (\bibinfo {year} {2016})}\BibitemShut {NoStop}%
\bibitem [{\citenamefont {Niu}\ \emph {et~al.}(2020)\citenamefont {Niu},
  \citenamefont {Wang}, \citenamefont {Mao}, \citenamefont {Huang},
  \citenamefont {Mokrousov},\ and\ \citenamefont {Dai}}]{AFM2D_Niu_PRL2020}%
  \BibitemOpen
  \bibfield  {author} {\bibinfo {author} {\bibfnamefont {C.}~\bibnamefont
  {Niu}}, \bibinfo {author} {\bibfnamefont {H.}~\bibnamefont {Wang}}, \bibinfo
  {author} {\bibfnamefont {N.}~\bibnamefont {Mao}}, \bibinfo {author}
  {\bibfnamefont {B.}~\bibnamefont {Huang}}, \bibinfo {author} {\bibfnamefont
  {Y.}~\bibnamefont {Mokrousov}}, \ and\ \bibinfo {author} {\bibfnamefont
  {Y.}~\bibnamefont {Dai}},\ }\href {\doibase 10.1103/PhysRevLett.124.066401}
  {\bibfield  {journal} {\bibinfo  {journal} {Phys. Rev. Lett.}\ }\textbf
  {\bibinfo {volume} {124}},\ \bibinfo {pages} {066401} (\bibinfo {year}
  {2020})}\BibitemShut {NoStop}%
\bibitem [{\citenamefont {Wan}\ \emph {et~al.}(2011)\citenamefont {Wan},
  \citenamefont {Turner}, \citenamefont {Vishwanath},\ and\ \citenamefont
  {Savrasov}}]{PyrochloreIridates_Wan_PRB2011}%
  \BibitemOpen
  \bibfield  {author} {\bibinfo {author} {\bibfnamefont {X.}~\bibnamefont
  {Wan}}, \bibinfo {author} {\bibfnamefont {A.~M.}\ \bibnamefont {Turner}},
  \bibinfo {author} {\bibfnamefont {A.}~\bibnamefont {Vishwanath}}, \ and\
  \bibinfo {author} {\bibfnamefont {S.~Y.}\ \bibnamefont {Savrasov}},\ }\href
  {\doibase 10.1103/PhysRevB.83.205101} {\bibfield  {journal} {\bibinfo
  {journal} {Phys. Rev. B}\ }\textbf {\bibinfo {volume} {83}},\ \bibinfo
  {pages} {205101} (\bibinfo {year} {2011})}\BibitemShut {NoStop}%
\bibitem [{\citenamefont {Xu}\ \emph {et~al.}(2011)\citenamefont {Xu},
  \citenamefont {Weng}, \citenamefont {Wang}, \citenamefont {Dai},\ and\
  \citenamefont {Fang}}]{MagneticWeylHgCrSe_Xu_PRL2011}%
  \BibitemOpen
  \bibfield  {author} {\bibinfo {author} {\bibfnamefont {G.}~\bibnamefont
  {Xu}}, \bibinfo {author} {\bibfnamefont {H.}~\bibnamefont {Weng}}, \bibinfo
  {author} {\bibfnamefont {Z.}~\bibnamefont {Wang}}, \bibinfo {author}
  {\bibfnamefont {X.}~\bibnamefont {Dai}}, \ and\ \bibinfo {author}
  {\bibfnamefont {Z.}~\bibnamefont {Fang}},\ }\href {\doibase
  10.1103/PhysRevLett.107.186806} {\bibfield  {journal} {\bibinfo  {journal}
  {Phys. Rev. Lett.}\ }\textbf {\bibinfo {volume} {107}},\ \bibinfo {pages}
  {186806} (\bibinfo {year} {2011})}\BibitemShut {NoStop}%
\bibitem [{\citenamefont {Yang}\ \emph {et~al.}(2017)\citenamefont {Yang},
  \citenamefont {Sun}, \citenamefont {Zhang}, \citenamefont {Shi},
  \citenamefont {Parkin},\ and\ \citenamefont {Yan}}]{Mn3Sn_AFMWeyl_NJP2017}%
  \BibitemOpen
  \bibfield  {author} {\bibinfo {author} {\bibfnamefont {H.}~\bibnamefont
  {Yang}}, \bibinfo {author} {\bibfnamefont {Y.}~\bibnamefont {Sun}}, \bibinfo
  {author} {\bibfnamefont {Y.}~\bibnamefont {Zhang}}, \bibinfo {author}
  {\bibfnamefont {W.-J.}\ \bibnamefont {Shi}}, \bibinfo {author} {\bibfnamefont
  {S.~S.~P.}\ \bibnamefont {Parkin}}, \ and\ \bibinfo {author} {\bibfnamefont
  {B.}~\bibnamefont {Yan}},\ }\href {\doibase 10.1088/1367-2630/aa5487}
  {\bibfield  {journal} {\bibinfo  {journal} {New Journal of Physics}\ }\textbf
  {\bibinfo {volume} {19}},\ \bibinfo {pages} {015008} (\bibinfo {year}
  {2017})}\BibitemShut {NoStop}%
\bibitem [{\citenamefont {Xu}\ \emph {et~al.}(2020)\citenamefont {Xu},
  \citenamefont {Elcoro}, \citenamefont {Song}, \citenamefont {Wieder},
  \citenamefont {Vergniory}, \citenamefont {Regnault}, \citenamefont {Chen},
  \citenamefont {Felser},\ and\ \citenamefont
  {Bernevig}}]{HighThroughput_Xu_Nature2020}%
  \BibitemOpen
  \bibfield  {author} {\bibinfo {author} {\bibfnamefont {Y.}~\bibnamefont
  {Xu}}, \bibinfo {author} {\bibfnamefont {L.}~\bibnamefont {Elcoro}}, \bibinfo
  {author} {\bibfnamefont {Z.-D.}\ \bibnamefont {Song}}, \bibinfo {author}
  {\bibfnamefont {B.~J.}\ \bibnamefont {Wieder}}, \bibinfo {author}
  {\bibfnamefont {M.~G.}\ \bibnamefont {Vergniory}}, \bibinfo {author}
  {\bibfnamefont {N.}~\bibnamefont {Regnault}}, \bibinfo {author}
  {\bibfnamefont {Y.}~\bibnamefont {Chen}}, \bibinfo {author} {\bibfnamefont
  {C.}~\bibnamefont {Felser}}, \ and\ \bibinfo {author} {\bibfnamefont {B.~A.}\
  \bibnamefont {Bernevig}},\ }\href {https://doi.org/10.1038/s41586-020-2837-0}
  {\bibfield  {journal} {\bibinfo  {journal} {Nature}\ }\textbf {\bibinfo
  {volume} {586}},\ \bibinfo {pages} {702} (\bibinfo {year}
  {2020})}\BibitemShut {NoStop}%
\bibitem [{\citenamefont {Zou}\ \emph {et~al.}(2019)\citenamefont {Zou},
  \citenamefont {He},\ and\ \citenamefont {Xu}}]{ReviewMTSM_Zho_npjCM2019}%
  \BibitemOpen
  \bibfield  {author} {\bibinfo {author} {\bibfnamefont {J.}~\bibnamefont
  {Zou}}, \bibinfo {author} {\bibfnamefont {Z.}~\bibnamefont {He}}, \ and\
  \bibinfo {author} {\bibfnamefont {G.}~\bibnamefont {Xu}},\ }\href {\doibase
  10.1038/s41524-019-0237-5} {\bibfield  {journal} {\bibinfo  {journal} {npj
  Computational Materials}\ }\textbf {\bibinfo {volume} {5}},\ \bibinfo {pages}
  {96} (\bibinfo {year} {2019})}\BibitemShut {NoStop}%
\bibitem [{\citenamefont {Wadley}\ \emph {et~al.}(2016)\citenamefont {Wadley},
  \citenamefont {Howells}, \citenamefont {{\v Z}elezn{\'y}}, \citenamefont
  {Andrews}, \citenamefont {Hills}, \citenamefont {Campion}, \citenamefont
  {Nov{\'a}k}, \citenamefont {Olejn{\'\i}k}, \citenamefont {Maccherozzi},
  \citenamefont {Dhesi}, \citenamefont {Martin}, \citenamefont {Wagner},
  \citenamefont {Wunderlich}, \citenamefont {Freimuth}, \citenamefont
  {Mokrousov}, \citenamefont {Kune{\v s}}, \citenamefont {Chauhan},
  \citenamefont {Grzybowski}, \citenamefont {Rushforth}, \citenamefont
  {Edmonds}, \citenamefont {Gallagher},\ and\ \citenamefont
  {Jungwirth}}]{SwitchingAFM_Wadley_Science2016}%
  \BibitemOpen
  \bibfield  {author} {\bibinfo {author} {\bibfnamefont {P.}~\bibnamefont
  {Wadley}}, \bibinfo {author} {\bibfnamefont {B.}~\bibnamefont {Howells}},
  \bibinfo {author} {\bibfnamefont {J.}~\bibnamefont {{\v Z}elezn{\'y}}},
  \bibinfo {author} {\bibfnamefont {C.}~\bibnamefont {Andrews}}, \bibinfo
  {author} {\bibfnamefont {V.}~\bibnamefont {Hills}}, \bibinfo {author}
  {\bibfnamefont {R.~P.}\ \bibnamefont {Campion}}, \bibinfo {author}
  {\bibfnamefont {V.}~\bibnamefont {Nov{\'a}k}}, \bibinfo {author}
  {\bibfnamefont {K.}~\bibnamefont {Olejn{\'\i}k}}, \bibinfo {author}
  {\bibfnamefont {F.}~\bibnamefont {Maccherozzi}}, \bibinfo {author}
  {\bibfnamefont {S.~S.}\ \bibnamefont {Dhesi}}, \bibinfo {author}
  {\bibfnamefont {S.~Y.}\ \bibnamefont {Martin}}, \bibinfo {author}
  {\bibfnamefont {T.}~\bibnamefont {Wagner}}, \bibinfo {author} {\bibfnamefont
  {J.}~\bibnamefont {Wunderlich}}, \bibinfo {author} {\bibfnamefont
  {F.}~\bibnamefont {Freimuth}}, \bibinfo {author} {\bibfnamefont
  {Y.}~\bibnamefont {Mokrousov}}, \bibinfo {author} {\bibfnamefont
  {J.}~\bibnamefont {Kune{\v s}}}, \bibinfo {author} {\bibfnamefont {J.~S.}\
  \bibnamefont {Chauhan}}, \bibinfo {author} {\bibfnamefont {M.~J.}\
  \bibnamefont {Grzybowski}}, \bibinfo {author} {\bibfnamefont {A.~W.}\
  \bibnamefont {Rushforth}}, \bibinfo {author} {\bibfnamefont {K.~W.}\
  \bibnamefont {Edmonds}}, \bibinfo {author} {\bibfnamefont {B.~L.}\
  \bibnamefont {Gallagher}}, \ and\ \bibinfo {author} {\bibfnamefont
  {T.}~\bibnamefont {Jungwirth}},\ }\href {\doibase 10.1126/science.aab1031}
  {\bibfield  {journal} {\bibinfo  {journal} {Science}\ }\textbf {\bibinfo
  {volume} {351}},\ \bibinfo {pages} {587} (\bibinfo {year} {2016})},\ \Eprint
  {http://arxiv.org/abs/https://www.science.org/doi/pdf/10.1126/science.aab1031}
  {https://www.science.org/doi/pdf/10.1126/science.aab1031} \BibitemShut
  {NoStop}%
\bibitem [{\citenamefont {Salemi}\ \emph {et~al.}(2019)\citenamefont {Salemi},
  \citenamefont {Berritta}, \citenamefont {Nandy},\ and\ \citenamefont
  {Oppeneer}}]{RashbaEdelsteinEffect_NatureComm_Salemi2019}%
  \BibitemOpen
  \bibfield  {author} {\bibinfo {author} {\bibfnamefont {L.}~\bibnamefont
  {Salemi}}, \bibinfo {author} {\bibfnamefont {M.}~\bibnamefont {Berritta}},
  \bibinfo {author} {\bibfnamefont {A.~K.}\ \bibnamefont {Nandy}}, \ and\
  \bibinfo {author} {\bibfnamefont {P.~M.}\ \bibnamefont {Oppeneer}},\ }\href
  {\doibase 10.1038/s41467-019-13367-z} {\bibfield  {journal} {\bibinfo
  {journal} {Nature Communications}\ }\textbf {\bibinfo {volume} {10}},\
  \bibinfo {pages} {5381} (\bibinfo {year} {2019})}\BibitemShut {NoStop}%
\bibitem [{\citenamefont {\ifmmode~\check{S}\else \v{S}\fi{}mejkal}\ \emph
  {et~al.}(2017)\citenamefont {\ifmmode~\check{S}\else \v{S}\fi{}mejkal},
  \citenamefont {\ifmmode~\check{Z}\else \v{Z}\fi{}elezn\'y}, \citenamefont
  {Sinova},\ and\ \citenamefont
  {Jungwirth}}]{SpinOrbitTorqueCuMnAs_Jungwirth_PRL2017}%
  \BibitemOpen
  \bibfield  {author} {\bibinfo {author} {\bibfnamefont {L.}~\bibnamefont
  {\ifmmode~\check{S}\else \v{S}\fi{}mejkal}}, \bibinfo {author} {\bibfnamefont
  {J.}~\bibnamefont {\ifmmode~\check{Z}\else \v{Z}\fi{}elezn\'y}}, \bibinfo
  {author} {\bibfnamefont {J.}~\bibnamefont {Sinova}}, \ and\ \bibinfo {author}
  {\bibfnamefont {T.}~\bibnamefont {Jungwirth}},\ }\href {\doibase
  10.1103/PhysRevLett.118.106402} {\bibfield  {journal} {\bibinfo  {journal}
  {Phys. Rev. Lett.}\ }\textbf {\bibinfo {volume} {118}},\ \bibinfo {pages}
  {106402} (\bibinfo {year} {2017})}\BibitemShut {NoStop}%
\bibitem [{\citenamefont {Shao}\ \emph {et~al.}(2019)\citenamefont {Shao},
  \citenamefont {Gurung}, \citenamefont {Zhang},\ and\ \citenamefont
  {Tsymbal}}]{DNLSpintronics_Shao_PRL2019}%
  \BibitemOpen
  \bibfield  {author} {\bibinfo {author} {\bibfnamefont {D.-F.}\ \bibnamefont
  {Shao}}, \bibinfo {author} {\bibfnamefont {G.}~\bibnamefont {Gurung}},
  \bibinfo {author} {\bibfnamefont {S.-H.}\ \bibnamefont {Zhang}}, \ and\
  \bibinfo {author} {\bibfnamefont {E.~Y.}\ \bibnamefont {Tsymbal}},\ }\href
  {\doibase 10.1103/PhysRevLett.122.077203} {\bibfield  {journal} {\bibinfo
  {journal} {Phys. Rev. Lett.}\ }\textbf {\bibinfo {volume} {122}},\ \bibinfo
  {pages} {077203} (\bibinfo {year} {2019})}\BibitemShut {NoStop}%
\bibitem [{\citenamefont {Young}\ and\ \citenamefont
  {Kane}(2015)}]{2DDirac_YoungKane_PRL2015}%
  \BibitemOpen
  \bibfield  {author} {\bibinfo {author} {\bibfnamefont {S.~M.}\ \bibnamefont
  {Young}}\ and\ \bibinfo {author} {\bibfnamefont {C.~L.}\ \bibnamefont
  {Kane}},\ }\href {\doibase 10.1103/PhysRevLett.115.126803} {\bibfield
  {journal} {\bibinfo  {journal} {Phys. Rev. Lett.}\ }\textbf {\bibinfo
  {volume} {115}},\ \bibinfo {pages} {126803} (\bibinfo {year}
  {2015})}\BibitemShut {NoStop}%
\bibitem [{\citenamefont {Schoop}\ \emph {et~al.}(2016)\citenamefont {Schoop},
  \citenamefont {Ali}, \citenamefont {Straßer}, \citenamefont {Topp},
  \citenamefont {Varykhalov}, \citenamefont {Marchenko}, \citenamefont
  {Duppel}, \citenamefont {Parkin}, \citenamefont {Lotsch},\ and\ \citenamefont
  {Ast}}]{ZrSiS_Schoop_NatureComm2016}%
  \BibitemOpen
  \bibfield  {author} {\bibinfo {author} {\bibfnamefont {L.~M.}\ \bibnamefont
  {Schoop}}, \bibinfo {author} {\bibfnamefont {M.~N.}\ \bibnamefont {Ali}},
  \bibinfo {author} {\bibfnamefont {C.}~\bibnamefont {Straßer}}, \bibinfo
  {author} {\bibfnamefont {A.}~\bibnamefont {Topp}}, \bibinfo {author}
  {\bibfnamefont {A.}~\bibnamefont {Varykhalov}}, \bibinfo {author}
  {\bibfnamefont {D.}~\bibnamefont {Marchenko}}, \bibinfo {author}
  {\bibfnamefont {V.}~\bibnamefont {Duppel}}, \bibinfo {author} {\bibfnamefont
  {S.~S.~P.}\ \bibnamefont {Parkin}}, \bibinfo {author} {\bibfnamefont {B.~V.}\
  \bibnamefont {Lotsch}}, \ and\ \bibinfo {author} {\bibfnamefont {C.~R.}\
  \bibnamefont {Ast}},\ }\href {https://doi.org/10.1038/ncomms11696} {\bibfield
   {journal} {\bibinfo  {journal} {Nature Communications}\ }\textbf {\bibinfo
  {volume} {7}},\ \bibinfo {pages} {11696} (\bibinfo {year}
  {2016})}\BibitemShut {NoStop}%
\bibitem [{\citenamefont {Klemenz}\ \emph {et~al.}(2019)\citenamefont
  {Klemenz}, \citenamefont {Lei},\ and\ \citenamefont
  {Schoop}}]{TopologySquareNets_AnnualReview_Schoop2019}%
  \BibitemOpen
  \bibfield  {author} {\bibinfo {author} {\bibfnamefont {S.}~\bibnamefont
  {Klemenz}}, \bibinfo {author} {\bibfnamefont {S.}~\bibnamefont {Lei}}, \ and\
  \bibinfo {author} {\bibfnamefont {L.~M.}\ \bibnamefont {Schoop}},\ }\href
  {\doibase 10.1146/annurev-matsci-070218-010114} {\bibfield  {journal}
  {\bibinfo  {journal} {Annual Review of Materials Research}\ }\textbf
  {\bibinfo {volume} {49}},\ \bibinfo {pages} {185} (\bibinfo {year} {2019})},\
  \Eprint
  {http://arxiv.org/abs/https://doi.org/10.1146/annurev-matsci-070218-010114}
  {https://doi.org/10.1146/annurev-matsci-070218-010114} \BibitemShut {NoStop}%
\bibitem [{\citenamefont {Schoop}\ \emph {et~al.}(2018)\citenamefont {Schoop},
  \citenamefont {Topp}, \citenamefont {Lippmann}, \citenamefont {Orlandi},
  \citenamefont {Müchler}, \citenamefont {Vergniory}, \citenamefont {Sun},
  \citenamefont {Rost}, \citenamefont {Duppel}, \citenamefont {Krivenkov},
  \citenamefont {Sheoran}, \citenamefont {Manuel}, \citenamefont {Varykhalov},
  \citenamefont {Yan}, \citenamefont {Kremer}, \citenamefont {Ast},\ and\
  \citenamefont {Lotsch}}]{CeSbTe_AFM_Schoop_Science2018}%
  \BibitemOpen
  \bibfield  {author} {\bibinfo {author} {\bibfnamefont {L.~M.}\ \bibnamefont
  {Schoop}}, \bibinfo {author} {\bibfnamefont {A.}~\bibnamefont {Topp}},
  \bibinfo {author} {\bibfnamefont {J.}~\bibnamefont {Lippmann}}, \bibinfo
  {author} {\bibfnamefont {F.}~\bibnamefont {Orlandi}}, \bibinfo {author}
  {\bibfnamefont {L.}~\bibnamefont {Müchler}}, \bibinfo {author}
  {\bibfnamefont {M.~G.}\ \bibnamefont {Vergniory}}, \bibinfo {author}
  {\bibfnamefont {Y.}~\bibnamefont {Sun}}, \bibinfo {author} {\bibfnamefont
  {A.~W.}\ \bibnamefont {Rost}}, \bibinfo {author} {\bibfnamefont
  {V.}~\bibnamefont {Duppel}}, \bibinfo {author} {\bibfnamefont
  {M.}~\bibnamefont {Krivenkov}}, \bibinfo {author} {\bibfnamefont
  {S.}~\bibnamefont {Sheoran}}, \bibinfo {author} {\bibfnamefont
  {P.}~\bibnamefont {Manuel}}, \bibinfo {author} {\bibfnamefont
  {A.}~\bibnamefont {Varykhalov}}, \bibinfo {author} {\bibfnamefont
  {B.}~\bibnamefont {Yan}}, \bibinfo {author} {\bibfnamefont {R.~K.}\
  \bibnamefont {Kremer}}, \bibinfo {author} {\bibfnamefont {C.~R.}\
  \bibnamefont {Ast}}, \ and\ \bibinfo {author} {\bibfnamefont {B.~V.}\
  \bibnamefont {Lotsch}},\ }\href {\doibase 10.1126/sciadv.aar2317} {\bibfield
  {journal} {\bibinfo  {journal} {Science Advances}\ }\textbf {\bibinfo
  {volume} {4}},\ \bibinfo {pages} {eaar2317} (\bibinfo {year}
  {2018})}\BibitemShut {NoStop}%
\bibitem [{\citenamefont {Hosen}\ \emph {et~al.}(2018)\citenamefont {Hosen},
  \citenamefont {Dhakal}, \citenamefont {Dimitri}, \citenamefont {Maldonado},
  \citenamefont {Aperis}, \citenamefont {Kabir}, \citenamefont {Sims},
  \citenamefont {Riseborough}, \citenamefont {Oppeneer}, \citenamefont
  {Kaczorowski}, \citenamefont {Durakiewicz},\ and\ \citenamefont
  {Neupane}}]{GdSbTe_HosenNeupane_SciReports2018}%
  \BibitemOpen
  \bibfield  {author} {\bibinfo {author} {\bibfnamefont {M.~M.}\ \bibnamefont
  {Hosen}}, \bibinfo {author} {\bibfnamefont {G.}~\bibnamefont {Dhakal}},
  \bibinfo {author} {\bibfnamefont {K.}~\bibnamefont {Dimitri}}, \bibinfo
  {author} {\bibfnamefont {P.}~\bibnamefont {Maldonado}}, \bibinfo {author}
  {\bibfnamefont {A.}~\bibnamefont {Aperis}}, \bibinfo {author} {\bibfnamefont
  {F.}~\bibnamefont {Kabir}}, \bibinfo {author} {\bibfnamefont
  {C.}~\bibnamefont {Sims}}, \bibinfo {author} {\bibfnamefont {P.}~\bibnamefont
  {Riseborough}}, \bibinfo {author} {\bibfnamefont {P.~M.}\ \bibnamefont
  {Oppeneer}}, \bibinfo {author} {\bibfnamefont {D.}~\bibnamefont
  {Kaczorowski}}, \bibinfo {author} {\bibfnamefont {T.}~\bibnamefont
  {Durakiewicz}}, \ and\ \bibinfo {author} {\bibfnamefont {M.}~\bibnamefont
  {Neupane}},\ }\href {https://doi.org/10.1038/s41598-018-31296-7} {\bibfield
  {journal} {\bibinfo  {journal} {Scientific Reports}\ }\textbf {\bibinfo
  {volume} {8}},\ \bibinfo {pages} {13283} (\bibinfo {year}
  {2018})}\BibitemShut {NoStop}%
\bibitem [{\citenamefont {Wang}\ \emph {et~al.}(2016)\citenamefont {Wang},
  \citenamefont {Zaliznyak}, \citenamefont {Ren}, \citenamefont {Wu},
  \citenamefont {Graf}, \citenamefont {Garlea}, \citenamefont {Warren},
  \citenamefont {Bozin}, \citenamefont {Zhu},\ and\ \citenamefont
  {Petrovic}}]{YbMnBiDiracAFM_Wang_PRB2016}%
  \BibitemOpen
  \bibfield  {author} {\bibinfo {author} {\bibfnamefont {A.}~\bibnamefont
  {Wang}}, \bibinfo {author} {\bibfnamefont {I.}~\bibnamefont {Zaliznyak}},
  \bibinfo {author} {\bibfnamefont {W.}~\bibnamefont {Ren}}, \bibinfo {author}
  {\bibfnamefont {L.}~\bibnamefont {Wu}}, \bibinfo {author} {\bibfnamefont
  {D.}~\bibnamefont {Graf}}, \bibinfo {author} {\bibfnamefont {V.~O.}\
  \bibnamefont {Garlea}}, \bibinfo {author} {\bibfnamefont {J.~B.}\
  \bibnamefont {Warren}}, \bibinfo {author} {\bibfnamefont {E.}~\bibnamefont
  {Bozin}}, \bibinfo {author} {\bibfnamefont {Y.}~\bibnamefont {Zhu}}, \ and\
  \bibinfo {author} {\bibfnamefont {C.}~\bibnamefont {Petrovic}},\ }\href
  {\doibase 10.1103/PhysRevB.94.165161} {\bibfield  {journal} {\bibinfo
  {journal} {Phys. Rev. B}\ }\textbf {\bibinfo {volume} {94}},\ \bibinfo
  {pages} {165161} (\bibinfo {year} {2016})}\BibitemShut {NoStop}%
\bibitem [{\citenamefont {Kealhofer}\ \emph {et~al.}(2018)\citenamefont
  {Kealhofer}, \citenamefont {Jang}, \citenamefont {Griffin}, \citenamefont
  {John}, \citenamefont {Benavides}, \citenamefont {Doyle}, \citenamefont
  {Helm}, \citenamefont {Moll}, \citenamefont {Neaton}, \citenamefont {Chan},
  \citenamefont {Denlinger},\ and\ \citenamefont
  {Analytis}}]{YbMnSbAFMDirac_PRB2018}%
  \BibitemOpen
  \bibfield  {author} {\bibinfo {author} {\bibfnamefont {R.}~\bibnamefont
  {Kealhofer}}, \bibinfo {author} {\bibfnamefont {S.}~\bibnamefont {Jang}},
  \bibinfo {author} {\bibfnamefont {S.~M.}\ \bibnamefont {Griffin}}, \bibinfo
  {author} {\bibfnamefont {C.}~\bibnamefont {John}}, \bibinfo {author}
  {\bibfnamefont {K.~A.}\ \bibnamefont {Benavides}}, \bibinfo {author}
  {\bibfnamefont {S.}~\bibnamefont {Doyle}}, \bibinfo {author} {\bibfnamefont
  {T.}~\bibnamefont {Helm}}, \bibinfo {author} {\bibfnamefont {P.~J.~W.}\
  \bibnamefont {Moll}}, \bibinfo {author} {\bibfnamefont {J.~B.}\ \bibnamefont
  {Neaton}}, \bibinfo {author} {\bibfnamefont {J.~Y.}\ \bibnamefont {Chan}},
  \bibinfo {author} {\bibfnamefont {J.~D.}\ \bibnamefont {Denlinger}}, \ and\
  \bibinfo {author} {\bibfnamefont {J.~G.}\ \bibnamefont {Analytis}},\ }\href
  {\doibase 10.1103/PhysRevB.97.045109} {\bibfield  {journal} {\bibinfo
  {journal} {Phys. Rev. B}\ }\textbf {\bibinfo {volume} {97}},\ \bibinfo
  {pages} {045109} (\bibinfo {year} {2018})}\BibitemShut {NoStop}%
\bibitem [{\citenamefont {May}\ \emph {et~al.}(2014)\citenamefont {May},
  \citenamefont {McGuire},\ and\ \citenamefont
  {Sales}}]{EuMnBiDiracAFM_May_PRB2014}%
  \BibitemOpen
  \bibfield  {author} {\bibinfo {author} {\bibfnamefont {A.~F.}\ \bibnamefont
  {May}}, \bibinfo {author} {\bibfnamefont {M.~A.}\ \bibnamefont {McGuire}}, \
  and\ \bibinfo {author} {\bibfnamefont {B.~C.}\ \bibnamefont {Sales}},\ }\href
  {\doibase 10.1103/PhysRevB.90.075109} {\bibfield  {journal} {\bibinfo
  {journal} {Phys. Rev. B}\ }\textbf {\bibinfo {volume} {90}},\ \bibinfo
  {pages} {075109} (\bibinfo {year} {2014})}\BibitemShut {NoStop}%
\bibitem [{\citenamefont {Yi}\ \emph {et~al.}(2017)\citenamefont {Yi},
  \citenamefont {Yang}, \citenamefont {Yang}, \citenamefont {Wang},
  \citenamefont {Matsushita}, \citenamefont {Miao}, \citenamefont {Jiao},
  \citenamefont {Cheng}, \citenamefont {Li}, \citenamefont {Yamaura},
  \citenamefont {Shi},\ and\ \citenamefont {Luo}}]{EuMnSbDiracAFM_Yi_PRB2017}%
  \BibitemOpen
  \bibfield  {author} {\bibinfo {author} {\bibfnamefont {C.}~\bibnamefont
  {Yi}}, \bibinfo {author} {\bibfnamefont {S.}~\bibnamefont {Yang}}, \bibinfo
  {author} {\bibfnamefont {M.}~\bibnamefont {Yang}}, \bibinfo {author}
  {\bibfnamefont {L.}~\bibnamefont {Wang}}, \bibinfo {author} {\bibfnamefont
  {Y.}~\bibnamefont {Matsushita}}, \bibinfo {author} {\bibfnamefont
  {S.}~\bibnamefont {Miao}}, \bibinfo {author} {\bibfnamefont {Y.}~\bibnamefont
  {Jiao}}, \bibinfo {author} {\bibfnamefont {J.}~\bibnamefont {Cheng}},
  \bibinfo {author} {\bibfnamefont {Y.}~\bibnamefont {Li}}, \bibinfo {author}
  {\bibfnamefont {K.}~\bibnamefont {Yamaura}}, \bibinfo {author} {\bibfnamefont
  {Y.}~\bibnamefont {Shi}}, \ and\ \bibinfo {author} {\bibfnamefont
  {J.}~\bibnamefont {Luo}},\ }\href {\doibase 10.1103/PhysRevB.96.205103}
  {\bibfield  {journal} {\bibinfo  {journal} {Phys. Rev. B}\ }\textbf {\bibinfo
  {volume} {96}},\ \bibinfo {pages} {205103} (\bibinfo {year}
  {2017})}\BibitemShut {NoStop}%
\bibitem [{\citenamefont {Shao}\ \emph {et~al.}(2020)\citenamefont {Shao},
  \citenamefont {Zhang}, \citenamefont {Gurung}, \citenamefont {Yang},\ and\
  \citenamefont {Tsymbal}}]{NLAHENeel_Shao_PRL2020}%
  \BibitemOpen
  \bibfield  {author} {\bibinfo {author} {\bibfnamefont {D.-F.}\ \bibnamefont
  {Shao}}, \bibinfo {author} {\bibfnamefont {S.-H.}\ \bibnamefont {Zhang}},
  \bibinfo {author} {\bibfnamefont {G.}~\bibnamefont {Gurung}}, \bibinfo
  {author} {\bibfnamefont {W.}~\bibnamefont {Yang}}, \ and\ \bibinfo {author}
  {\bibfnamefont {E.~Y.}\ \bibnamefont {Tsymbal}},\ }\href {\doibase
  10.1103/PhysRevLett.124.067203} {\bibfield  {journal} {\bibinfo  {journal}
  {Phys. Rev. Lett.}\ }\textbf {\bibinfo {volume} {124}},\ \bibinfo {pages}
  {067203} (\bibinfo {year} {2020})}\BibitemShut {NoStop}%
\bibitem [{\citenamefont {N\'u\~nez}\ \emph {et~al.}(2006)\citenamefont
  {N\'u\~nez}, \citenamefont {Duine}, \citenamefont {Haney},\ and\
  \citenamefont {MacDonald}}]{AFMSpintronics_Nunez_PRB2006}%
  \BibitemOpen
  \bibfield  {author} {\bibinfo {author} {\bibfnamefont {A.~S.}\ \bibnamefont
  {N\'u\~nez}}, \bibinfo {author} {\bibfnamefont {R.~A.}\ \bibnamefont
  {Duine}}, \bibinfo {author} {\bibfnamefont {P.}~\bibnamefont {Haney}}, \ and\
  \bibinfo {author} {\bibfnamefont {A.~H.}\ \bibnamefont {MacDonald}},\ }\href
  {\doibase 10.1103/PhysRevB.73.214426} {\bibfield  {journal} {\bibinfo
  {journal} {Phys. Rev. B}\ }\textbf {\bibinfo {volume} {73}},\ \bibinfo
  {pages} {214426} (\bibinfo {year} {2006})}\BibitemShut {NoStop}%
\bibitem [{\citenamefont {Shick}\ \emph {et~al.}(2010)\citenamefont {Shick},
  \citenamefont {Khmelevskyi}, \citenamefont {Mryasov}, \citenamefont
  {Wunderlich},\ and\ \citenamefont
  {Jungwirth}}]{AFMSpintronics_Shick_PRB2010}%
  \BibitemOpen
  \bibfield  {author} {\bibinfo {author} {\bibfnamefont {A.~B.}\ \bibnamefont
  {Shick}}, \bibinfo {author} {\bibfnamefont {S.}~\bibnamefont {Khmelevskyi}},
  \bibinfo {author} {\bibfnamefont {O.~N.}\ \bibnamefont {Mryasov}}, \bibinfo
  {author} {\bibfnamefont {J.}~\bibnamefont {Wunderlich}}, \ and\ \bibinfo
  {author} {\bibfnamefont {T.}~\bibnamefont {Jungwirth}},\ }\href {\doibase
  10.1103/PhysRevB.81.212409} {\bibfield  {journal} {\bibinfo  {journal} {Phys.
  Rev. B}\ }\textbf {\bibinfo {volume} {81}},\ \bibinfo {pages} {212409}
  (\bibinfo {year} {2010})}\BibitemShut {NoStop}%
\bibitem [{\citenamefont {Kontani}\ \emph {et~al.}(2008)\citenamefont
  {Kontani}, \citenamefont {Tanaka}, \citenamefont {Naito}, \citenamefont
  {S.~Hirashima}, \citenamefont {Yamada},\ and\ \citenamefont
  {Inoue}}]{SHE_Kontani_JPSJ2008}%
  \BibitemOpen
  \bibfield  {author} {\bibinfo {author} {\bibfnamefont {H.}~\bibnamefont
  {Kontani}}, \bibinfo {author} {\bibfnamefont {T.}~\bibnamefont {Tanaka}},
  \bibinfo {author} {\bibfnamefont {M.}~\bibnamefont {Naito}}, \bibinfo
  {author} {\bibfnamefont {D.}~\bibnamefont {S.~Hirashima}}, \bibinfo {author}
  {\bibfnamefont {K.}~\bibnamefont {Yamada}}, \ and\ \bibinfo {author}
  {\bibfnamefont {J.}~\bibnamefont {Inoue}},\ }\href {\doibase
  10.1143/JPSJS.77SA.275} {\bibfield  {journal} {\bibinfo  {journal} {Journal
  of the Physical Society of Japan}\ }\textbf {\bibinfo {volume} {77}},\
  \bibinfo {pages} {275} (\bibinfo {year} {2008})},\ \Eprint
  {http://arxiv.org/abs/https://doi.org/10.1143/JPSJS.77SA.275}
  {https://doi.org/10.1143/JPSJS.77SA.275} \BibitemShut {NoStop}%
\bibitem [{\citenamefont {Peters}\ and\ \citenamefont
  {Yanase}(2018)}]{Edelstein_Robert_PRB2018}%
  \BibitemOpen
  \bibfield  {author} {\bibinfo {author} {\bibfnamefont {R.}~\bibnamefont
  {Peters}}\ and\ \bibinfo {author} {\bibfnamefont {Y.}~\bibnamefont
  {Yanase}},\ }\href {\doibase 10.1103/PhysRevB.97.115128} {\bibfield
  {journal} {\bibinfo  {journal} {Phys. Rev. B}\ }\textbf {\bibinfo {volume}
  {97}},\ \bibinfo {pages} {115128} (\bibinfo {year} {2018})}\BibitemShut
  {NoStop}%
\bibitem [{\citenamefont {Lei}\ \emph {et~al.}(2019)\citenamefont {Lei},
  \citenamefont {Duppel}, \citenamefont {Lippmann}, \citenamefont {Nuss},
  \citenamefont {Lotsch},\ and\ \citenamefont {Schoop}}]{CDWMag_Lei_AQT2019}%
  \BibitemOpen
  \bibfield  {author} {\bibinfo {author} {\bibfnamefont {S.}~\bibnamefont
  {Lei}}, \bibinfo {author} {\bibfnamefont {V.}~\bibnamefont {Duppel}},
  \bibinfo {author} {\bibfnamefont {J.~M.}\ \bibnamefont {Lippmann}}, \bibinfo
  {author} {\bibfnamefont {J.}~\bibnamefont {Nuss}}, \bibinfo {author}
  {\bibfnamefont {B.~V.}\ \bibnamefont {Lotsch}}, \ and\ \bibinfo {author}
  {\bibfnamefont {L.~M.}\ \bibnamefont {Schoop}},\ }\href@noop {} {\bibfield
  {journal} {\bibinfo  {journal} {Advanced Quantum Technologies}\ }\textbf
  {\bibinfo {volume} {2}},\ \bibinfo {pages} {1900045} (\bibinfo {year}
  {2019})}\BibitemShut {NoStop}%
\bibitem [{\citenamefont {Blaha}\ \emph {et~al.}(2020)\citenamefont {Blaha},
  \citenamefont {Schwarz}, \citenamefont {Tran}, \citenamefont {Laskowski},
  \citenamefont {Madsen},\ and\ \citenamefont {Marks}}]{Wien2k2021}%
  \BibitemOpen
  \bibfield  {author} {\bibinfo {author} {\bibfnamefont {P.}~\bibnamefont
  {Blaha}}, \bibinfo {author} {\bibfnamefont {K.}~\bibnamefont {Schwarz}},
  \bibinfo {author} {\bibfnamefont {F.}~\bibnamefont {Tran}}, \bibinfo {author}
  {\bibfnamefont {R.}~\bibnamefont {Laskowski}}, \bibinfo {author}
  {\bibfnamefont {G.~K.~H.}\ \bibnamefont {Madsen}}, \ and\ \bibinfo {author}
  {\bibfnamefont {L.~D.}\ \bibnamefont {Marks}},\ }\href {\doibase
  10.1063/1.5143061} {\bibfield  {journal} {\bibinfo  {journal} {The Journal of
  Chemical Physics}\ }\textbf {\bibinfo {volume} {152}},\ \bibinfo {pages}
  {074101} (\bibinfo {year} {2020})}\BibitemShut {NoStop}%
\bibitem [{\citenamefont {Giannozzi}\ \emph {et~al.}(2009)\citenamefont
  {Giannozzi}, \citenamefont {Baroni},\ and\ \citenamefont {et~al}}]{QE}%
  \BibitemOpen
  \bibfield  {author} {\bibinfo {author} {\bibfnamefont {P.}~\bibnamefont
  {Giannozzi}}, \bibinfo {author} {\bibfnamefont {S.}~\bibnamefont {Baroni}}, \
  and\ \bibinfo {author} {\bibfnamefont {N.~B.}\ \bibnamefont {et~al}},\ }\href
  {http://stacks.iop.org/0953-8984/21/i=39/a=395502} {\bibfield  {journal}
  {\bibinfo  {journal} {Journal of Physics: Condensed Matter}\ }\textbf
  {\bibinfo {volume} {21}},\ \bibinfo {pages} {395502} (\bibinfo {year}
  {2009})}\BibitemShut {NoStop}%
\bibitem [{\citenamefont {Perdew}\ \emph {et~al.}(1996)\citenamefont {Perdew},
  \citenamefont {Burke},\ and\ \citenamefont {Ernzerhof}}]{PBE}%
  \BibitemOpen
  \bibfield  {author} {\bibinfo {author} {\bibfnamefont {J.~P.}\ \bibnamefont
  {Perdew}}, \bibinfo {author} {\bibfnamefont {K.}~\bibnamefont {Burke}}, \
  and\ \bibinfo {author} {\bibfnamefont {M.}~\bibnamefont {Ernzerhof}},\ }\href
  {\doibase 10.1103/PhysRevLett.77.3865} {\bibfield  {journal} {\bibinfo
  {journal} {Phys. Rev. Lett.}\ }\textbf {\bibinfo {volume} {77}},\ \bibinfo
  {pages} {3865} (\bibinfo {year} {1996})}\BibitemShut {NoStop}%
\bibitem [{\citenamefont {Johannes}\ and\ \citenamefont
  {Pickett}(2005)}]{EuN_Pickett_PRB2005}%
  \BibitemOpen
  \bibfield  {author} {\bibinfo {author} {\bibfnamefont {M.~D.}\ \bibnamefont
  {Johannes}}\ and\ \bibinfo {author} {\bibfnamefont {W.~E.}\ \bibnamefont
  {Pickett}},\ }\href {\doibase 10.1103/PhysRevB.72.195116} {\bibfield
  {journal} {\bibinfo  {journal} {Phys. Rev. B}\ }\textbf {\bibinfo {volume}
  {72}},\ \bibinfo {pages} {195116} (\bibinfo {year} {2005})}\BibitemShut
  {NoStop}%
\bibitem [{\citenamefont {Kunes}\ \emph {et~al.}(2005)\citenamefont {Kunes},
  \citenamefont {Ku},\ and\ \citenamefont
  {E.~Pickett}}]{EuExchange_Pickett_JPSJ2005}%
  \BibitemOpen
  \bibfield  {author} {\bibinfo {author} {\bibfnamefont {J.}~\bibnamefont
  {Kunes}}, \bibinfo {author} {\bibfnamefont {W.}~\bibnamefont {Ku}}, \ and\
  \bibinfo {author} {\bibfnamefont {W.}~\bibnamefont {E.~Pickett}},\ }\href
  {\doibase 10.1143/JPSJ.74.1408} {\bibfield  {journal} {\bibinfo  {journal}
  {Journal of the Physical Society of Japan}\ }\textbf {\bibinfo {volume}
  {74}},\ \bibinfo {pages} {1408} (\bibinfo {year} {2005})},\ \Eprint
  {http://arxiv.org/abs/https://doi.org/10.1143/JPSJ.74.1408}
  {https://doi.org/10.1143/JPSJ.74.1408} \BibitemShut {NoStop}%
\bibitem [{\citenamefont {Larson}\ and\ \citenamefont
  {Lambrecht}(2006)}]{Eu_Larson_IOP2006}%
  \BibitemOpen
  \bibfield  {author} {\bibinfo {author} {\bibfnamefont {P.}~\bibnamefont
  {Larson}}\ and\ \bibinfo {author} {\bibfnamefont {W.~R.~L.}\ \bibnamefont
  {Lambrecht}},\ }\href {\doibase 10.1088/0953-8984/18/49/024} {\bibfield
  {journal} {\bibinfo  {journal} {Journal of Physics: Condensed Matter}\
  }\textbf {\bibinfo {volume} {18}},\ \bibinfo {pages} {11333} (\bibinfo {year}
  {2006})}\BibitemShut {NoStop}%
\bibitem [{\citenamefont {Zhang}\ \emph {et~al.}(2017)\citenamefont {Zhang},
  \citenamefont {Aryal}, \citenamefont {Huang}, \citenamefont {Chen},
  \citenamefont {Lai}, \citenamefont {Graf}, \citenamefont {Besara},
  \citenamefont {Siegrist}, \citenamefont {Manousakis},\ and\ \citenamefont
  {Baumbach}}]{Ce_PRM_Zhang2017}%
  \BibitemOpen
  \bibfield  {author} {\bibinfo {author} {\bibfnamefont {S.}~\bibnamefont
  {Zhang}}, \bibinfo {author} {\bibfnamefont {N.}~\bibnamefont {Aryal}},
  \bibinfo {author} {\bibfnamefont {K.}~\bibnamefont {Huang}}, \bibinfo
  {author} {\bibfnamefont {K.-W.}\ \bibnamefont {Chen}}, \bibinfo {author}
  {\bibfnamefont {Y.}~\bibnamefont {Lai}}, \bibinfo {author} {\bibfnamefont
  {D.}~\bibnamefont {Graf}}, \bibinfo {author} {\bibfnamefont {T.}~\bibnamefont
  {Besara}}, \bibinfo {author} {\bibfnamefont {T.}~\bibnamefont {Siegrist}},
  \bibinfo {author} {\bibfnamefont {E.}~\bibnamefont {Manousakis}}, \ and\
  \bibinfo {author} {\bibfnamefont {R.~E.}\ \bibnamefont {Baumbach}},\ }\href
  {\doibase 10.1103/PhysRevMaterials.1.044404} {\bibfield  {journal} {\bibinfo
  {journal} {Phys. Rev. Materials}\ }\textbf {\bibinfo {volume} {1}},\ \bibinfo
  {pages} {044404} (\bibinfo {year} {2017})}\BibitemShut {NoStop}%
\bibitem [{\citenamefont {Mostofi}\ \emph {et~al.}(2014)\citenamefont
  {Mostofi}, \citenamefont {Yates}, \citenamefont {Pizzi}, \citenamefont {Lee},
  \citenamefont {Souza}, \citenamefont {Vanderbilt},\ and\ \citenamefont
  {Marzari}}]{Wannier902014}%
  \BibitemOpen
  \bibfield  {author} {\bibinfo {author} {\bibfnamefont {A.~A.}\ \bibnamefont
  {Mostofi}}, \bibinfo {author} {\bibfnamefont {J.~R.}\ \bibnamefont {Yates}},
  \bibinfo {author} {\bibfnamefont {G.}~\bibnamefont {Pizzi}}, \bibinfo
  {author} {\bibfnamefont {Y.-S.}\ \bibnamefont {Lee}}, \bibinfo {author}
  {\bibfnamefont {I.}~\bibnamefont {Souza}}, \bibinfo {author} {\bibfnamefont
  {D.}~\bibnamefont {Vanderbilt}}, \ and\ \bibinfo {author} {\bibfnamefont
  {N.}~\bibnamefont {Marzari}},\ }\href {\doibase
  https://doi.org/10.1016/j.cpc.2014.05.003} {\bibfield  {journal} {\bibinfo
  {journal} {Computer Physics Communications}\ }\textbf {\bibinfo {volume}
  {185}},\ \bibinfo {pages} {2309 } (\bibinfo {year} {2014})}\BibitemShut
  {NoStop}%
\bibitem [{\citenamefont {Qiao}\ \emph {et~al.}(2018)\citenamefont {Qiao},
  \citenamefont {Zhou}, \citenamefont {Yuan},\ and\ \citenamefont
  {Zhao}}]{SHCWannier_Qiao_Zhao_PRB2018}%
  \BibitemOpen
  \bibfield  {author} {\bibinfo {author} {\bibfnamefont {J.}~\bibnamefont
  {Qiao}}, \bibinfo {author} {\bibfnamefont {J.}~\bibnamefont {Zhou}}, \bibinfo
  {author} {\bibfnamefont {Z.}~\bibnamefont {Yuan}}, \ and\ \bibinfo {author}
  {\bibfnamefont {W.}~\bibnamefont {Zhao}},\ }\href {\doibase
  10.1103/PhysRevB.98.214402} {\bibfield  {journal} {\bibinfo  {journal} {Phys.
  Rev. B}\ }\textbf {\bibinfo {volume} {98}},\ \bibinfo {pages} {214402}
  (\bibinfo {year} {2018})}\BibitemShut {NoStop}%
\bibitem [{\citenamefont {Tremel}\ and\ \citenamefont
  {Hoffmann}(1987)}]{squarenets_TremelJACS1987}%
  \BibitemOpen
  \bibfield  {author} {\bibinfo {author} {\bibfnamefont {W.}~\bibnamefont
  {Tremel}}\ and\ \bibinfo {author} {\bibfnamefont {R.}~\bibnamefont
  {Hoffmann}},\ }\href {\doibase 10.1021/ja00235a021} {\bibfield  {journal}
  {\bibinfo  {journal} {Journal of the American Chemical Society}\ }\textbf
  {\bibinfo {volume} {109}},\ \bibinfo {pages} {124} (\bibinfo {year}
  {1987})}\BibitemShut {NoStop}%
\bibitem [{\citenamefont {Klemenz}\ \emph {et~al.}(2020)\citenamefont
  {Klemenz}, \citenamefont {Schoop},\ and\ \citenamefont
  {Cano}}]{StackedSqNets_Klemenz_PRB2020}%
  \BibitemOpen
  \bibfield  {author} {\bibinfo {author} {\bibfnamefont {S.}~\bibnamefont
  {Klemenz}}, \bibinfo {author} {\bibfnamefont {L.}~\bibnamefont {Schoop}}, \
  and\ \bibinfo {author} {\bibfnamefont {J.}~\bibnamefont {Cano}},\ }\href
  {\doibase 10.1103/PhysRevB.101.165121} {\bibfield  {journal} {\bibinfo
  {journal} {Phys. Rev. B}\ }\textbf {\bibinfo {volume} {101}},\ \bibinfo
  {pages} {165121} (\bibinfo {year} {2020})}\BibitemShut {NoStop}%
\bibitem [{\citenamefont {Hua}\ \emph {et~al.}(2018)\citenamefont {Hua},
  \citenamefont {Nie}, \citenamefont {Song}, \citenamefont {Yu}, \citenamefont
  {Xu},\ and\ \citenamefont {Yao}}]{TypeIVMSG_Hau_PRB2018}%
  \BibitemOpen
  \bibfield  {author} {\bibinfo {author} {\bibfnamefont {G.}~\bibnamefont
  {Hua}}, \bibinfo {author} {\bibfnamefont {S.}~\bibnamefont {Nie}}, \bibinfo
  {author} {\bibfnamefont {Z.}~\bibnamefont {Song}}, \bibinfo {author}
  {\bibfnamefont {R.}~\bibnamefont {Yu}}, \bibinfo {author} {\bibfnamefont
  {G.}~\bibnamefont {Xu}}, \ and\ \bibinfo {author} {\bibfnamefont
  {K.}~\bibnamefont {Yao}},\ }\href {\doibase 10.1103/PhysRevB.98.201116}
  {\bibfield  {journal} {\bibinfo  {journal} {Phys. Rev. B}\ }\textbf {\bibinfo
  {volume} {98}},\ \bibinfo {pages} {201116} (\bibinfo {year}
  {2018})}\BibitemShut {NoStop}%
\bibitem [{\citenamefont {Sinova}\ \emph {et~al.}(2015)\citenamefont {Sinova},
  \citenamefont {Valenzuela}, \citenamefont {Wunderlich}, \citenamefont
  {Back},\ and\ \citenamefont {Jungwirth}}]{SHE_Sinova_JungwirthRMP2015}%
  \BibitemOpen
  \bibfield  {author} {\bibinfo {author} {\bibfnamefont {J.}~\bibnamefont
  {Sinova}}, \bibinfo {author} {\bibfnamefont {S.~O.}\ \bibnamefont
  {Valenzuela}}, \bibinfo {author} {\bibfnamefont {J.}~\bibnamefont
  {Wunderlich}}, \bibinfo {author} {\bibfnamefont {C.~H.}\ \bibnamefont
  {Back}}, \ and\ \bibinfo {author} {\bibfnamefont {T.}~\bibnamefont
  {Jungwirth}},\ }\href {\doibase 10.1103/RevModPhys.87.1213} {\bibfield
  {journal} {\bibinfo  {journal} {Rev. Mod. Phys.}\ }\textbf {\bibinfo {volume}
  {87}},\ \bibinfo {pages} {1213} (\bibinfo {year} {2015})}\BibitemShut
  {NoStop}%
\bibitem [{\citenamefont {Schrieffer}(1967)}]{SchriefferKondo1967}%
  \BibitemOpen
  \bibfield  {author} {\bibinfo {author} {\bibfnamefont {J.~R.}\ \bibnamefont
  {Schrieffer}},\ }\href {\doibase 10.1063/1.1709517} {\bibfield  {journal}
  {\bibinfo  {journal} {Journal of Applied Physics}\ }\textbf {\bibinfo
  {volume} {38}},\ \bibinfo {pages} {1143} (\bibinfo {year} {1967})},\ \Eprint
  {http://arxiv.org/abs/https://doi.org/10.1063/1.1709517}
  {https://doi.org/10.1063/1.1709517} \BibitemShut {NoStop}%
\end{thebibliography}

%

\appendix

\section{Non-magnetic phase and Wannier tightbinding}
\label{appendixA}

\subsection{Band characters}
\begin{figure}[htb]
    \begin{center}
       \subfigure[]{
            \includegraphics[width=0.45\textwidth]{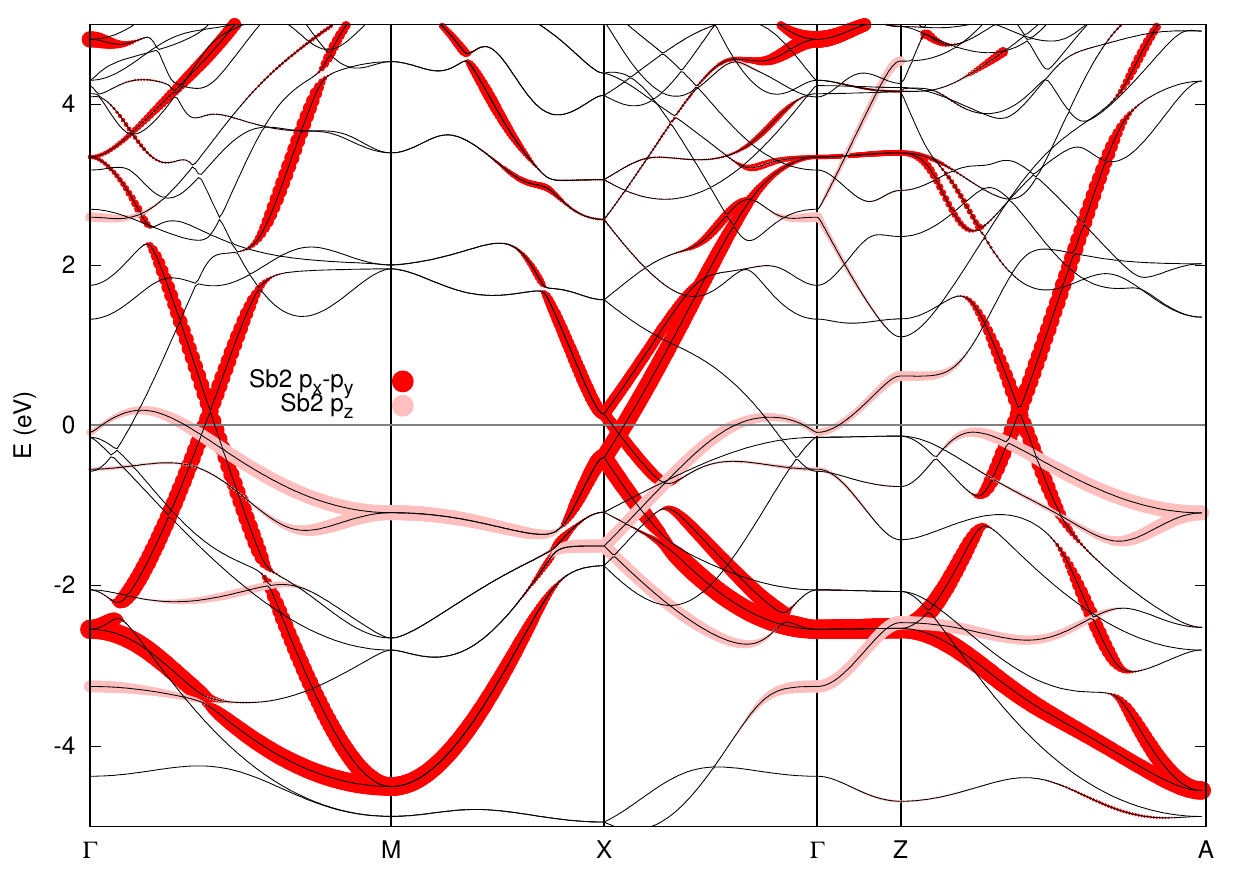}
        }
\hskip -0.05 in
        \subfigure[]{
            \includegraphics[width=0.45\textwidth]{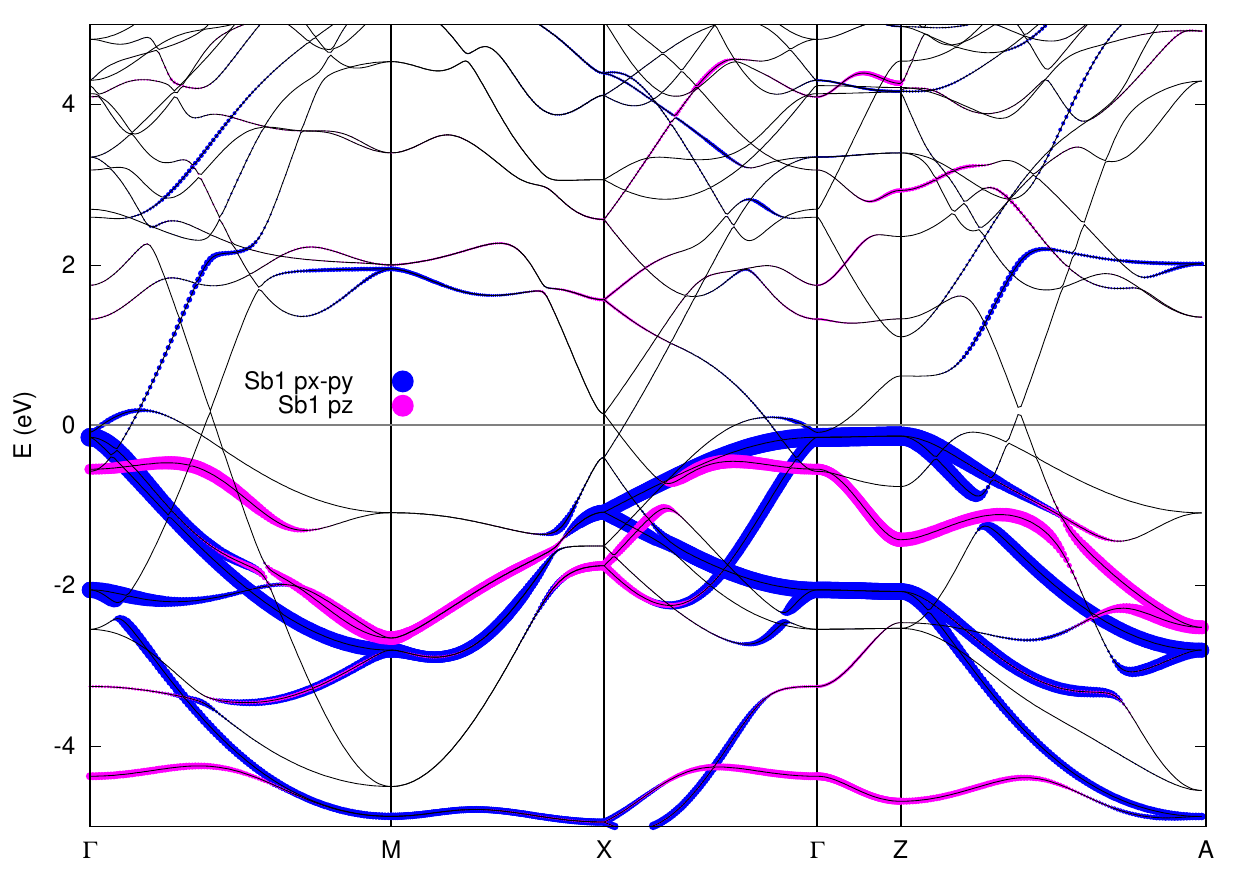}
        } 
            \end{center}
            \caption{Character of the bands in the vicinity of the Fermi level for non-magnetic phase in the absence of SOC: (a) Sb2 $p_x$-$p_y$ and $p_z$ character and  (b) Sb1 $p_x$-$p_y$ and $p_z$ character .The size of the dots is proportional to the orbital composition of the particular state.
        }
        \label{fig:appen-bandcharacter} 
\end{figure}

In Fig.~\ref{fig:appen-bandcharacter}, we show the orbital composition of the bands (or band characters) in the vicinity of the Fermi level for the non-magnetic phase.

\subsection{Wannier bands}
\begin{figure}[htb]
    \begin{center}
       \subfigure[]{
            \includegraphics[width=0.40\textwidth]{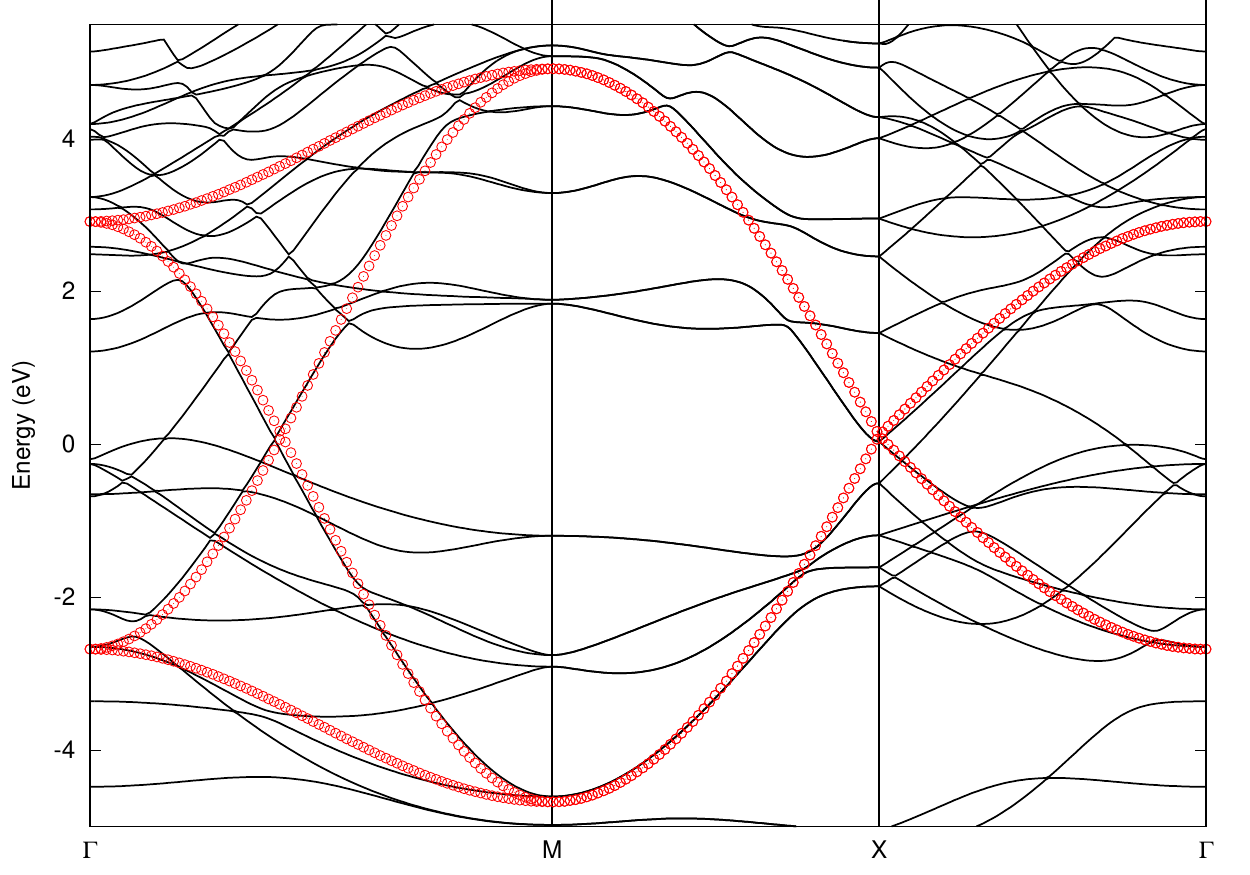}
        }
\hskip -0.05 in
        \subfigure[]{
            \includegraphics[width=0.40\textwidth]{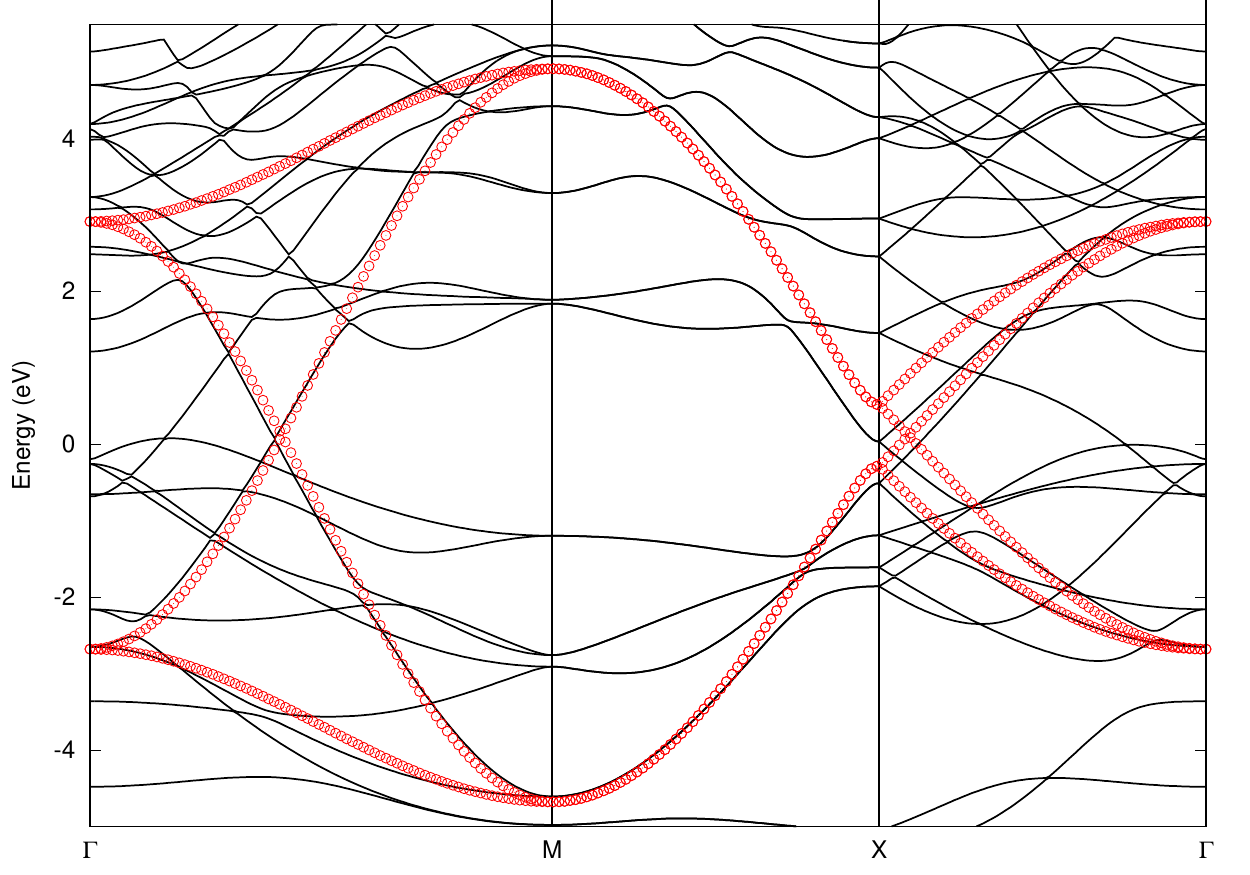}
        } 
            \end{center}
            \caption{Wannier bands (red dots) from 4$\times$4 Hamiltonian formed by the $p_x$-$p_y$ orbitals of the Sb2 square lattice by considering upto the (a) nn and (b) nnn hopping terms given in Table~\ref{table:TightBindingParams}.
        }
        \label{fig:appen-WannierComparison-4band} 
\end{figure}

\begin{figure}[htb]
    \begin{center}
        \subfigure[]{
            \includegraphics[width=0.40\textwidth]{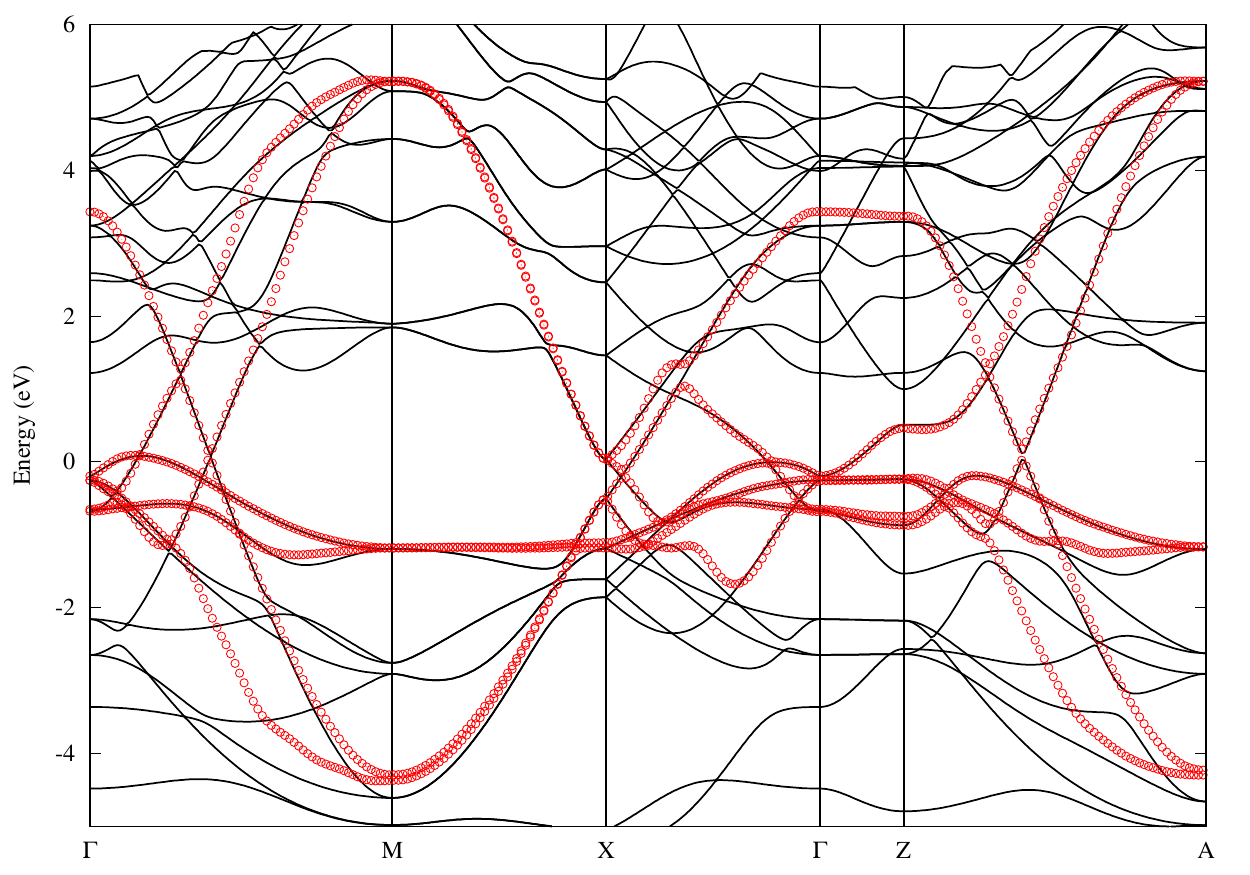}
        }
\hskip -0.05 in
       \subfigure[]{
            \includegraphics[width=0.40\textwidth]{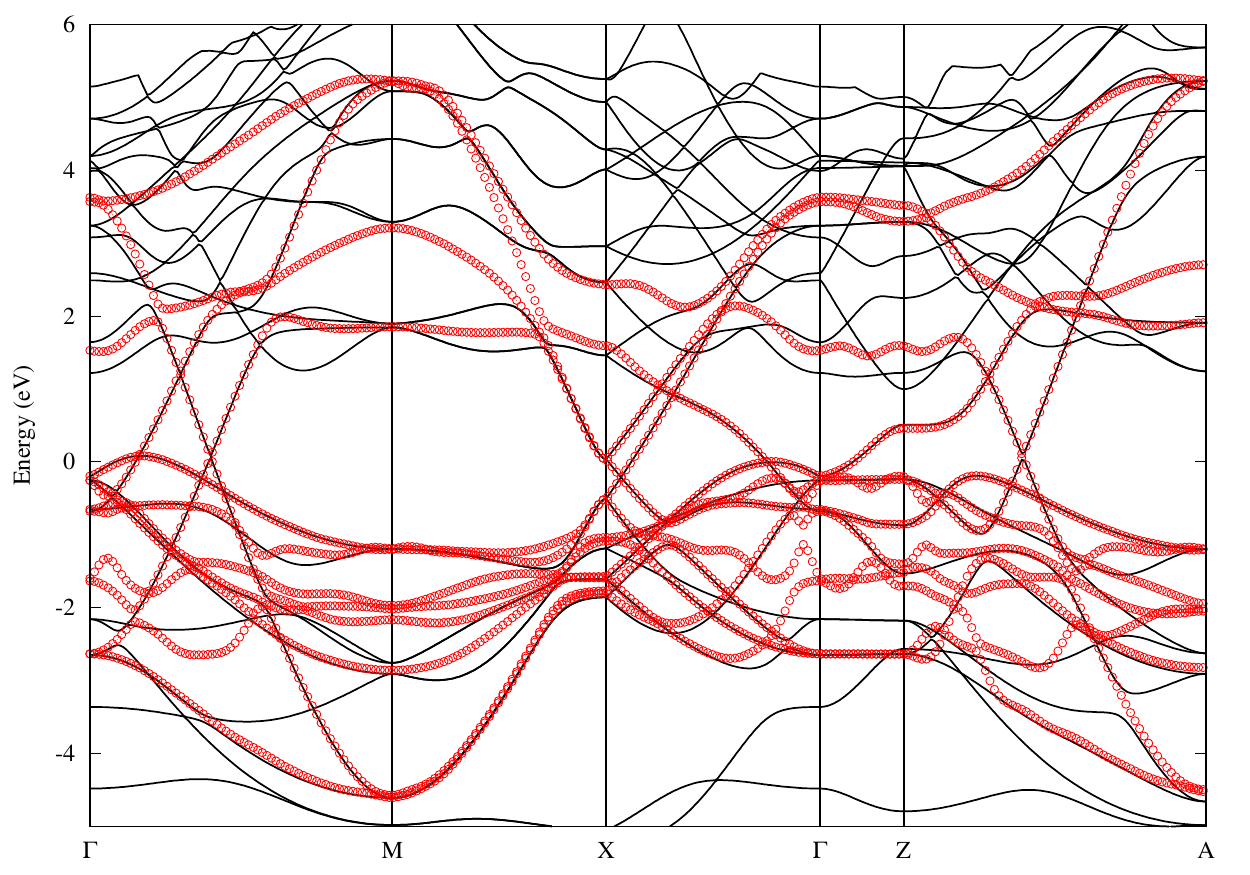}

        } 
            \end{center}
            \caption{ (a) 6$\times$6 and (b)   12$\times$12 Wannier bands (red dots) superimposed onto the DFT bands (black lines). Long range hooping terms are included in the Wannier Hamiltonian to reproduce the bands exactly.
        }
        \label{fig:appen-WannierComparison-6band} 
\end{figure}

In Figs.~\ref{fig:appen-WannierComparison-4band} and ~\ref{fig:appen-WannierComparison-6band}, we compare the eigenvalues obtained from different tightbinding Wannier Hamiltonian forms with DFT calculated bands for the non-magnetic phase.

\subsection{Analytical expression of the eigenvalues for the 4$\times$4 TB Hamiltonian}

\begin{figure*}[htb]
    \begin{center}
            \subfigure[]{
            \includegraphics[width=0.31\textwidth]{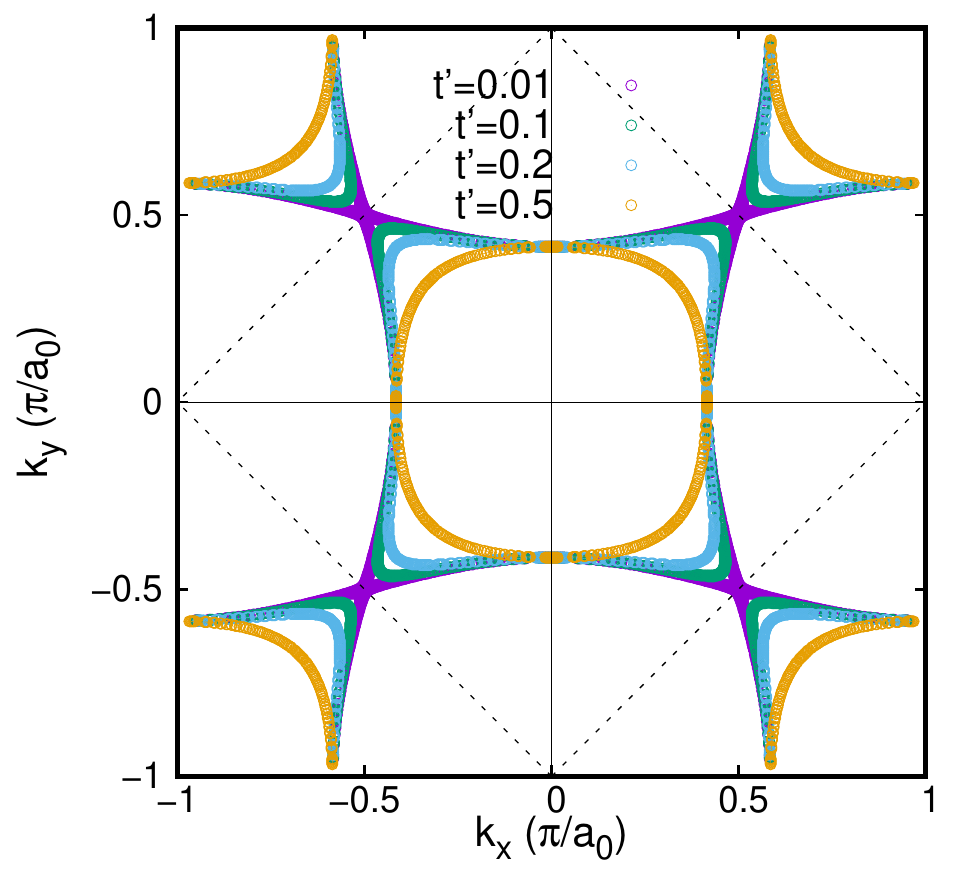}
        }
\hskip -0.05 in
       \subfigure[]{
            \includegraphics[width=0.31\textwidth]{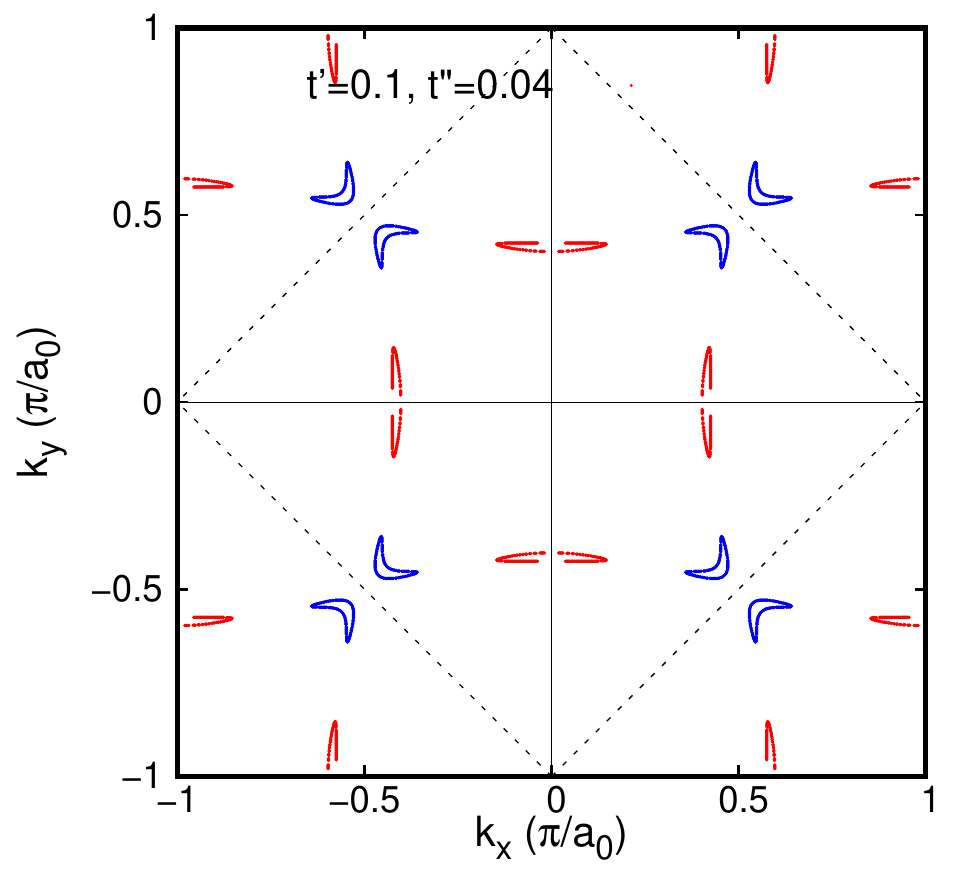}
        } 
\hskip -0.05 in
       \subfigure[]{
            \includegraphics[width=0.31\textwidth]{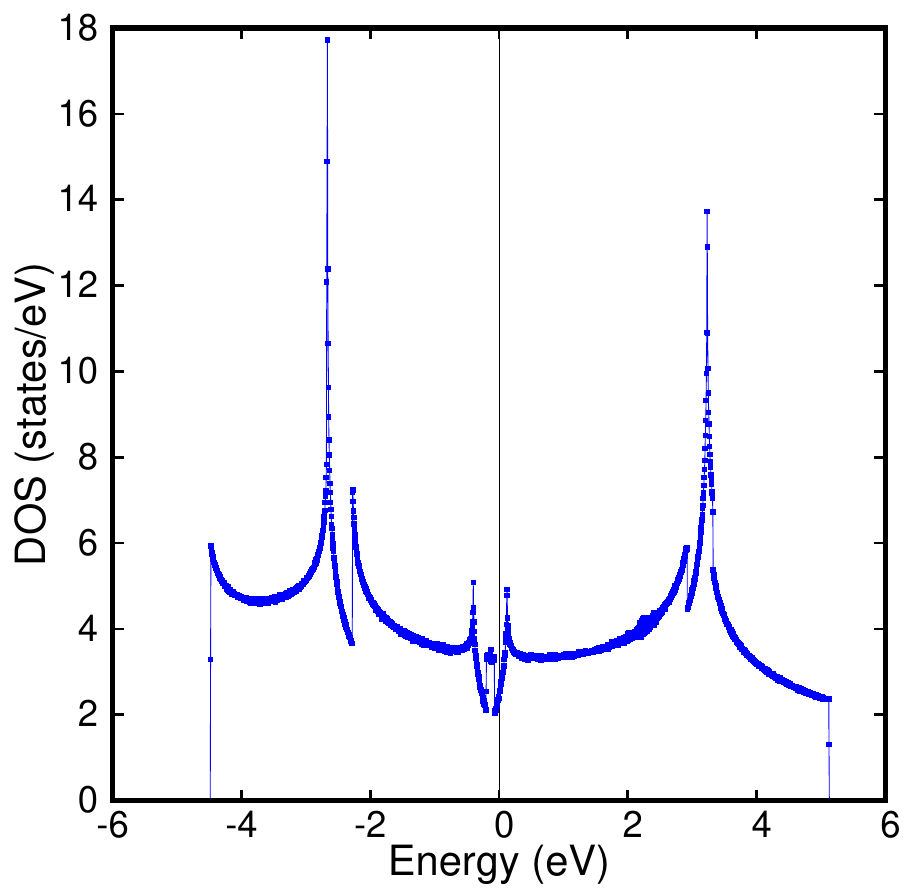}
        } 
            \end{center}
            \caption{ (a)  Fermi surface (FS) of the 4$\times4$ tightbinding Hamiltonian in the extended zone scheme. $t''=0$ at half-filling without SOC. SOC opens gap for this case.  (b) and (c) FS and density of states respectively with SOC for  $t''=0.04$ eV.  The chemical potential is tuned to make the system compensated which corresponds to E$\sim0.014$eV in Fig.(c). 
            The red and blue colored FS pieces  denote hole and  electron pockets in Fig. (b).
            The dashed lines in Fig. (a) and (b) show the reduced Brillouin zone. 
        }
        \label{fig:appen-nodaline_analytical}
\end{figure*}

In this Appendix, we derive  the 4$\times$4 TB Hamiltonian for noninteracting band electrons whose parameters are given in the main text.
For concreteness, we define a $\sqrt{2}\times\sqrt{2}$ square lattice containing 2 atoms in the unit cell with lattice vectors: $\vec{a}=a_0(\hat{x},-\hat{y})$, $\vec{b}=a_0(\hat{x},\hat{y})$, where $a_0$ is the distance between the nearest neighbour atoms of the primitive square lattice unit cell. {\color{blue} Picture?} 
The tight binding Hamiltonian in $\bf{k}$-space for such 2-atom and 2-orbital system without a spin-orbit coupling is:
\begin{eqnarray}
\hat{H_0} = \sum_{\bf{k}}  c^{\dagger}_{\bf{k}} H_0({\bf{k}}) c_{\bf{k}}
\end{eqnarray}
where, $c^{\dagger}_{\bf{k}}= (c^{1^{\dagger}}_{\bf{k}_{px}}, c^{1^{\dagger}}_{\bf{k}_{py}}, c^{2^{\dagger}}_{\bf{k}_{px}}, c^{2^{\dagger}}_{\bf{k}_{py}}) $ such that $c^{1^{\dagger}}_{\bf{k}_{px}}$ creates an  electron at the $p_x$ Wannier orbital located at site 1 and so on. 
Similarly, the Hamiltonian matrix in this basis is given by:
\begin{equation}
H_0({\bf k})=
\begin{pmatrix}
  t''_{\bf k} & t'_{\bf k} & t_{{\bf k}_{\sigma\pi}} &  0 \\
  t'_{\bf k} & t''_{\bf k} & 0 & t_{{\bf k}_{\pi\sigma}}  \\
  t_{{\bf k}_{\sigma\pi}} & 0 & t''_{\bf k} & t'_{\bf k}  \\  
  0 & t_{{\bf k}_{\pi\sigma}} & t''_{\bf k} & t"_{\bf k}\\
\end{pmatrix},
\label{eqn:Ham}
\end{equation}
where we have defined the hopping matrix elements as follows: $t_{{\bf k}_{\sigma \pi}}=-2(t_{\sigma}\cos(k_xa_0)+t_{\pi}\cos(k_ya_0))$, $t_{{\bf k}_{\pi\sigma}}=-2(t_{\pi}\cos(k_xa_0)+t_{\sigma}\cos(k_ya_0))$ and $t'_{\bf k}=4t'\sin(k_xa_0)\sin(k_ya_0), ~~ t''_{\bf k} = 2t''(\cos(2k_xa_0) + \cos(2k_ya_0))$ and $t_{\sigma}=-1.9eV$, $t_{\pi}=0.5eV$, $t'=0.1eV$, and $t''\sim 0.04eV$. $t_{\sigma}$ and $t_{\pi}$ are the nearest-neighbour $\sigma$ and $\pi$ hoppings respectively and $t'$, $t''$ are the next nearest and the 3$^{rd}$ neighbour hopping matrix elements. 
Note that the $t''_{\bf k}$ term, which introduces ${\bf k}$- dependent shift of the bands, is included here for better comparison with the DFT results and does not change the main conclusions.
The eigenvalues  of $H({\bf k})$ are:
\begin{eqnarray}
&& E_0({\bf k}) = t''_{\bf k}\label{eqn:evalsH0}\\
&& \pm \frac{1}{2}\Big\{t_{{\bf k}_{\sigma \pi}} + t_{{\bf k}_{\pi \sigma}} \pm [(t_{{\bf k}_{\sigma \pi}} - t_{{\bf k}_{\pi \sigma}})^2 + 4(t'_{\bf k})^2]^{1/2}\Big\}\nonumber
\end{eqnarray}

{First we discuss the condition when $t''=0$ where particle-hole symmetry is preserved. The situation of 1/2-filling corresponds  to zero chemical potential. Then the Fermi surface is determined by the  zero energy solution of Eq.~(\ref{eqn:evalsH0}): 
\begin{equation}
(t_{{\bf k}_{\sigma \pi}} t_{{\bf k}_{\pi \sigma}} - t'^2_{{\bf k}})=0. 
\label{eqn:zeroenergy}
\end{equation}
This condition gives a diamond-shaped Fermi surface as shown in Fig.~\ref{fig:appen-nodaline_analytical}(a). When the nnn hopping strength is increased, the Fermi surface becomes more circular and shifts away from the zone boundary (X-point).

For $t'=0$, the eigenvalues have a trivial form given by:
\begin{eqnarray}
E_0({\bf k}) &=& \pm   t_{{\bf k}_{\sigma \pi}} \& ~  t_{{\bf k}_{\pi \sigma}}
\end{eqnarray}
When $|k_x|$=$|k_y|$=$k_0$, the eigenvalues are $\pm2(t_{\sigma}+t_{\pi})cos(k_0a_0)$. 
This gives degenerate bands along this line as seen along the $\Gamma$-X and M-X line in Fig.~\ref{fig:appen-WannierComparison-4band}(a).

 We see that in the absence of further interactions we have a metal with a half filled conduction band. Such situation is unstable with respect to unit cell doubling which can gap out (at least partially) the Fermi surface. Such unit cell  doubling can occur already on the single particle level or be a consequence of the interactions. Below we consider the former mechanism first.

The spin orbit coupling (SOC) is of the form 
\begin{equation}
H_{SOC} = \sigma^z(\lambda \tau_y\otimes I + \delta \tau_z\otimes\gamma_z), \label{eq:SOC}
\end{equation}
where $\lambda$ and $\delta$ are constants and $\gamma$, $\sigma$, $\tau$ are the Pauli matrices acting on the site, spin, and orbital indices, respectively.
Such form of SOC does not couple the $|\uparrow\rangle$ and  $|\downarrow\rangle$ spin sectors; hence we can still diagonalize the Hamiltonian analytically.
The eigenvalue for each spin sector in the presence of SOC is given by:
\begin{eqnarray}
&& E_\mathrm{SOC}({\bf k}) = \pm\sqrt{X(\bf k)^2+\delta^2} + t''(\bf k),\\
&& X^2({\bf k}) = \nonumber\\
    && \frac{1}{4}\Big\{t_{{\bf k}_{\sigma \pi}} + t_{{\bf k}_{\pi \sigma}} \pm [(t_{{\bf k}_{\sigma \pi}} - t_{{\bf k}_{\pi \sigma}})^2 + 4(\lambda^2 + t'^2_{\bf k}])^{1/2}\Big\}^2\nonumber.
\label{eqn:evalsH0SOC}
\end{eqnarray}
The estimate is $\delta \approx 0.1$eV. For $t''=0$, the FS is fully gapped at 1/2 filling as discussed in the main text.
However, for $t''\ne 0$ case, the FS is partially gapped.  In Fig.~\ref{fig:appen-nodaline_analytical}(b), we show the FS in the presence of SOC for non-zero $t''$ at the value of chemical potential when the system is fully compensated. As seen in the figure, there are electron and hole pockets at the X point and along $\Gamma$-M direction, respectively, similar to the FS of real system with small DOS at the Fermi level [Fig.~\ref{fig:appen-nodaline_analytical}(c)].

\section{Derivation of the exchange Hamiltonian}
\label{appendixB}

\begin{figure*}[htb]
    \begin{center}
     \subfigure[]{
            \includegraphics[width=0.45\textwidth]{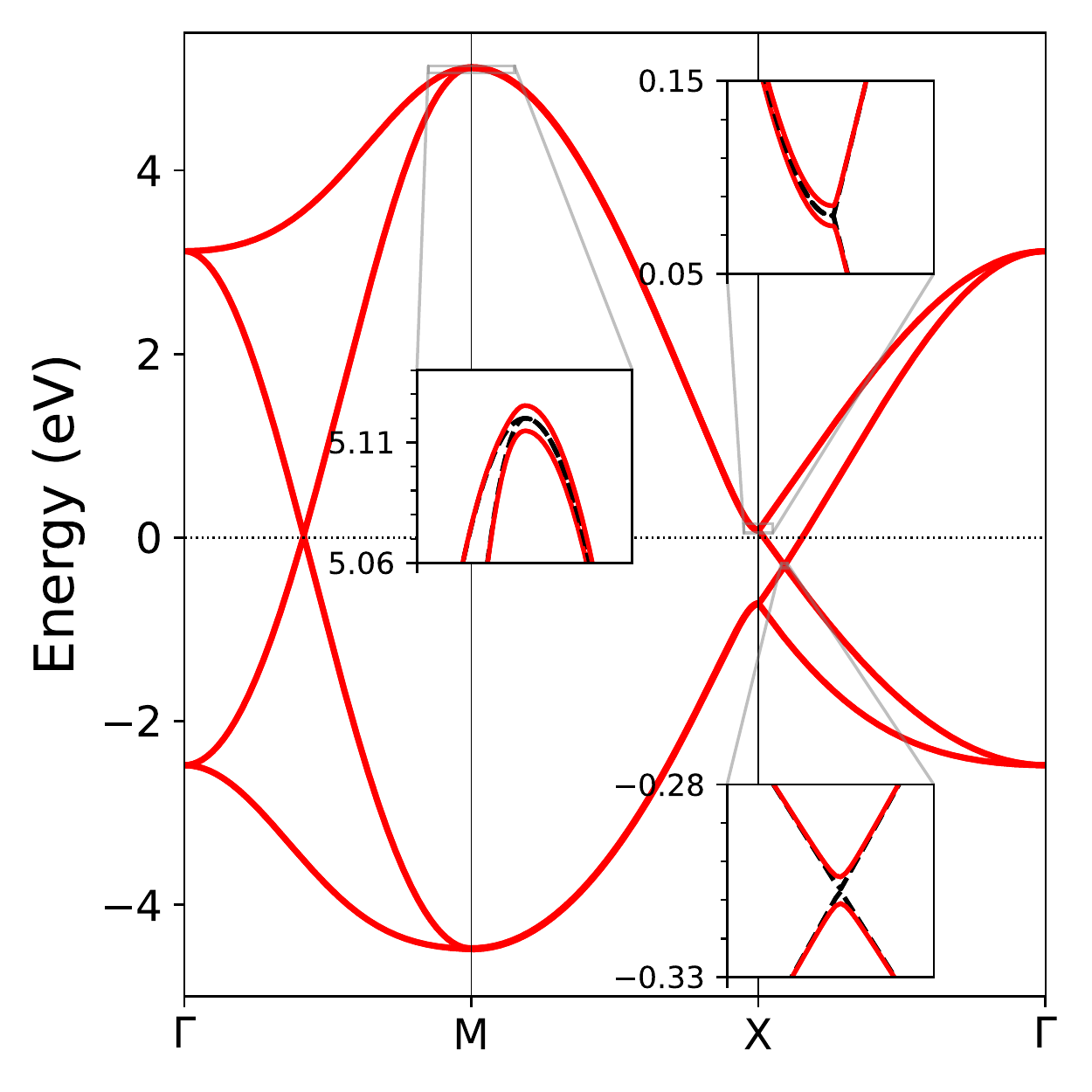}
        }
     \subfigure[]{
            \includegraphics[width=0.45\textwidth]{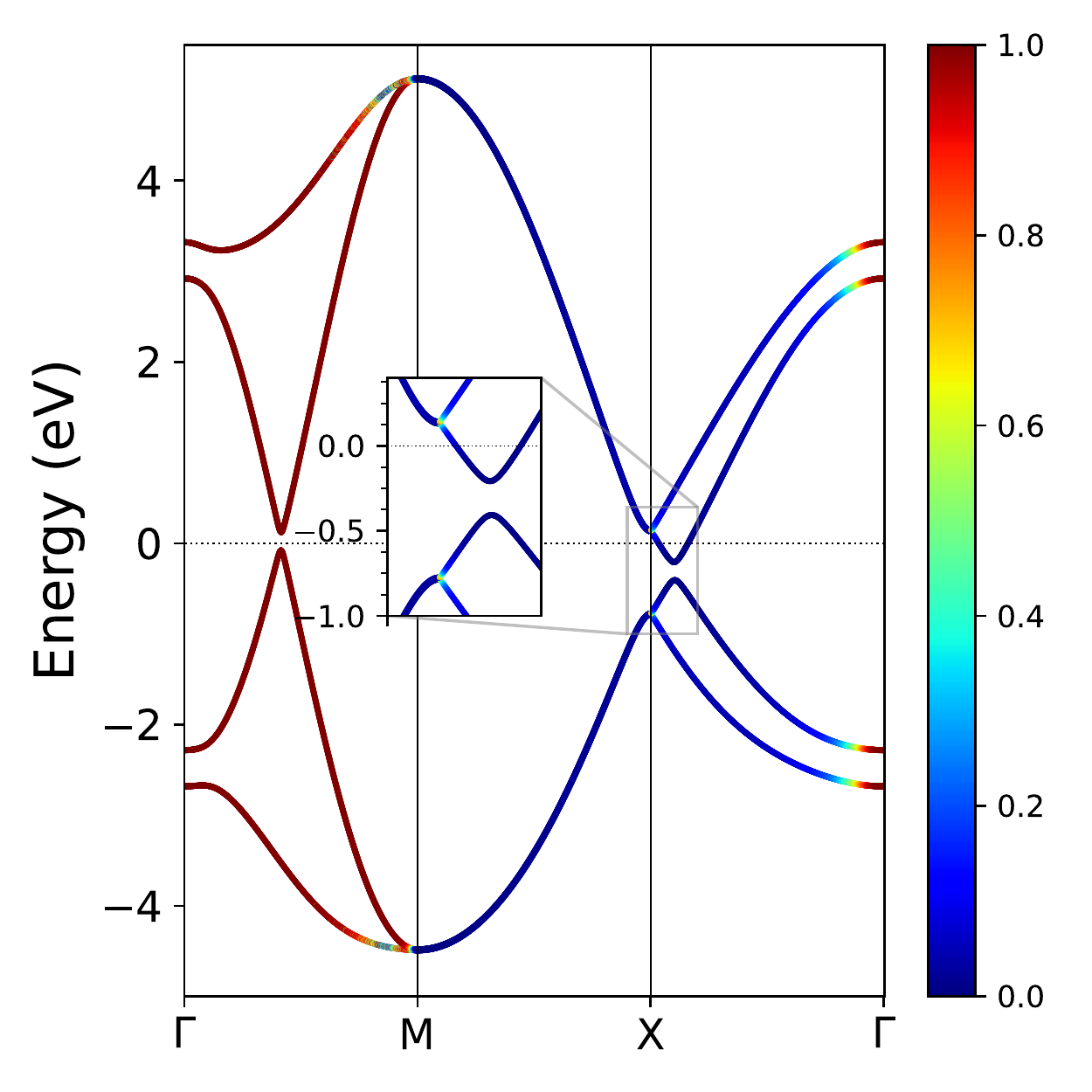}
        }
    \end{center}
            \caption{ A band structure of the nnn 4$\times$4 tight binding Hamiltonian in the presence of the effective Kondo exchange Hamiltonian for the AFM arrangement of the Eu-$f$ spins. Fig. (a) shows a comparison with the eigenvalues without the exchange Hamiltonian (black dots) at zero spin-orbit coupling (SOC).
            Notice that the Kondo exchange splits the band degeneracies by $\sim J_K<S>$  at the M, X and along  X-$\Gamma$ direction. The crossing along $\Gamma-M$ is preserved by Kondo exchange in the absence of SOC.
            The dispersion is identical for any orientation of the Neel magnetization. 
            (b) Band dispersion in the presence of Kondo exchange with SOC when the spin quantization axis is along the \xhat~ direction. The color palette shows the magnitude of the $\langle\hat{\sigma_z}\rangle$.  
        }
        \label{fig:appen-bandswex}
\end{figure*}

In order to understand how the Eu-4$f$ states may affect the band structure and the Fermi surface properties, we have also derived an effective Kondo exchange Hamiltonian  within the second order perturbation theory in $V$. The 4$f$-shell of Eu ions is half filled. The Hund's rule dictates that at the ground state Eu ion has  zero  angular momentum and the total spin is S=7/2. Then according to Ref.~\onlinecite{SchriefferKondo1967}, the effective Kondo Hamiltonian for an Eu ion located at the origin is the multichannel Kondo model. For a spherically symmetric case it would have the following form:
\begin{eqnarray}
H_{exchange} = J_K\sum_{m}c^{\dagger}_{m\alpha}(|k|)({\vec\sigma}_{\alpha\beta}\vec S)c_{m\beta}(|k|),
\end{eqnarray}
where $m$ are the electron orbital quantum numbers. In the limit when the Hund's coupling exceeds the Kondo temperature ~\cite{SchriefferKondo1967}, the Kondo coupling is given by:
\begin{equation}
    J_K = \frac{V^2}{2S}\Big[-1/\epsilon_f + 1/(\epsilon_f+U)\Big],
\end{equation}
 with $V$ being the overall scale of the hybridization strength between the Eu-$f$ and $p$ electrons and $\epsilon_f$ being the energy levels of the $f$ states. We estimate $J_K \sim 16$K. 
 

 The exchange Hamiltonian is derived under assumption  that the hybridization matrix elements are equal for all orbitals involved in the screening of the spin. For the model on a lattice the latter assumption no longer holds. One has   to recalculate the spherical orbitals into the Wannier functions basis, $m$-th orbital will have its own exchange $\sim |V_m|^2$ which will lead to  exchange anisotropy.

\begin{table*}[!tht]
\caption{\label{table:Hybridization}
Hybridization matrix element (in eV) between the occupied $|Euf\sigma\rangle$ and the $|p\sigma\rangle$-states from the Sb2 square nets in the antiferromagnetic configuration without SOC obtained from the Wannier function analysis.
For Eu1 atoms, the $|\uparrow\rangle$ states are occupied whereas for Eu2 atoms, $|\downarrow\rangle$ states are occupied.
The matrix elements between the $|p^2\sigma\rangle$ and $|Eu\sigma\rangle$ states are similar with $p_x \leftrightarrow  p_y$. The sign of the matrix elements are fixed by the symmetry of the corresponding hopping integral.
Note that the $p$-orbitals in this table are aligned along Eu-atoms and vice-versa.
}
\begin{center}
\begin{tabular}{p{65pt}p{55pt}p{55pt}p{55pt}p{55pt}p{55pt}p{55pt}p{55pt}} 
\hline\hline
 $\langle WFs|H|WFs\rangle$ & Eu1-$f_{xz^2}\uparrow$ & Eu1-$f_{yz^2}\uparrow$  & Eu1-$f_{z^3}\uparrow$ & Eu1-$f_{x(x^2-3y^2)}\uparrow$ & Eu1-$f_{y(3x^2-y^2)}\uparrow$ & Eu1-$f_{z(x^2-y^2)}\uparrow$ & Eu1-$f_{xyz}\uparrow$\\ \hline 
   $p_x^1\uparrow$ & -0.13 & 0 & -0.05 & -0.03 & 0 & -0.11 & 0.0 \\ 
   $p_y^1\uparrow$ & 0 & 0.01 & 0 & 0 & 0.07 & 0 & 0.07 \\ 
   $p_z^1\uparrow$ & -0.11 & 0 & 0.06 & -0.06 & 0 & -0.13 & 0\\ 
   \hline
 $\langle WFs|H|WFs\rangle$ & Eu2-$f_{xz^2}\downarrow$ & Eu2-$f_{yz^2}\downarrow$  & Eu2-$f_{z^3}\downarrow$ & Eu2-$f_{x(x^2-3y^2)}\downarrow$ & Eu2-$f_{y(3x^2-y^2)}\downarrow$ & Eu2-$f_{z(x^2-y^2)}\downarrow$ & Eu2-$f_{xyz}\downarrow$\\ \hline
   $p_x^1\downarrow$ & 0.01 & 0 & 0 & -0.07 & 0 & 0 & 0.07 \\ 
   $p_y^1\downarrow$ & 0 & -0.13 & -0.05 & 0 & 0.03 & 0.11 & 0 \\ 
   $p_z^1\downarrow$ & 0 & -0.11 & 0.06 & 0 & 0.06 & 0.13 & 0 \\ 
\hline\hline 
\end{tabular}
\end{center}
\end{table*}

By using the hybridization matrix elements between the localized Eu-4$f$  and the itinerant  $p_x$-$p_y$-orbitals from the \textit{ab-initio} calculations as shown in Table. ~\ref{table:Hybridization}, we obtain the following effective Kondo exchange Hamiltonian in the basis of the $p$ orbitals $c^{\dagger}_{\bf k \sigma} = (c^{1^{\dagger}}_{\bf{k}_{px}\sigma}, c^{1^{\dagger}}_{\bf{k}_{py}\sigma}, c^{2^{\dagger}}_{\bf{k}_{px}\sigma}, c^{2^{\dagger}}_{\bf{k}_{py}\sigma})$:
\begin{eqnarray}
&& H^{ex}_{\sigma\sigma'} = J_K\sum_{k,k', R_1} c^+_{{\bf k}\sigma} \hat g_+(\bf k,\bf k') ({\vec\s}\vec S_{\bf R_1})_{\sigma,\sigma'}\re^{-\ri({\bf k}-{\bf k}'){\bf R}_1} c_{{\bf k}'{\sigma'}} + \nonumber\\
    && J_K\sum_{k,k', R_2}c^+_{{\bf k}{\sigma}} \hat g_-(\bf k,\bf k') ({\vec\s}\vec S_{\bf R_2})_{\sigma \sigma'}\re^{-\ri({\bf k}-{\bf k}'){\bf R}_2} c_{{\bf k}'{\sigma'}},
\label{eq:exchange}
\end{eqnarray}
where, $\vec S_{\bf R_1}$ and  $\vec S_{\bf R_2}$ are the spins localized on the $\bf R_1$ and $\bf R_2$ sites. $g_+$ and $g_-$ are the effective exchange matrices given by:
\bea
&& g_+(\bf k,\bf k') = \left(
\begin{array}{cccccc}
g_1  & -g_2& g_3-g_4 & -g_3 -g_4\\
-g_2 & g_1 & g_3+g_4 & -g_3+g_4 \\
h.c. & h.c. & g_1 & g_2\\
h.c. & h.c. & g_2 & g_1
\end{array}
\right)\nonumber,
\eea
\bea
&& g_-(\bf k,\bf k') = \left(
\begin{array}{cccccc}
g_1  & g_2 & g_3-g_4^* & g_3 + g_4^*\\
g_2 & g_1 & -g_3-g_4^* & -g_3 +g_4^* \\
h.c. & h.c. & g_1 & -g_2\\
h.c. & h.c. & -g_2 & g_1
\end{array}
\right)\nonumber,\\
\eea
and, $g_1 = 1, ~~ g_2 = 1/2, ~~ g_3=A s_x^*(\bf k)s_y^*(\bf k')$, $g_4=B s_x^*(\bf k)s_y(\bf k')$, 
 $s_a =1-e^{-ik_a}$,  
$A = 1/16, ~~ B = 3/16$. 

{\bf The role of the Eu magnetic moments}--
  The interaction between conduction electrons and magnetic moments may lead to many different effects. Depending on how strong is the coupling between these two subsystems, the net result many be very different. Below we will consider two scenario. 
  
 In both cases we will treat the Eu spins in the mean field approximation, that is  as classical vectors $\vec S = \pm <S>{\bf n}$, where  ${\bf n}$ is a unit vector and $<S> <7/2$. 
 
 First, we consider the case with zero SOC. Then, we will discuss the case with  strong SOC $\delta$ (we will formulate the precise criterion later). In the latter case, the band electrons are {partially gapped with small density of states at the Fermi level and the influence of the spin order on their spectrum is small. The nature of the spin order is determined by the interaction between the spins through the conduction electrons.  

 Consider spins  with opposite directions on ${\bf R}_1$ and ${\bf R}_2$ sites: $\vec S = \pm <S>{\bf n}$, where ${\bf n}$ is a unit vector and $<S> <7/2$ is the average magnetization which has to be determined self-consistently. Then the Kondo exchange (\ref{eq:exchange}) becomes an effective modification of the spin-orbit coupling:
\begin{eqnarray}
&& H^{ex}_{MF} = 2J_K<S> \sum_{k}c^{\dagger}_{\bf k}(\vec\sigma\vec n)[g_2\tau^x\otimes\gamma^z + \Im m g_4\tau^z\otimes\gamma^y \nonumber \\
&&+ \Im m g_3 \tau^y\otimes\gamma_x - (\Re e g_3 + \Re e g_4) \tau^y\otimes\gamma^y]c_{\bf k}. \label{eq:exMF}
\end{eqnarray}
In Fig.~\ref{fig:appen-bandswex}(a), we show the band dispersion of the $p$-electrons in the presence of the Kondo exchange term for zero SOC.
The antiferromagnetic order splits the band dispersion at several places highlighted in the figure insets by magnitude of $\sim J_K<S>$.

In order to study the differences in the spectrum due to the spin orientation of the \Neel vector, we will need to combine (\ref{eq:SOC}) and (\ref{eq:exMF}).
Close to the X-point ($\pi/2,\pi/2)$, $t_{\sigma\pi} \sim t_{\pi\sigma} \sim 0$ i.e. there is no coupling between the two sublattices and one can derive the effective Hamiltonian for just one of the sublattice:
\begin{eqnarray}
&& H_1 = H_2 = \mathcal{O}(J_K^2<S>^2) + t''_{\bf k} \nonumber\\
&&\tau^x (t'_{\bf k} -2J_K<S>g_2\sigma^z) +(\vec\sigma\vec n) (\delta\tau^z + \lambda\tau^y).
\end{eqnarray}

If the Neel magnetization is along the $z$-axis, the spectrum is:
\begin{eqnarray}
&& E = \mathcal{O}(J_K^2<S>^2) + t''_{\bf k} \pm \nonumber\\
&& [\delta^2 +\lambda^2 +(t'_{\bf{k}} -2J_K<S>g_2\sigma^z )^2]^{1/2}.
\end{eqnarray}
If it is perpendicular to $z$-axis then,
\begin{eqnarray}
&& E = \mathcal{O}(J_K^2<S>^2) + t''_{\bf k} \pm \nonumber\\
&& \Big(\sqrt{t'^2_{\bf k} +\lambda^2 +\delta^2} \pm 2J_K<S>g_2\Big).
\end{eqnarray}

With our estimates of $J_K$ and $\delta$, we conclude that the first term can be neglected.  
Thus, we find that there is a small effect in the spectrum due to the rotation of the Eu spins which nevertheless remains partially gapped due to the SOC. More importantly, 
if ${\bf n}$ is directed along the $\hat{z}$-axis, the $z$-projection of electron spin is a good quantum number. For all other directions it is not. 
This can be seen from the spin-texture of the bands for the \xhat-phase along the high symmetry directions in Fig.~\ref{fig:appen-bandswex}(b).
Such differences in the spin-texture between the \zhat~ and \xhat~phases can introduce differences in the spin transport properties like spin Hall conductivity which is proportional to the matrix element of the spin operator.

\section{Electronic dispersion and spin Hall conductivity results for magnetic phases}
\label{appendixC}

\subsection{DFT Bands and DOS }
Fig.~\ref{fig:appen-bandsdos_soz} shows the DFT calculated electronic bands and DOS for AAF-\zhat~ magnetic phase in the presence of SOC and U of 6 eV.

\begin{figure*}[htb]
    \begin{center}
            \includegraphics[width=0.90\textwidth]{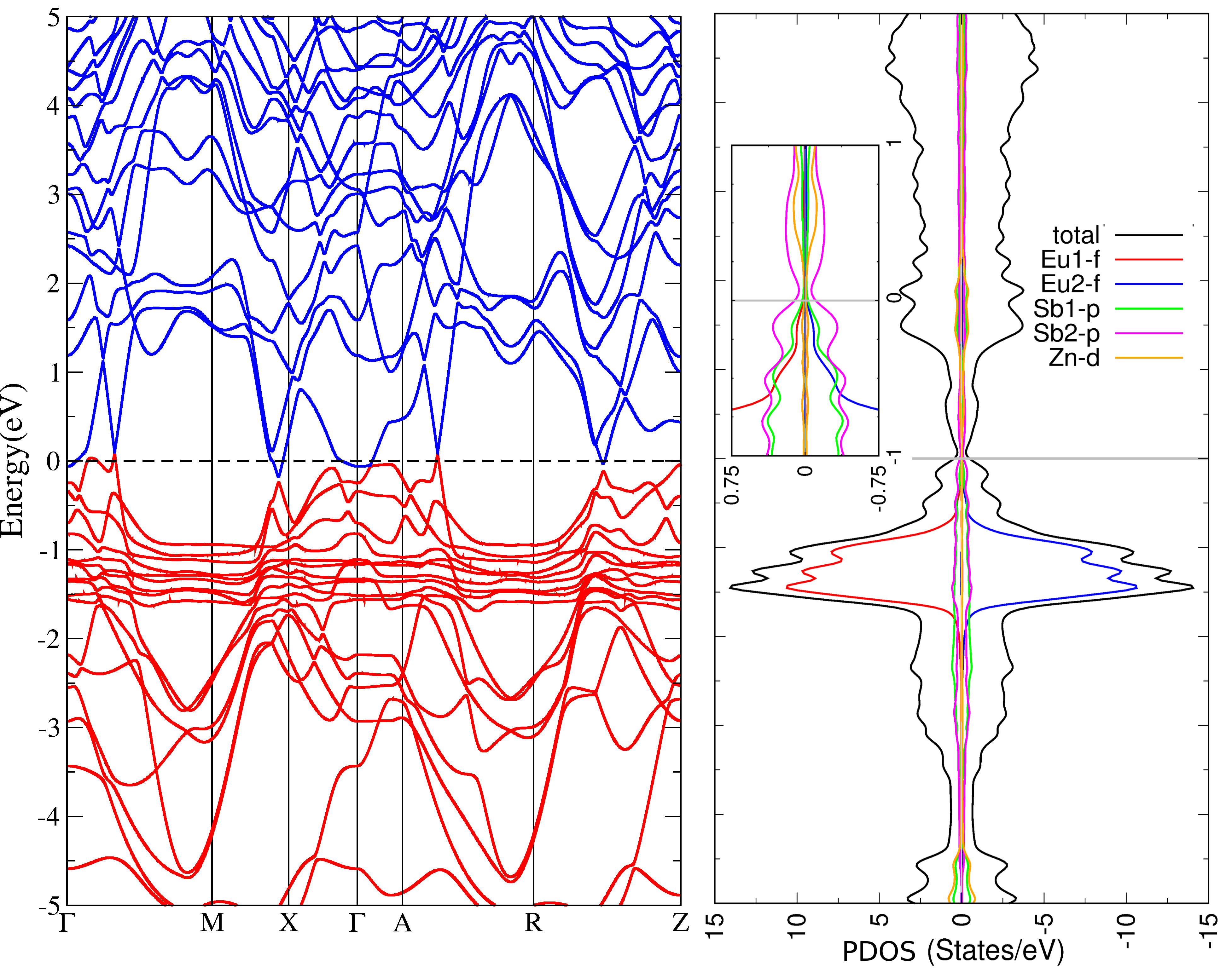}
            \end{center}
            \caption{ DFT bands and projected density of states (PDOS) for AAF-$\hat{z}$ magnetic pattern. Positive (negative) PDOS values indicate up (down) states. The minority spin states for AFM aligned Eu-4$f$ electrons are $\sim$10 eV above E$_F$, hence not shown here. The small DOS at  E$_F$ comes from Sb2 $p_x$-$p_y$ orbitals. 
        }
        \label{fig:appen-bandsdos_soz} 
\end{figure*}

\subsection{Comparison with other components of the Spin hall conductivity}
In the main text, we showed the comparison between the spin hall conductivity (SHC) response as a function of the chemical potential between the \xhat~ and \zhat~ phase for just one component of the SHC tensor. Here, we also show the comparison with other components of the SHC tensor.
\begin{figure*}[htb]
    \begin{center}
        \subfigure[]{
            \includegraphics[width=0.31\textwidth]{fig8a.pdf}
        } 
\hskip -0.10 in
        \subfigure[]{
            \includegraphics[width=0.31\textwidth]{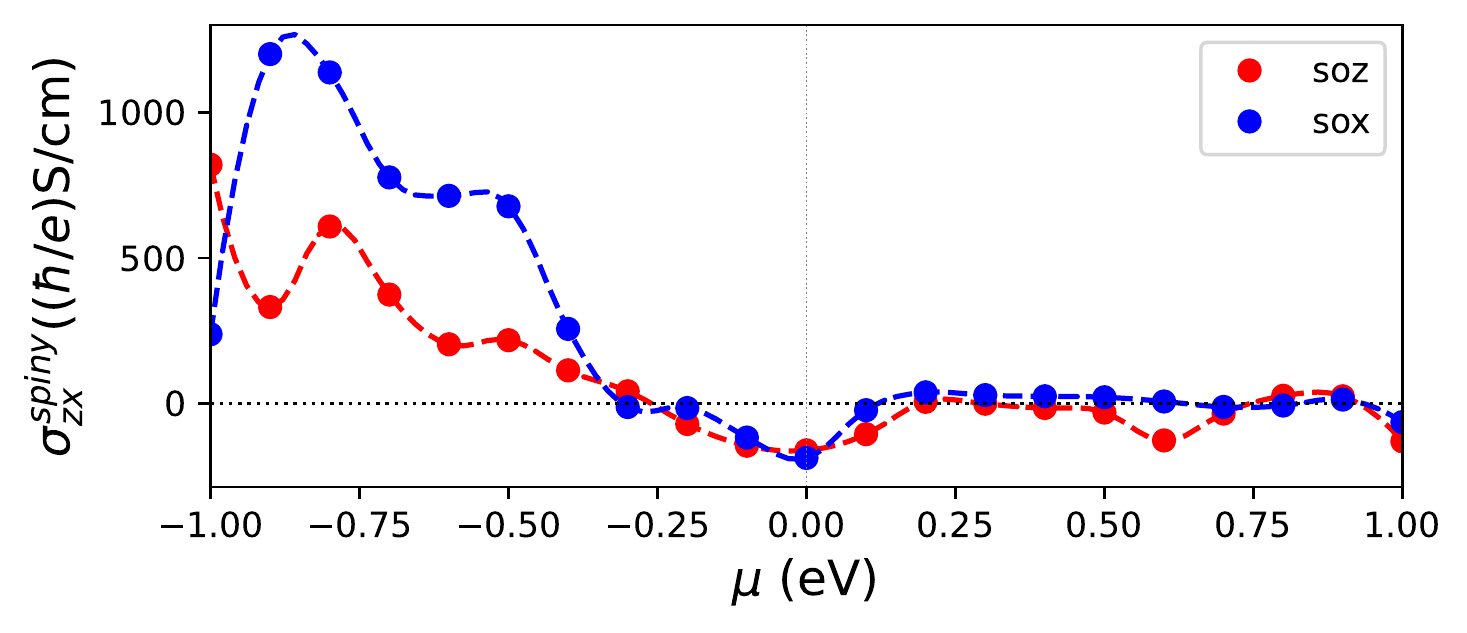}
         }
\hskip -0.10 in
        \subfigure[]{
            \includegraphics[width=0.31\textwidth]{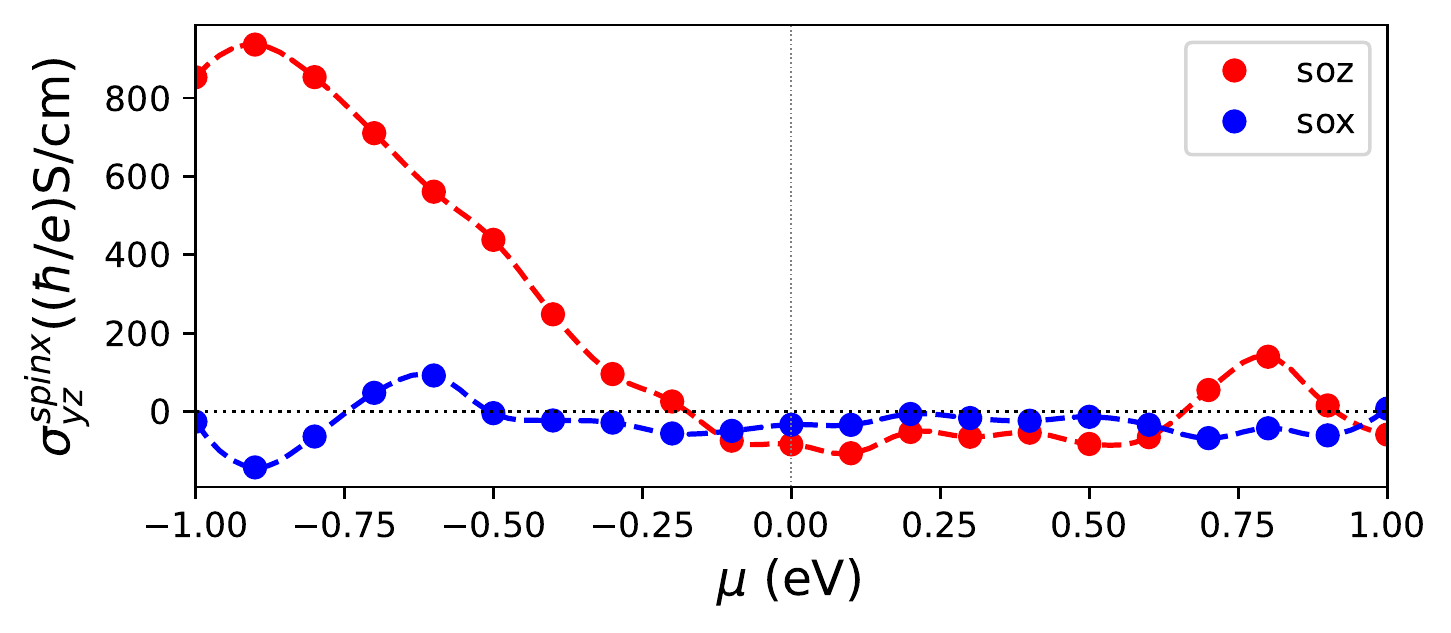}
         }
     \end{center}
            \caption{Comparison of the (a) $\sigma_{xy}^z$, (b) $\sigma_{yz}^x$ and (c) $\sigma_{zx}^y$component of the SHC tensor as a function of the chemical potential between the AAF-x and z patterns.
        }
        \label{fig:appen-AAF-SHC} 
\end{figure*}

\end{document}